**A genomic dominion with regulatory dependencies on human-specific single-nucleotide changes in Modern Humans**

Gennadi V. Glinsky[1]

[1] Institute of Engineering in Medicine

University of California, San Diego

9500 Gilman Dr. MC 0435

La Jolla, CA 92093-0435, USA

Correspondence: gglinskii@ucsd.edu

Web: http://iem.ucsd.edu/people/profiles/guennadi-v-glinskii.html

**Running title:** Human-specific regulatory control of an offspring survival genomic dominion

**Abstract**

Gene set enrichment analyses of 8,405 genes linked with 35,074 human-specific (hs) regulatory single-nucleotide changes (SNCs) revealed the staggering breadth of significant associations with morphological structures, physiological processes, and pathological conditions of Modern Humans. Significant enrichment traits include more than 1,000 anatomically-distinct regions of the adult human brain, many different types of human cells and tissues, more than 200 common human disorders and more than 1,000 records of rare diseases. Thousands of genes connected with regulatory hsSNCs have been identified in this contribution, which represent essential genetic elements of the autosomal inheritance and survival of species phenotypes: a total of 1,494 genes linked with either autosomal dominant or recessive inheritance as well as 2,273 genes associated with premature death, embryonic lethality, as well as pre-, peri-, neo-, and post-natal lethality of both complete and incomplete penetrance. Therefore, thousands of heritable traits and critical genes impacting the offspring survival appear under the human-specific regulatory control in genomes of Modern Humans. These observations highlight the remarkable translational opportunities afforded by the discovery of genetic regulatory loci harboring hsSNCs that are fixed in humans, distinct from other primates, and located in differentially-accessible (DA) chromatin regions during human brain development.

## Introduction

DNA sequences of coding genes defining the structure of macromolecules comprising the essential building blocks of life at the cellular and organismal levels remain highly conserved during the evolution of humans and other Great Apes (Chimpanzee Sequencing and Analysis Consortium, 2005; Kronenberg et al., 2018). In striking contrast, a compendium of nearly hundred thousand candidate human-specific regulatory sequences (HSRS) has been assembled in recent years (Glinsky et al., 2015-2019; Kanton et al., 2019), thus validating the idea that unique to human phenotypes may result from human-specific changes to genomic regulatory sequences defined as “regulatory mutations” (King and Wilson, 1975). The best evidence of the exquisite degree of accuracy of the contemporary molecular definition of human-specific regulatory sequences is identification of 35,074 single nucleotide changes (SNCs) that are fixed in humans, distinct from other primates, and located within differentially-accessible (DA) chromatin regions during the human brain development in cerebral organoids (Kanton et al., 2019). However, only a small fraction of identified DA chromatin peaks (600 of 17,935 DA peaks; 3.3%) manifest associations with differential expression in human versus chimpanzee cerebral organoids model of brain development, consistent with the hypothesis that regulatory effects on gene expression of these chromatin regions are not restricted to the early stages of brain development. Annotation of SNCs derived and fixed in modern humans that overlap DA chromatin regions during brain development revealed that essentially all candidate regulatory human-specific SNCs are shared with the archaic humans (35,010 SNCs; 99.8%) and only 64 SNCs are unique to modern humans (Kanton et al., 2019). This remarkable conservation on the human lineage of human-specific SNCs associated with human brain development sows the seed of interest for in-depth exploration of coding genes expression of which may be affected by genetic regulatory loci harboring human-specific SNCs. The GREAT algorithm (McLean et al., 2010, 2011) was utilized to identify 8,405 genes expression of which might be affected by 35,074 human-specific SNCs located in DA chromatin regions during brain development. Comprehensive gene set enrichment analyses of these genes revealed the staggering breadth of associations with physiological processes and pathological conditions of *H. sapiens*, including more than 1,000 anatomically-distinct regions of the adult human brain, many human tissues and cell types, more than 200 common human disorders and 1,116 rare diseases.

## Results and discussion

### Identification and characterization of putative genetic regulatory targets associated with human-specific single nucleotide changes (SNCs) in in differentially accessible (DA) chromatin regions during brain development

To identify and characterize human genes associated with 35,074 human-specific single nucleotide changes (SNCs) in differentially accessible (DA) chromatin regions during human and chimpanzee neurogenesis in cerebral organoids (Kanton et al., 2019), the GREAT algorithm (McLean et al., 2011) have been employed. These analyses identified 8,405 genes with putative regulatory connections to human-specific SNCs (Figure 1) and revealed a remarkable breadth of highly significant associations with a multitude of biological processes, molecular functions, genetic and metabolic pathways, cellular compartments, and gene expression perturbations (Supplemental Table Set S1).

It has been noted that particularly striking numbers of significant associations were uncovered by the GREAT algorithm during the analyses of two databases:

1) The Human Phenotype Ontology containing over 13,000 terms describing clinical phenotypic abnormalities that have been observed in human diseases, including hereditary disorders (326 significant records with binominal FDR Q-Value < 0.05);

2) The MGI Expression Detected ontology referencing genes expressed in specific anatomical structures at specific developmental stages (Theiler stages) in the mouse (370 significant records with binominal FDR Q-Value < 0.05).

These observations support the hypothesis that biological functions of genes under the putative regulatory control of human-specific SNCs in DA chromatin regions during brain development are not limited to the contribution to the early stages of neuro- and corticogenesis. Collectively, findings reported in the Supplemental Table Set S1 argue that genes expression of which is affected by human-specific SNCs may represent a genomic dominion of putative regulatory dependency from HSRS that is likely to play an important role in a broad spectrum of physiological processes and pathological conditions of Modern Humans.

**Identification of genes expression of which distinguishes thousands of anatomically distinct areas of the adult human brain, various regions of the central nervous system, and many different cell types and tissues in the human body**

To validate and extend these observations, next the comprehensive gene set enrichment analyses were performed employing the web-based Enrichr API protocols (Chen et al., 2013; Kuleshov et al., 2016), which interrogated nearly 200,000 gene sets from more than 100 gene set libraries. The results of these analyses are summarized in the Table 1 and reported in details in the Supplemental Table Set S2. Genes expression of which were placed during evolution under the regulatory control of ~ 35,000 human-specific SNCs demonstrate a staggering breadth of significant associations with a broad spectrum of anatomically distinct regions, various cell and tissue types, a multitude of physiological processes, and a numerous pathological conditions of *H. sapiens*.

Of particular interest is the apparent significant enrichment of human-specific SNCs-associated genes among both up-regulated and down-regulated genes, expression of which discriminates thousands of anatomically distinct areas of the adult human brain defined in the Allen Brain Atlas (Figure 2; Supplemental Table Set S2). Notably, genes expressed in various thalamus regions appear frequently among the top-scored anatomical areas of the human brain (Figure 2; Supplemental Table Set S2). These observations support the hypothesis that genetic loci harboring human-specific SNCs may exert regulatory effects on structural and functional features of the adult human brain, thus, likely affecting the development and functions of the central nervous system in Modern Humans. Consistent with this idea, the examination of the enrichment patterns of human-specific SNCs-associated genes in the ARCHS4 Human Tissues' gene expression database revealed that top 10 most significantly enriched records overlapping a majority of region-specific marker genes constitute various anatomically-distinct regions of the central nervous system (Figure 2; Supplemental Table Set S2). However, results of gene set enrichment analyses convincingly demonstrate that inferred regulatory effects of genetic loci harboring human-specific SNCs are not restricted only to the various regions of the central nervous system, they appear to affect gene expression profiles of many different cell types and tissues in the human body (Table 1; Supplemental Table Set S2).

**Identification and characterization of genes expression of which is altered during aging of humans, rats, and mice**

Genes altered expression of which is implicated in the aging of various tissues and organs of humans, rats, and mice are significantly enriched among 8,405 genes associated with human-specific regulatory SNCs (Figure 3; Supplemental Table Set S2). Aging of the hippocampus was implicated most frequently among genes manifesting increased expression with age, while among genes exhibiting aging-associated decreased expression the hippocampus and frontal cortex were identified repeatedly (Figure 3). Overall, twice as many significant association records were observed among aging-associated down-regulated genes compared to up-regulated genes (Table 1). Collectively, these observations indicate that genes changes in expression of which were associated with aging in mammals, in particular, hippocampal and frontal cortex aging, represent important elements of a genomic dominion that was placed under regulatory control of genetic loci harboring human-specific SNCs.

**Identification of genes implicated in development and manifestations of hundreds physiological and pathological phenotypes and autosomal inheritance in Modern Humans**

Interrogations of the Human Phenotype Ontology database (298 significantly enriched records identified), the Genome-Wide Association Study (GWAS) Catalogue (241 significantly enriched records identified), and the database of Human Genotypes and Phenotypes (136 significantly enriched records identified) revealed several hundred physiological and pathological phenotypes and thousands of genes manifesting significant enrichment patterns defined at the adjusted p value $< 0.05$ (Figure 4; Table 1; Supplemental Table Set S2). Interestingly, 645 and 849 genes implicated in the autosomal dominant (HP:0000006) and recessive (HP:0000007) inheritance were identified amongst genes associated with human-specific regulatory SNCs (Figure 4; Supplemental Table Set S2). Notable pathological conditions among top-scored records identified in the database of Human Genotypes and Phenotypes are stroke, myocardial infarction, coronary artery disease, and heart failure (Figure 4).

A total of 241 significantly enriched records (Table 1) were documented by gene set enrichment analyses of the GWAS catalogue (2019), among which a highly diverse spectrum of pathological conditions linked to genes associated with human-specific regulatory SNCs was identified, including obesity, type 2 diabetes, amyotrophic lateral sclerosis, autism spectrum disorders, attention deficit hyperactivity disorder, bipolar disorder, major depressive disorder, schizophrenia, Alzheimer's disease, malignant melanoma, diverticular disease, asthma, coronary artery disease, glaucoma, as well as breast, prostate and colorectal cancers (Figure 4; Supplemental Table Set S2). These observations indicate that thousands of genes putatively associated with genetic regulatory loci harboring human-specific SNCs affect risk of developing numerous pathological conditions in Modern Humans.

**Identification of genes expression of which is altered in several hundred common human disorders**

Gene set enrichment analyses-guided interrogation of the Gene Expression Omnibus (GEO) database revealed the remarkably diverse spectrum of human diseases with the etiologic origins in multiple organs and tissues and highly heterogeneous pathophysiological trajectories of their pathogenesis (Figure 5; Supplemental Table Set S2). Overlapping gene sets between disease-associated genes and human-specific SNCs-linked genes comprise of hundreds of genes that were either up-regulated (204 significant disease records) or down-regulated (240 significant disease records) in specific pathological conditions, including schizophrenia, bipolar disorder, various types of malignant tumors, Crohn's disease, ulcerative colitis, Down syndrome, Alzheimer's disease, spinal muscular atrophy, multiple sclerosis, autism spectrum disorders, type 2 diabetes mellitus, morbid obesity, cardiomyopathy (Figure 5; Supplemental Table Set S2). These observations demonstrate that thousands of genes expression of which is altered in a myriad of human diseases appear associated with genetic regulatory loci harboring human-specific SNCs.

**Identification of genes implicated in more than 1,000 records classified as human rare diseases**

Present analyses demonstrate that thousands of genes associated with human-specific regulatory SNCs have been previously identified as genetic elements affecting the likelihood of development a multitude of common human disorders. Similarly, thousands of genes expression of which is altered during development and manifestation of multiple common human disorders appear linked to genetic regulatory loci harboring human-specific SNCs. Remarkably, interrogations of the Enrichr's libraries of genes associated with Modern Humans' rare diseases identified 473, 603, 641, and 1,116 significantly enriched records of various rare disorders employing the Rare Diseases GeneRIF gene lists library, the Rare Diseases GeneRIF ARCHS4 predictions library, the Rare Diseases AutoRIF ARCHS4 predictions library, and the Rare Diseases AutoRIF Gene lists library, respectively (Figure 6; Supplemental Table Set S2). Taken together, these observations demonstrate that thousands of genes associated with hundreds of human rare disorders appear linked with human-specific regulatory SNCs.

**Gene ontology analyses of putative regulatory targets of genetic loci harboring human-specific SNCs**

Gene Ontology (GO) analyses identified a constellation of biological processes (GO Biological Process: 308 significant records) supplemented with a multitude of molecular functions (GO Molecular Function: 81 significant records) that appear under the regulatory control of human-specific SNCs (Figure 7; Supplemental Table Set 2). Consistently, both databases identified frequently the components of transcriptional regulation and protein kinase activities among most significant records. Other significantly enriched records of interest are regulation of apoptosis, cell proliferation, migration, and various binding properties (cadherin binding; sequence-specific DNA binding; protein-kinase binding; amyloid-beta binding; actin binding; tubulin binding; microtubule binding; PDZ domain binding) which are often supplemented by references to the corresponding activity among the enriched records, for example, enriched records of both binding and activity of protein kinases.

Interrogation of GO Cellular Component database identified 29 significantly enriched records, among which nuclear chromatin as well as various cytoskeleton and membrane components appear noteworthy (Figure 7). Both GO Biological Process and GO Cellular Component database identified significantly enriched records associated with the central nervous system development and functions such as axonogenesis and axon guidance; generation of neurons, neuron differentiation, and neuron projection morphogenesis; cellular components of dendrites and dendrite's membrane; ionotropic glutamate receptor complex. In several instances biologically highly consistent enrichment records have been identified in different GO databases: cadherin binding (GO Molecular Function) and catenin complex (GO Cellular Component); actin binding (GO Molecular Function) and actin cytoskeleton, cortical actin cytoskeleton, actin-based cell projections (GO Cellular Component); microtubule motor activity, tubulin binding, microtubule binding (GO Molecular Function) and microtubule organizing center, microtubule cytoskeleton (GO Cellular Component).

Analyses of human and mouse databases of the Kyoto Encyclopedia of Genes and Genomes (KEGG; Figure 8) identified more than 100 significantly enriched records in each database (KEGG 2019 Human (2019): 129 significant records; KEGG 2019 Mouse: 106 significant records). Genes associated with human-specific regulatory SNCs were implicated in a remarkably broad spectrum of signaling pathways ranging from pathways regulating the pluripotency of stem cells to cell type-specific morphogenesis and differentiation pathways, for example, melanogenesis and adrenergic signaling in cardiomyocytes (Figure 8). Genes under putative regulatory control of human-specific SNCs include hundreds of genes contributing to specific functions of specialized differentiated cells (gastric acid secretion; insulin secretion; aldosterone synthesis and secretion), multiple receptor/ligand-specific signaling pathways, as well as genetic constituents of pathways commonly deregulated in cancer and linked to the organ-specific malignancies, for example, breast, colorectal, and small cell lung cancers (Figure 8). Other notable entries among most significantly enriched records include axon guidance; dopaminergic, glutamatergic, and cholinergic synapses; neuroactive receptor-ligand interactions; and AGE-RAGE signaling pathway in diabetic complications (Figure 8; Supplemental Table Set 2).

**Structurally, functionally, and evolutionary distinct classes of HSRS share the relatively restricted elite set of common genetic targets**

It has been suggested that unified activities of thousands candidate HSRS comprising a coherent compendium of genomic regulatory elements markedly distinct in their structure, function, and evolutionary origin may have contributed to development and manifestation of human-specific phenotypic traits (Glinsky, 2019). It was interest to determine whether genes previously linked to other classes of HSRS, which were identified without considerations of human-specific SNCs, overlap with genes associated in this contribution with genomic regulatory loci harboring human-specific SNCs. It was observed that the common gene set comprises of 7,406 coding genes (88% of all human-specific SNCs-associated genes), indicating that structurally-diverse HSRS, the evolutionary origin of which has been driven by mechanistically-distinct processes, appear to favor the regulatory alignment with the relatively restricted elite set of genetic targets (Figure 9).

Previous studies have identified stem cell-associated retroviral sequences (SCARS) encoded by human endogenous retroviruses LTR7/HERVH and LTR5_Hs/HERVK as one of the significant sources of the evolutionary origin of HSRS (Glinsky, 2015-2019), including human-specific transcription factor binding sites (TFBS) for NANOG, OCT4, and CTCF (Glinsky, 2015). Next, the common sets of genetic regulatory targets were identified for genes expression of which is regulated by SCARS and genes associated in this study with human-specific regulatory SNCs (Figure 9). It has been determined that each of the structurally-distinct families of SCARS appears to share a common set of genetic regulatory targets with human-specific SNCs (Figure 9). Overall, expression of nearly half (4,029 genes; 48%) of all genes identified as putative regulatory targets of human-specific SNCs is regulated by SCARS (Figure 9). Consistent with the idea that structurally-diverse HSRS may favor the relatively restricted elite set of genetic targets, the common gene set of regulatory targets for HSRS, SCARS, and SNCs comprises of 7,833 coding genes or 93% of all genes associated in this contribution with human-specific regulatory SNCs (Figure 9).

To gain insights into mechanisms of SCARS-mediated effects on expression of 4,029 genes linked to human-specific regulatory SNCs, the numbers of genes expression of which was either activated (down-regulated following SCARS silencing) or inhibited (up-regulated following SCARS silencing) by SCARS have been determined. It was observed that SCARS exert the predominantly inhibitory effect on expression of genes associated with human-specific regulatory SNCs, which is exemplified by activated expression of as many as 87% of genes affected by SCARS silencing (Figure 9).

**Identification of 2,273 genes associated with human-specific SNCs and implicated in premature death and embryonic, prenatal, perinatal, neonatal, and postnatal lethality phenotypes**

Interrogation of MGI Mammalian Phenotype databases revealed several hundred mammalian phenotypes affected by thousands of genes associated with genomic regulatory regions harboring human-specific SNCs: the MGI Mammalian Phenotype (2017) database identified 749 significant enrichment records, while the MGI Mammalian Phenotype Level 4 (2019) database identified 407 significant enrichment records (Figure 10; Supplemental Table Set S2). Strikingly, present analyses identified a total of 2,273 genes that are associated with premature death, essential for embryonic survival, implicated in prenatal, perinatal, neonatal, and postnatal lethality phenotypes of both complete and incomplete penetrance (Figure 11) and appear under regulatory control of genetic loci harboring human-specific SNCs. A significant fraction of these 2,273 offspring survival genes were implicated in the autosomal dominant (389 genes) and recessive (426 genes) inheritance in Modern Humans (Figure 11). Based on these observations, it has been concluded that thousands of genes within the genomic dominion of putative regulatory dependency from human-specific SNCs represent the essential genetic elements of the survival of species phenotypes.

## Methods

### Data source and analytical protocols

*Candidate human-specific regulatory sequences and African Apes-specific retroviral insertions*

A total of 94,806 candidate HSRS, including 35,074 human-specific SNCs, detailed descriptions of which and corresponding references of primary original contributions are reported elsewhere (Glinsky et al., 2015-2019; Kanton et al., 2019). Solely publicly available datasets and resources were used in this contribution. The significance of the differences in the expected and observed numbers of events was calculated using two-tailed Fisher's exact test. Additional placement enrichment tests were performed for individual classes of HSRS taking into account the size in bp of corresponding genomic regions.

### Data analysis

### Categories of DNA sequence conservation

Identification of highly-conserved in primates (pan-primate), primate-specific, and human-specific sequences was performed as previously described (Glinsky, 2015-2019). In brief, all categories were defined by direct and reciprocal mapping using LiftOver. Specifically, the following categories of candidate regulatory sequences were distinguished:

- Highly conserved in primates' sequences: DNA sequences that have at least 95% of bases remapped during conversion from/to human (Homo sapiens, hg38), chimp (Pan troglodytes, v5), and bonobo (Pan paniscus, v2; in specifically designated instances, Pan paniscus, v1 was utilized for comparisons). Similarly, highly-conserved sequences were defined for hg38 and latest releases of genomes of Gorilla, Orangutan, Gibbon, and Rhesus.
- Primate-specific: DNA sequences that failed to map to the mouse genome (mm10).
- Human-specific: DNA sequences that failed to map at least 10% of bases from human to both chimpanzee and bonobo. All candidate HSRS identified based on the sequence alignments failures to genomes of both chimpanzee and bonobo were subjected to more stringent additional analyses requiring the mapping failures to genomes of Gorilla, Orangutan, Gibbon, and Rhesus. These loci were considered created *de novo* human-specific regulatory sequences (HSRS).

To infer the putative evolutionary origins, each evolutionary classification was defined independently by running the corresponding analyses on all candidate HSRS representing the specific category. For example, human-rodent conversion identify sequences that are absent in the mouse genome based on the sequence identity threshold of 10%). Additional comparisons were performed using the same methodology and exactly as stated in the manuscript text and described in details below.

**Genome-wide proximity placement analysis**

Genome-wide Proximity Placement Analysis (GPPA) of distinct genomic features co-localizing with HSRS was carried out as described previously (Glinsky, 2015-2019). Briefly, a typical example of the analytical protocol is described below. The significance of overlaps between hESC active enhances and human-specific transcription factor binding sites (hsTFBS) was examined by first identifying all hsTFBS that overlap with any of the genomic regions tested in the ChIP-STARR-seq dataset (Barakat etl, 2018; Glinsky et al., 2018-2019). Then, the relative frequency of active enhancers overlapping with hsTFBS was calculated. To assess the significance of the observed overlap of genomic coordinates, the values recorded for hsTFBS were compared with the expected frequency of active and non-active enhancers that overlap with all TFBS for NANOG (15%) and OCT4 (25%) as previously determined (Barakat et al 2018). The analyses demonstrated that more than 95% of hsTFBS co-localized with sequences in the tested regions of the hESC genome.

The Enrichr API (January 2018 through October 2019 releases) (Chen et al., 2013; Kuleshov et al., 2016) was used to test genes linked to HSRS of interest for significant enrichment in numerous functional categories. In all tables and plots (unless stated otherwise), the "combined score" calculated by Enrichr is reported, which is a product of the significance estimate and the magnitude of enrichment (combined score $c = log(p) * z$, where $p$ is the Fisher's exact test p-value and $z$ is the z-score deviation from the expected rank). When technically feasible, larger sets of genes comprising several thousand entries were analyzed. Regulatory connectivity maps between HSRS and coding genes and additional functional enrichment analyses were performed with the GREAT algorithm (McLean et al., 2010; 2011) at default settings.

*Statistical Analyses of the Publicly Available Datasets*

All statistical analyses of the publicly available genomic datasets, including error rate estimates, background and technical noise measurements and filtering, feature peak calling, feature selection, assignments of genomic coordinates to the corresponding builds of the reference human genome, and data visualization, were performed exactly as reported in the original publications and associated references linked to the corresponding data visualization tracks (http://genome.ucsc.edu/). Any modifications or new elements of statistical analyses are described in the corresponding sections of the Results. Statistical significance of the Pearson correlation coefficients was determined using GraphPad Prism version 6.00 software. Both nominal and Bonferroni adjusted p values were estimated. The significance of the differences in the numbers of events between the groups was calculated using two-sided Fisher's exact and Chi-square test, and the significance of the overlap between the events was determined using the hypergeometric distribution test (Tavazoie et al., 1999).

## Supplemental Information

Supplemental information includes Supplemental Tables S1 and S2, Supplemental Text, and Supplemental Figures.

## Author Contributions

This is a single author contribution. All elements of this work, including the conception of ideas, formulation, and development of concepts, execution of experiments, analysis of data, and writing of the paper, were performed by the author.

## Acknowledgements

This work was made possible by the open public access policies of major grant funding agencies and international genomic databases and the willingness of many investigators worldwide to share their primary research data. I would like to thank my anonymous colleagues for their valuable critical contributions during the peer review process of this work.

**Table 1.** Associations with human physiological processes and pathological conditions of 8,405 genes linked with 35,074 human-specific single nucleotide changes (SNC) within differentially-accessible (DA) chromatin regions identified during human and chimpanzee brain development in cerebral organoids.

| Database | Number of significant records* |
|---|---|
| **ARCHS4 Human Tissues** | 39 |
| **Human Brain Regions: Allen Brain Atlas (Up-regulated genes)** | 1,200 |
| **Human Brain Regions: Allen Brain Atlas (Down-regulated genes)** | 1,062 |
| **Aging Perturbations from GEO (Up-regulated genes)** | 34 |
| **Aging Perturbations from GEO (Down-regulated genes)** | 67 |
| **GO Biological Process** | 308 |
| **GO Molecular Function** | 81 |
| **GO Cellular Component** | 29 |
| **KEGG 2019 Human** | 129 |
| **KEGG 2019 Mouse** | 106 |
| **MGI Mammalian Phenotype 2017** | 749 |
| **MGI Mammalian Phenotype Level 4 2019** | 407 |
| **Human Phenotype Ontology** | 298 |
| **GWAS Catalog 2019** | 241 |
| **Database of Human Genotypes and Phenotypes (dbGaP)** | 136 |
| **Disease Perturbations from GEO (Up-regulated genes)** | 204 |
| **Disease Perturbations from GEO (Down-regulated genes)** | 240 |
| **Rare Diseases GeneRIF Gene Lists** | 473 |
| **Rare Diseases GeneRIF ARCHS4 Predictions** | 603 |
| **Rare Diseases AutoRIF ARCHS4 Predictions** | 641 |
| **Rare Diseases AutoRIF Gene Lists** | 1,116 |

Legend: *, defined at adjusted p-value < 0.05; GEO, gene expression omnibus; GO, gene ontologies; GWAS, genome-wide association studies; ARCHS4, all RNA-seq and ChIP-seq sample and signature search; KEGG, Kyoto Encyclopedia of Genes and Genomes; MGI, mouse genome informatics;

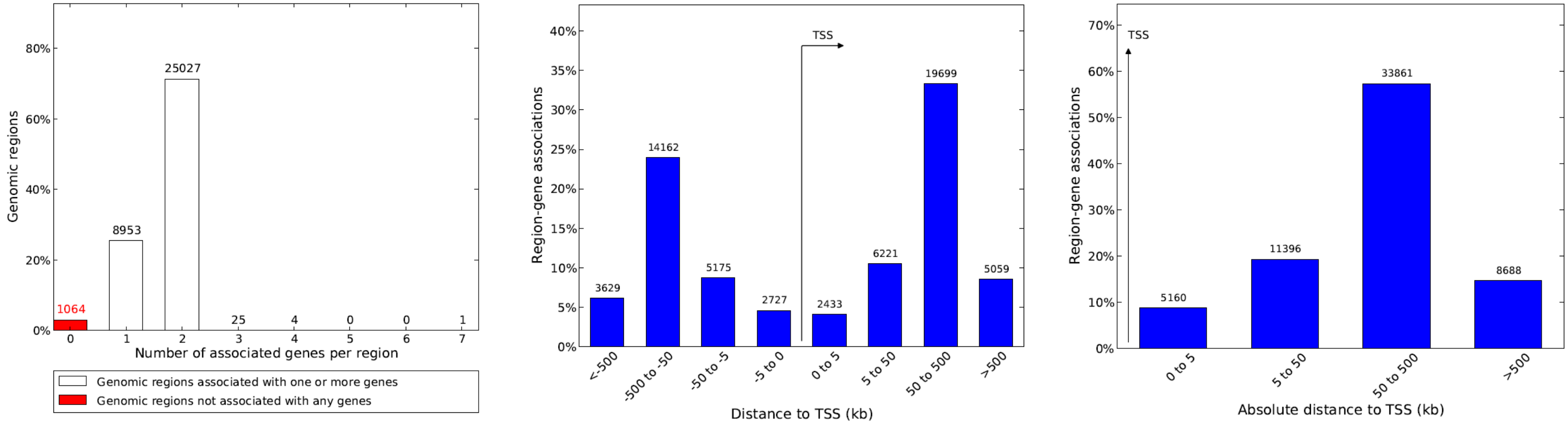

**Figure 1.** GREAT analysis identifies 8,405 human genes associated with 35,074 human-specific single nucleotide changes (SNCs) in differentially accessible (DA) chromatin regions during human and chimpanzee brain development in cerebral organoids. A total of 1,064 of all 35,074 SNCs (3%) are not associated with any genes in the human genome, while a total of 34,010 human-specific SNCs in DA regions appear associated with 8,405 human genes.

### ARCHS4 Human Tissues: 8,045 genes

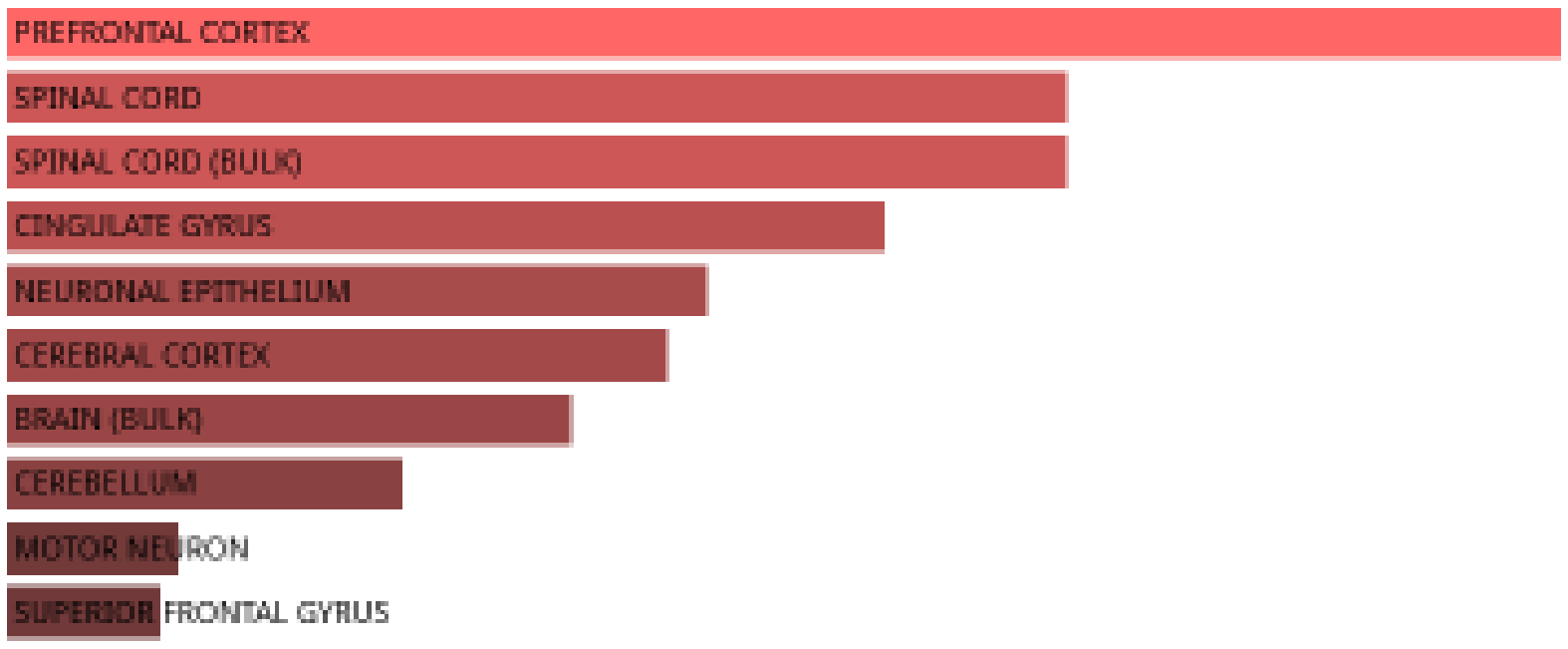

### ARCHS4 Human Tissues: 8,045 genes
### Top 10 of 39 significant records

| Term | Overlap | P-value | Adjusted P-value |
|---|---|---|---|
| PREFRONTAL CORTEX | 1348/2316 | 2.06686E-62 | 2.23E-60 |
| SPINAL CORD | 1299/2316 | 1.17222E-47 | 4.22E-46 |
| SPINAL CORD (BULK) | 1299/2316 | 1.17222E-47 | 4.22E-46 |
| CINGULATE GYRUS | 1279/2316 | 3.13703E-42 | 8.47E-41 |
| NEURONAL EPITHELIUM | 1258/2316 | 6.71385E-37 | 1.45E-35 |
| CEREBRAL CORTEX | 1253/2316 | 1.09815E-35 | 1.98E-34 |
| BRAIN (BULK) | 1241/2316 | 7.35673E-33 | 1.14E-31 |
| CEREBELLUM | 1218/2316 | 8.73598E-28 | 1.18E-26 |
| MOTOR NEURON | 1184/2316 | 4.19448E-21 | 5.03E-20 |
| SUPERIOR FRONTAL GYRUS | 1181/2316 | 1.4634E-20 | 1.58E-19 |

### Allen Brain Atlas Up: 8,045 genes

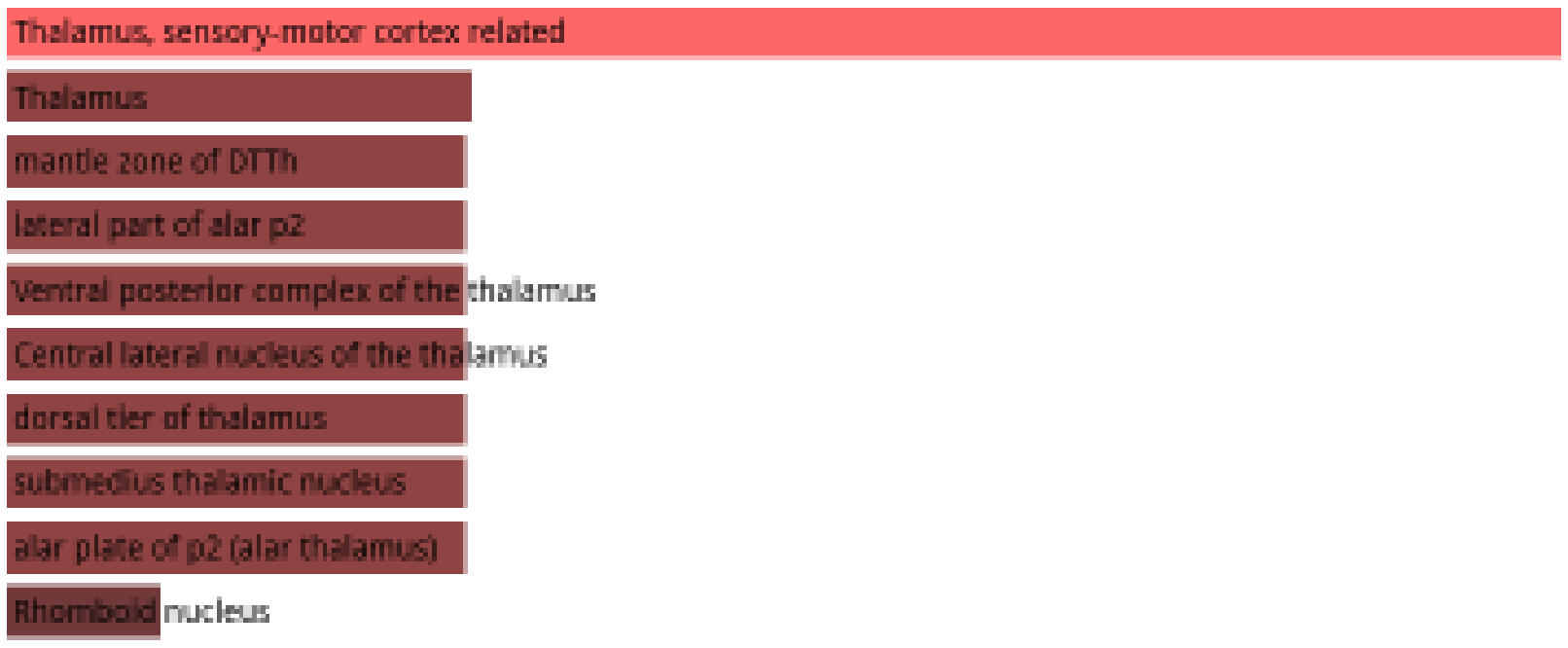

### Allen Brain Atlas Up: 8,045 genes
### Networks

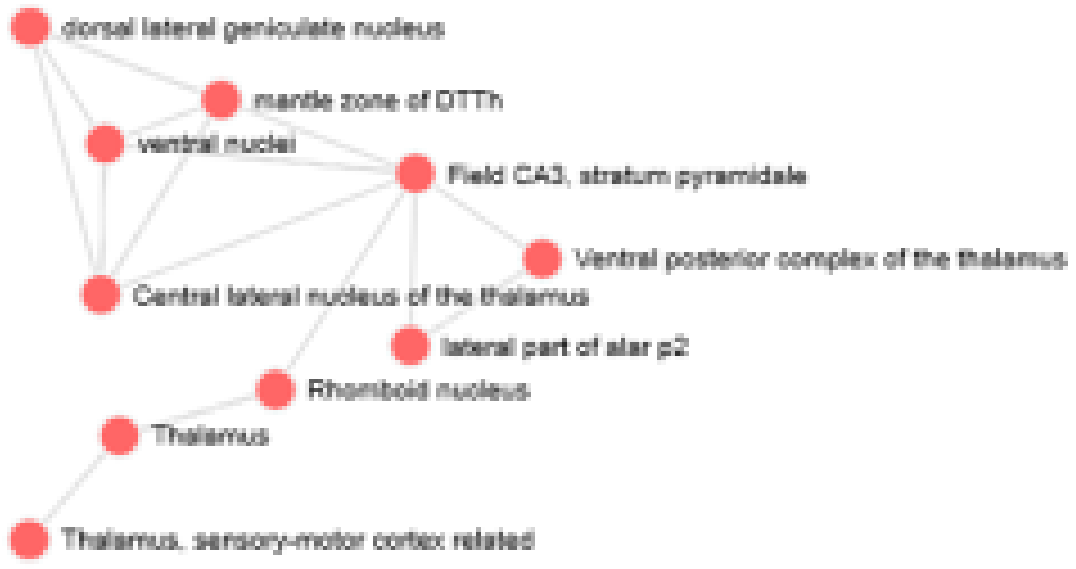

## Allen Brain Atlas Up: 8,045 genes

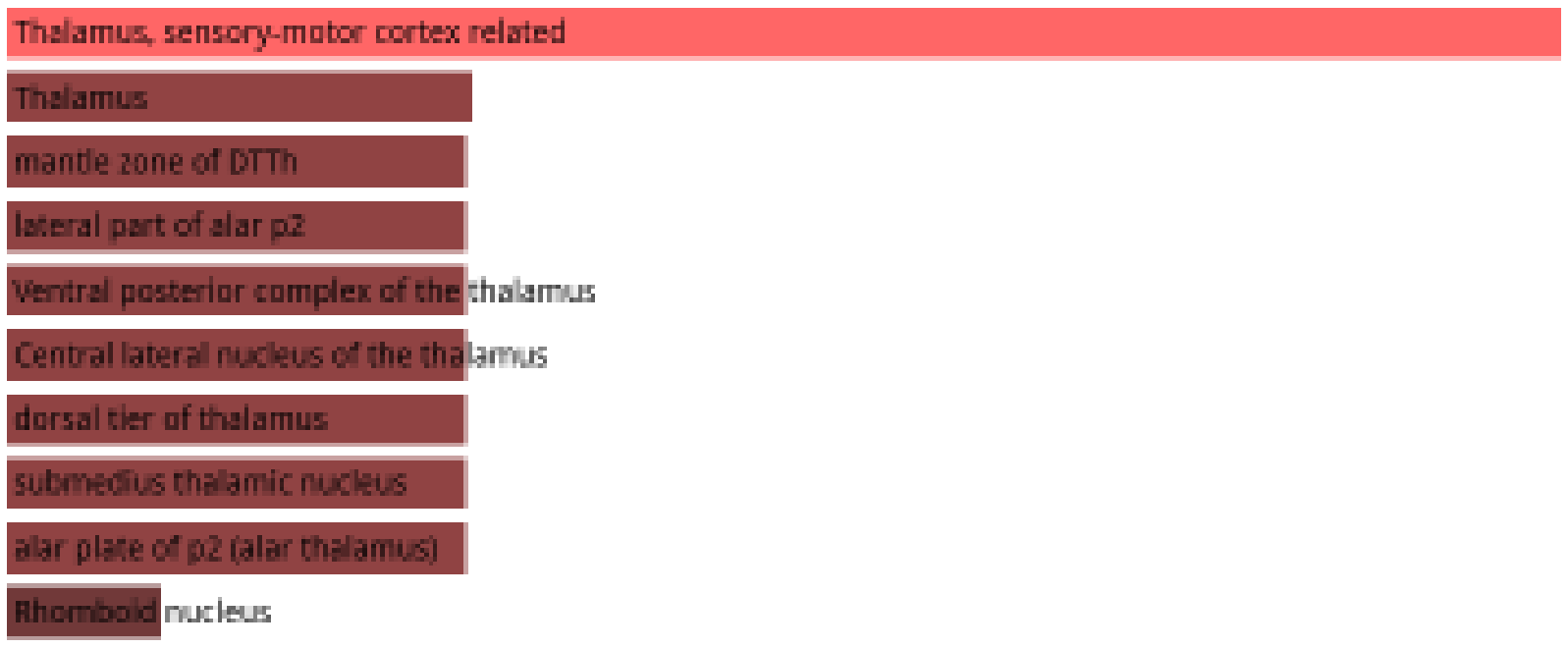

## Allen Brain Atlas Up: 8,045 genes Networks

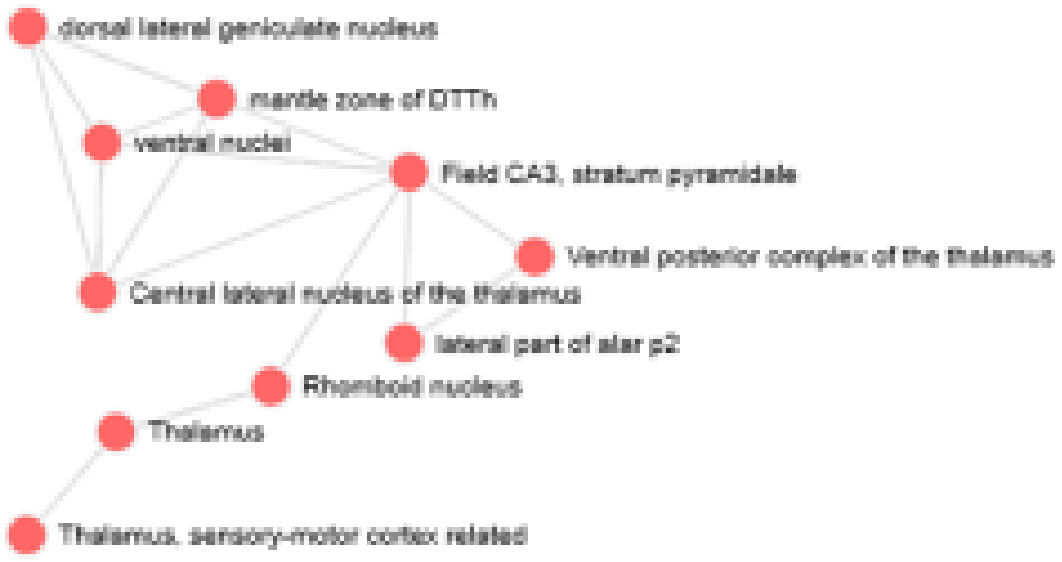

## Allen Brain Atlas Up: 8,045 genes Top 30 of 1,200 significant records

| Term | Overlap | P-value | Adjusted P-value |
|---|---|---|---|
| Thalamus, sensory-motor cortex related | 207/301 | 3.57E-21 | 7.82E-18 |
| Thalamus | 222/334 | 9.71E-20 | 2.42E-17 |
| mantle zone of DTTh | 204/301 | 9.94E-20 | 2.42E-17 |
| lateral part of alar p2 | 204/301 | 9.94E-20 | 2.42E-17 |
| Ventral posterior complex of the thalamus | 204/301 | 9.94E-20 | 2.42E-17 |
| Central lateral nucleus of the thalamus | 204/301 | 9.94E-20 | 2.42E-17 |
| alar plate of p2 (alar thalamus) | 204/301 | 9.94E-20 | 2.42E-17 |
| submedius thalamic nucleus | 204/301 | 9.94E-20 | 2.42E-17 |
| dorsal tier of thalamus | 204/301 | 9.94E-20 | 2.42E-17 |
| Rhomboid nucleus | 227/345 | 2.51E-19 | 5.34E-17 |
| ventral nuclei | 203/301 | 2.92E-19 | 5.34E-17 |
| Ventral group of the dorsal thalamus | 203/301 | 2.92E-19 | 5.34E-17 |
| Field CA3, stratum pyramidale | 202/301 | 8.47E-19 | 1.24E-16 |
| Ventral posteromedial nucleus of the thalamus | 202/301 | 8.47E-19 | 1.24E-16 |
| ventral posteromedial nucleus | 202/301 | 8.47E-19 | 1.24E-16 |
| intermediate stratum of DTTh | 201/301 | 2.42E-18 | 2.94E-16 |
| prosomere 2 | 201/301 | 2.42E-18 | 2.94E-16 |
| central lateral nucleus | 201/301 | 2.42E-18 | 2.94E-16 |
| dorsal lateral geniculate nucleus | 200/301 | 6.79E-18 | 7.84E-16 |
| hilus of the DG | 199/301 | 1.88E-17 | 2.06E-15 |
| lateral (parvicellular) part of MD | 198/301 | 5.13E-17 | 4.33E-15 |
| Dorsal part of the lateral geniculate complex | 198/301 | 5.13E-17 | 4.33E-15 |
| Ventral posterolateral nucleus of the thalamus | 198/301 | 5.13E-17 | 4.33E-15 |
| posterior (ventral) nucleus | 198/301 | 5.13E-17 | 4.33E-15 |
| Hippocampal formation | 198/301 | 5.13E-17 | 4.33E-15 |
| intralaminar nuclei | 198/301 | 5.13E-17 | 4.33E-15 |
| intermediate stratum of DG | 353/597 | 9.09E-17 | 7.38E-15 |
| Field CA3, stratum radiatum | 219/342 | 1.17E-16 | 9.12E-15 |
| mantle zone of CA | 197/301 | 1.38E-16 | 9.76E-15 |
| Thalamus, polymodal association cortex related | 197/301 | 1.38E-16 | 9.76E-15 |

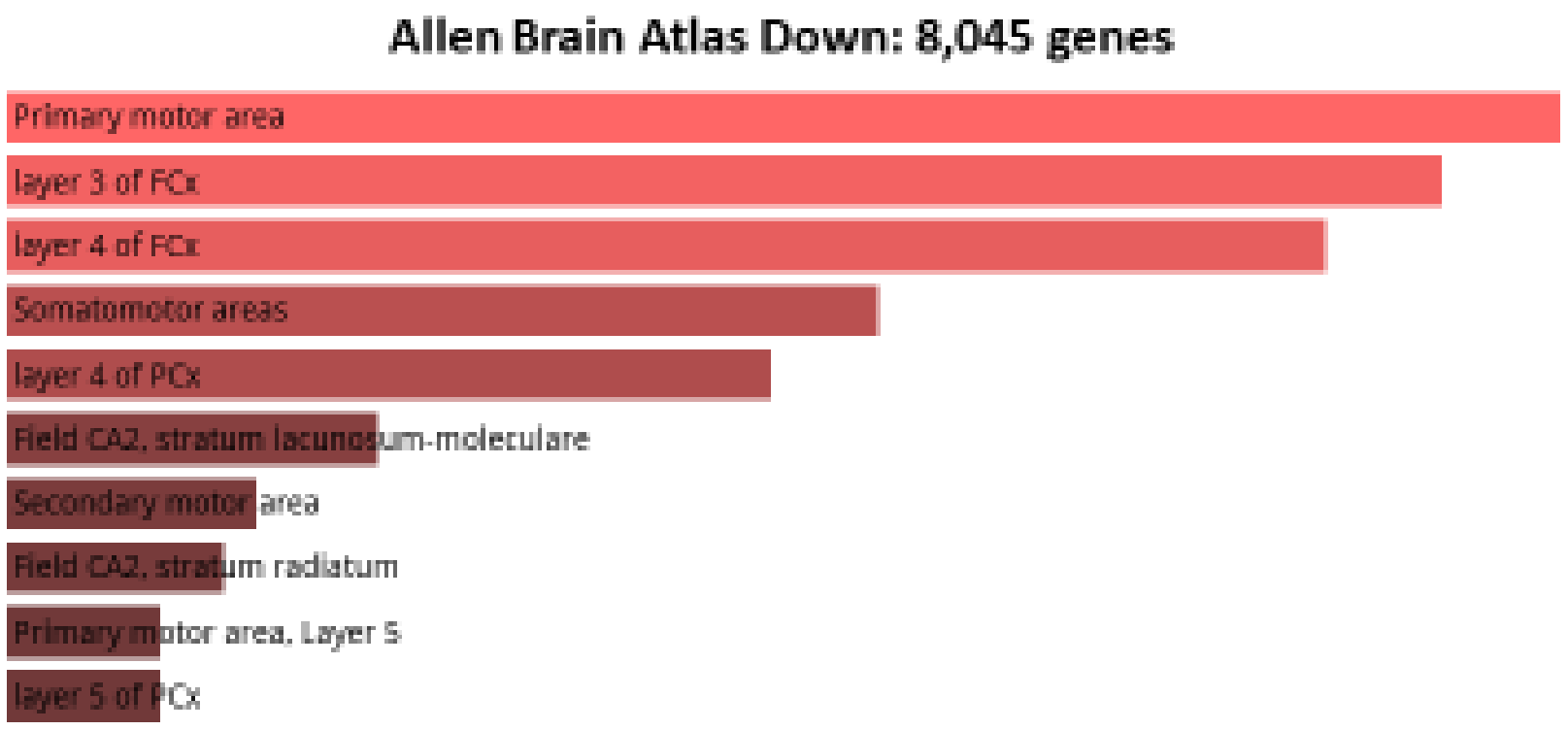
Allen Brain Atlas Down: 8,045 genes
Primary motor area
layer 3 of FCx
layer 4 of FCx
Somatomotor areas
layer 4 of PCx
Field CA2, stratum lacunosum-moleculare
Secondary motor area
Field CA2, stratum radiatum
Primary motor area, Layer 5
layer 5 of PCx

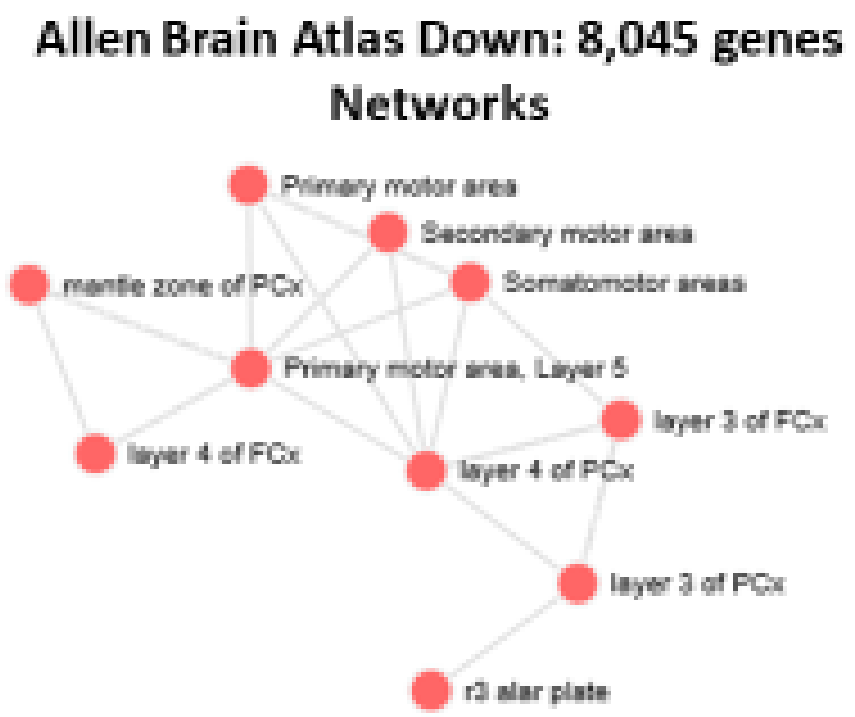
Allen Brain Atlas Down: 8,045 genes
Networks
Primary motor area
Secondary motor area
mantle zone of PCx
Somatomotor areas
Primary motor area, Layer 5
layer 3 of FCx
layer 4 of FCx
layer 4 of PCx
layer 3 of PCx
r3 alar plate

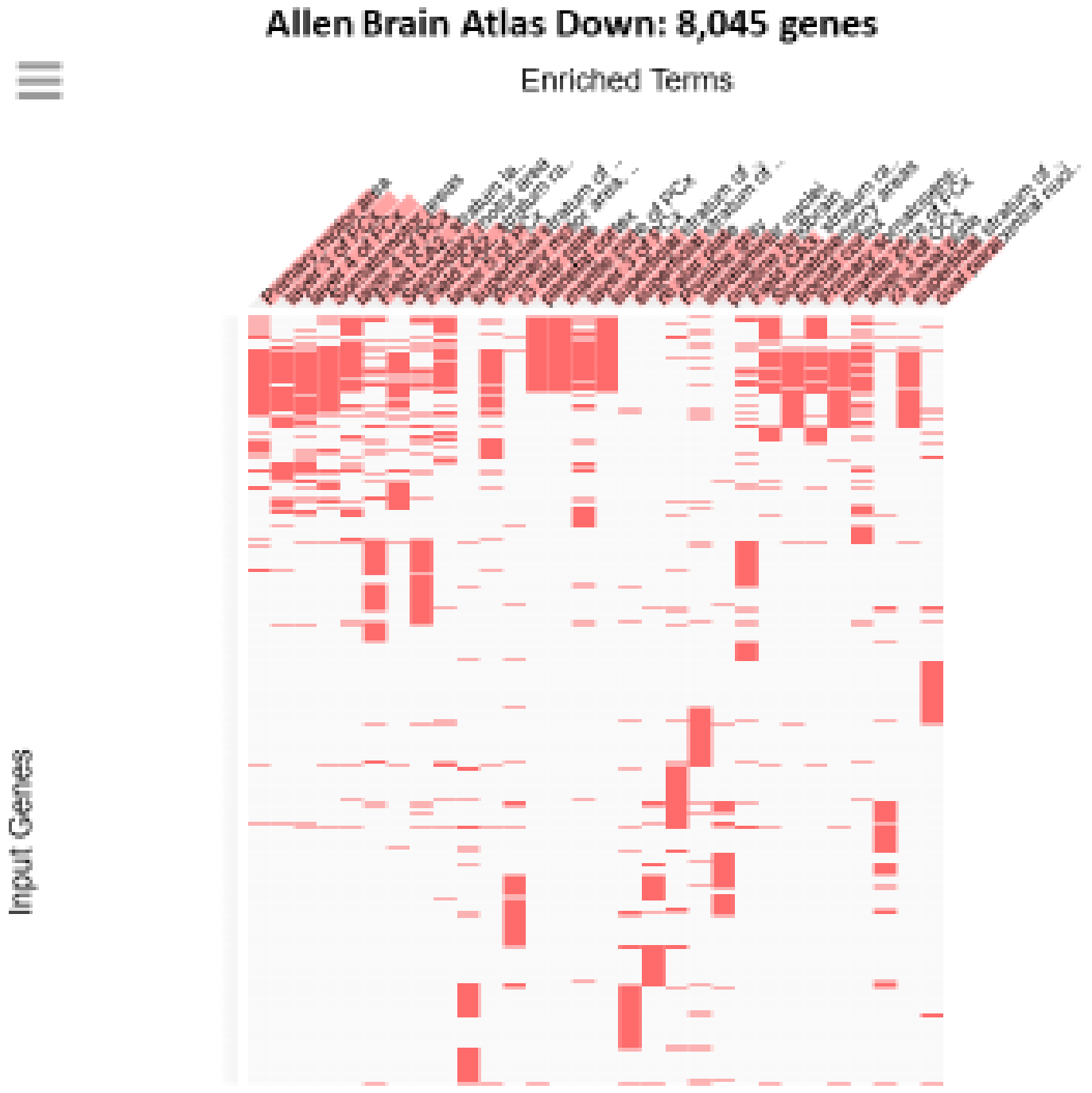
Allen Brain Atlas Down: 8,045 genes
Enriched Terms
Input Genes

**Allen Brain Atlas Down: 8,045 genes**
**Top 31 of 1,062 significant records**

| Term | Overlap | P-value | Adjusted P-value |
|---|---|---|---|
| Primary motor area | 195/300 | 5.73E-16 | 1.26E-12 |
| layer 3 of FCx | 194/300 | 1.49E-15 | 1.63E-12 |
| layer 4 of FCx | 193/300 | 3.81E-15 | 2.79E-12 |
| Somatomotor areas | 189/300 | 1.42E-13 | 7.77E-11 |
| layer 4 of PCx | 188/300 | 3.38E-13 | 1.48E-10 |
| Field CA2, stratum lacunosum-moleculare | 248/426 | 8.01E-12 | 2.93E-09 |
| Secondary motor area | 183/300 | 2.1E-11 | 6.57E-09 |
| Field CA2, stratum radiatum | 262/458 | 2.75E-11 | 7.54E-09 |
| Primary motor area, Layer 5 | 182/300 | 4.59E-11 | 9.16E-09 |
| layer 5 of PCx | 182/300 | 4.59E-11 | 9.16E-09 |
| superficial stratum of m1B | 182/300 | 4.59E-11 | 9.16E-09 |
| r3 alar plate | 181/300 | 9.92E-11 | 1.67E-08 |
| parietal cortex | 181/300 | 9.92E-11 | 1.67E-08 |
| mantle zone of PCx | 180/300 | 2.11E-10 | 2.44E-08 |
| layer 3 of PCx | 180/300 | 2.11E-10 | 2.44E-08 |
| Pontine gray | 180/300 | 2.11E-10 | 2.44E-08 |
| r6 alar plate | 183/300 | 2.11E-10 | 2.44E-08 |
| superficial stratum of PCx (cortical plate/marginal zone) | 180/300 | 2.11E-10 | 2.44E-08 |
| superficial stratum of p2B | 180/300 | 2.11E-10 | 2.44E-08 |
| intralaminar nuclei | 179/300 | 4.44E-10 | 4.64E-08 |
| pontine hindbrain | 182/300 | 4.44E-10 | 4.64E-08 |
| Field CA3, stratum radiatum | 211/364 | 4.83E-10 | 4.81E-08 |
| Primary somatosensory area | 178/300 | 9.2E-10 | 8.07E-08 |
| frontal cortex | 179/300 | 9.2E-10 | 8.07E-08 |
| Somatosensory areas | 178/300 | 9.2E-10 | 8.07E-08 |
| mantle zone of FCx | 178/300 | 1.88E-09 | 1.37E-07 |
| layer 3 of OCx | 177/300 | 1.88E-09 | 1.37E-07 |
| r3 basal plate | 177/300 | 1.88E-09 | 1.37E-07 |
| superficial stratum of FCx (cortical plate/marginal zone) | 178/300 | 1.88E-09 | 1.37E-07 |
| oval paracentral nucleus | 177/300 | 1.88E-09 | 1.37E-07 |
| Primary motor area, Layer 2/3 | 176/300 | 3.79E-09 | 2.6E-07 |

**Figure 2.** Identification of genes expression of which distinguishes thousands of anatomically distinct areas of the adult human brain, various regions of the central nervous system, and many different cell types and tissues in the human body.

## Aging Perturbations from GEO up: 8,405 genes

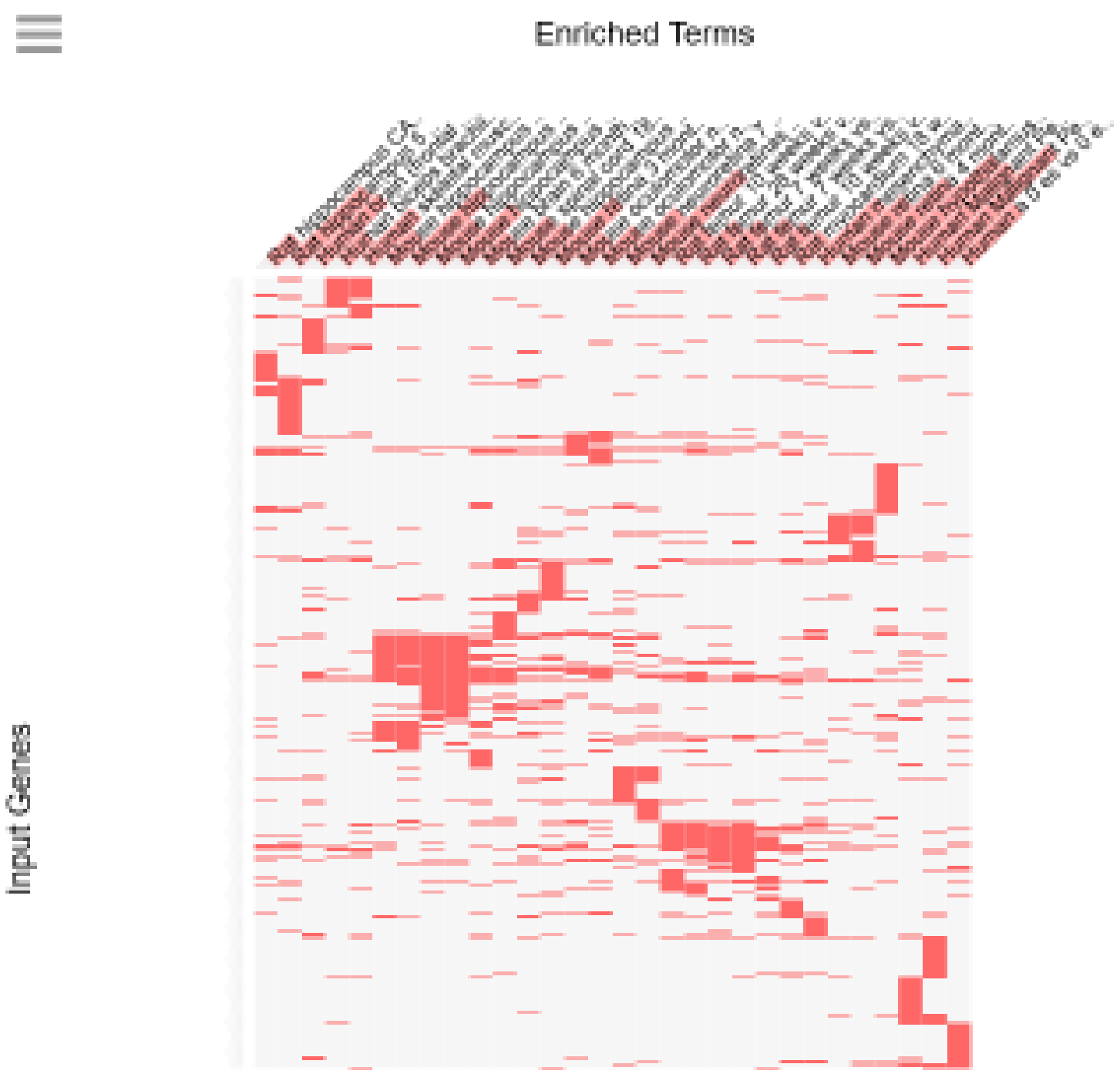

## Aging Perturbations from GEO up: 8,405 genes
### Top 30 significant records

| Term | Overlap | P-value | Adjusted P-value |
|---|---|---|---|
| Human_Malignant Peripheral Nerve Sheath Tumour_24 years vs 53 years_GSE17118_aging:364 | 197/337 | 6.9E-10 | 1.97E-07 |
| Mouse_hypothalamus_42 days vs 182 days_GDS3895_aging:107 | 201/349 | 2.59E-09 | 3.7E-07 |
| Human_Malignant Peripheral Nerve Sheath Tumour_27 years vs 61 years_GSE17118_aging:363 | 195/343 | 1.78E-08 | 1.5E-06 |
| Mouse_neuroretinas_7 weeks vs 64 weeks_GSE38671_aging:211 | 168/289 | 2.1E-08 | 1.5E-06 |
| Mouse_retina_4 months vs 10 months_GSE33674_aging:304 | 148/254 | 1.16E-07 | 6.63E-06 |
| Rat_hippocampus_5 months vs 24 months_GSE14505_aging:346 | 230/429 | 6.82E-07 | 3.25E-05 |
| Mouse_hippocampus_9 months vs 20 months_GSE48911_aging:391 | 243/459 | 1.23E-06 | 5.02E-05 |
| Mouse_retina_6 months vs 10 months_GSE33674_aging:305 | 125/218 | 3.39E-06 | 0.000121 |
| Human_brain (frontal cortex)_55 years vs 82 years_GSE53890_aging:229 | 171/313 | 4.05E-06 | 0.000129 |
| Mouse_retina_3 months vs 16 months_GDS2654_aging:66 | 199/372 | 4.53E-06 | 0.00013 |
| Human_mesenchymal stem cells (from bone marrow)_42 years vs 79 years_GSE35955_aging:293 | 133/238 | 1.03E-05 | 0.000255 |
| Human_a_0 es vs 0 es_GDS5077_aging:106 | 172/319 | 1.07E-05 | 0.000255 |
| Mouse_hippocampus_9 months vs 14 months_GSE48911_aging:390 | 254/494 | 1.28E-05 | 0.000283 |
| Mouse_neuroretina_7 weeks vs 64 weeks_GSE38671_aging:210 | 171/319 | 1.76E-05 | 0.00036 |
| Human_skeletal muscle_19 years vs 65 years_GDS4858_aging:10 | 140/259 | 5.77E-05 | 0.0011 |
| Mouse_spinal cord_18 months vs 30 months_GDS1280_aging:1 | 172/331 | 0.000151 | 0.002697 |
| Mouse_hippocampus_9 months vs 14 months_GSE48911_aging:384 | 246/494 | 0.000252 | 0.004247 |
| Rat_femur_7 weeks vs 53 weeks_GDS509_aging:264 | 197/389 | 0.000331 | 0.005257 |
| Rat_femur_28 weeks vs 54 weeks_GDS509_aging:271 | 193/383 | 0.000523 | 0.007575 |
| Rat_femur_7 weeks vs 27 weeks_GDS509_aging:258 | 155/301 | 0.00053 | 0.007575 |
| Mouse_hippocampus_2 months vs 15 months_GSE5078_aging:398 | 151/293 | 0.000593 | 0.008072 |
| Rat_femur_10 weeks vs 30 weeks_GDS509_aging:260 | 123/236 | 0.001057 | 0.013739 |
| Mouse_oculomotor nucleus_6 months vs 30 months_GDS1280_aging:6 | 140/273 | 0.001185 | 0.01473 |
| Rat_femur_10 weeks vs 56 weeks_GDS509_aging:266 | 191/384 | 0.001248 | 0.014873 |
| Mouse_hippocampus_9 months vs 20 months_GSE48911_aging:385 | 216/442 | 0.001959 | 0.022408 |
| Human_biceps brachii muscles_24 years vs 70 years_GDS4858_aging:33 | 119/233 | 0.003152 | 0.034649 |
| Rat_myocardium_18 weeks vs 22 weeks_GDS4025_aging:142 | 124/244 | 0.003271 | 0.034649 |
| Rat_hippocampus CA3 region_6 months vs 25 months_GSE14724_aging:347 | 170/345 | 0.003642 | 0.036658 |
| Mouse_brain_6 months vs 14 months_GSE15129_aging:313 | 189/387 | 0.003717 | 0.036658 |
| Mouse_spinal cord_6 months vs 30 months_GDS1280_aging:3 | 197/405 | 0.003882 | 0.037008 |

Aging Perturbations from GEO down: 8,405 genes

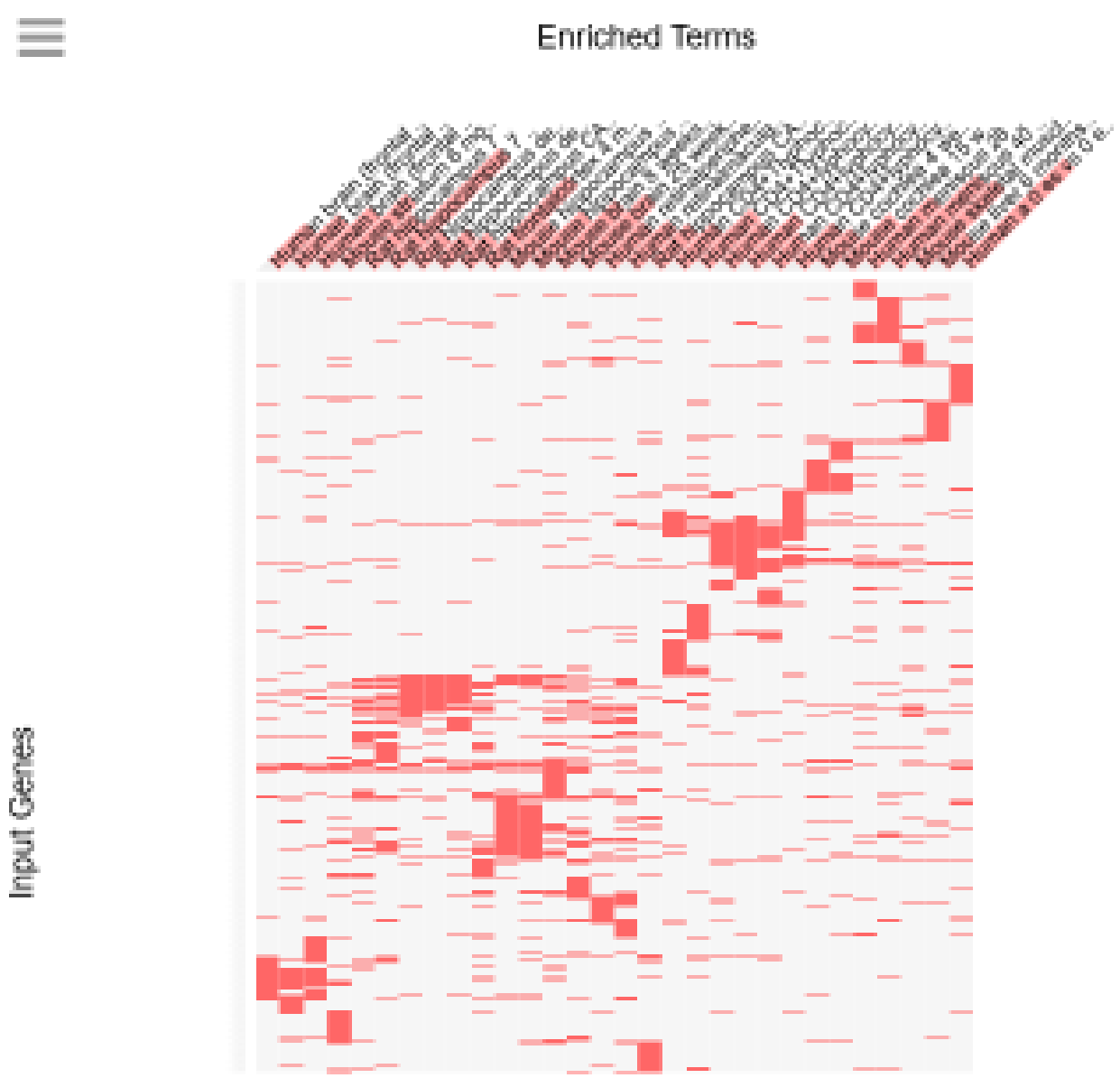

Aging Perturbations from GEO down: 8,405 genes
Top 30 significant records

| Term | Overlap | P-value | Adjusted P-value |
|---|---|---|---|
| Rat_hippocampus_7 months vs 21 months_GDS4019_aging:98 | 196/314 | 1.85E-13 | 5.28E-11 |
| Human_a_0 es vs 0 es_GDS5077_aging:106 | 175/281 | 4.78E-12 | 6.83E-10 |
| Human_meniscus_40 years vs 55 years_GSE45233_aging:209 | 189/320 | 4.76E-10 | 4.54E-08 |
| Rat_frontal cortex_4 months vs 22 months_GDS3939_aging:99 | 192/327 | 6.86E-10 | 4.9E-08 |
| Mouse_skeletal muscle precursor_12 months vs 24 months_GDS4892_aging:9 | 137/224 | 5.29E-09 | 3.02E-07 |
| Mouse_brain_25 weeks vs 100 weeks_GSE41018_aging:400 | 178/311 | 3.83E-08 | 1.68E-06 |
| Mouse_striatum_6 months vs 21 months_GDS4153_aging:78 | 164/283 | 4.12E-08 | 1.68E-06 |
| Human_meniscus_28 years vs 47 years_GSE45233_aging:208 | 165/286 | 5.7E-08 | 2.04E-06 |
| Mouse_brain_8 weeks vs 104 weeks_GSE20411_aging:311 | 224/408 | 8.36E-08 | 2.66E-06 |
| Human_brain (frontal cortex)_28 years vs 100 years_GSE53890_aging:228 | 197/356 | 2.32E-07 | 6.64E-06 |
| Human_CD4+Tlyphocytes_61 years vs 81 years_GSE62373_aging:185 | 193/349 | 3.26E-07 | 8.48E-06 |
| Human_CD4+lyphocytes_40 years vs 72 years_GSE62373_aging:163 | 203/373 | 7.65E-07 | 1.82E-05 |
| Mouse_hippocampus_4 months vs 9 months_GSE48911_aging:387 | 230/430 | 8.6E-07 | 1.89E-05 |
| Rat_hippocampus_7 months vs 21 months_GDS4019_aging:97 | 156/277 | 9.99E-07 | 2.04E-05 |
| Mouse_cochlea_15 weeks vs 45 weeks_GSE35234_aging:164 | 142/250 | 1.58E-06 | 3.01E-05 |
| Human_brain (frontal cortex)_55 years vs 82 years_GSE53890_aging:229 | 159/287 | 2.97E-06 | 5.31E-05 |
| Human_retinalmacula_18 years vs 74 years_GSE32614_aging:150 | 182/336 | 4.14E-06 | 6.96E-05 |
| Mouse_hippocampus_4 months vs 9 months_GSE48911_aging:381 | 226/429 | 4.52E-06 | 7.18E-05 |
| Human_CD4+lyphocytes_40 years vs 61 years_GSE62373_aging:160 | 192/358 | 5.34E-06 | 7.77E-05 |
| Human_CD4+Tlyphocytes_29 years vs 81 years_GSE62373_aging:176 | 201/377 | 5.44E-06 | 7.77E-05 |
| Human_CD4+lyphocytes_29 years vs 61 years_GSE62373_aging:154 | 184/342 | 6.43E-06 | 8.75E-05 |
| Human_retinalperiphery_32 years vs 74 years_GSE32614_aging:151 | 160/293 | 8.42E-06 | 0.000109 |
| Human_brain (frontal cortex)_28 years vs 82 years_GSE53890_aging:227 | 154/281 | 9.47E-06 | 0.000118 |
| Rat_hippocampus CA3 region_6 months vs 25 months_GSE14724_aging:347 | 141/255 | 1.2E-05 | 0.000143 |
| Mouse_cochlea_4 weeks vs 45 weeks_GSE35234_aging:161 | 145/268 | 4.04E-05 | 0.000463 |
| Rat_hippocampus dentate gyrus_18 months vs 28 months_GSE21681_aging:357 | 124/225 | 4.78E-05 | 0.000522 |
| Rat_hippocampus dentate gyrus_18 months vs 28 months_GSE21681_aging:358 | 147/273 | 4.93E-05 | 0.000522 |
| Human_CD4+lyphocytes_40 years vs 81 years_GSE62373_aging:165 | 197/380 | 6.24E-05 | 0.000638 |
| Human_CD4+lyphocytes_29 years vs 81 years_GSE62373_aging:156 | 215/419 | 6.73E-05 | 0.000664 |
| Rat_hippocampus CA1_18 months vs 28 months_GSE21681_aging:353 | 144/270 | 0.000107 | 0.001004 |

**Figure 3.** Identification and characterization of genes expression of which is altered during aging of humans, rats, and mice.

Human Phenotype Ontology: 8,405 genes

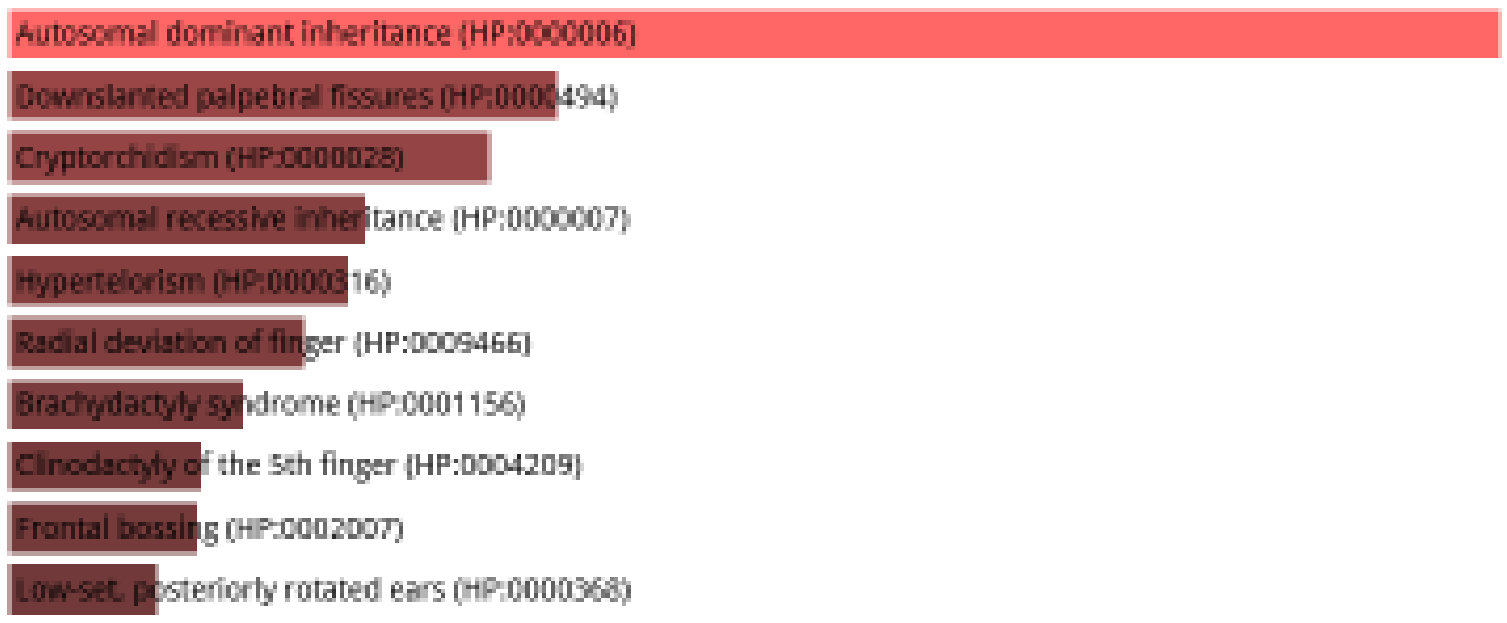

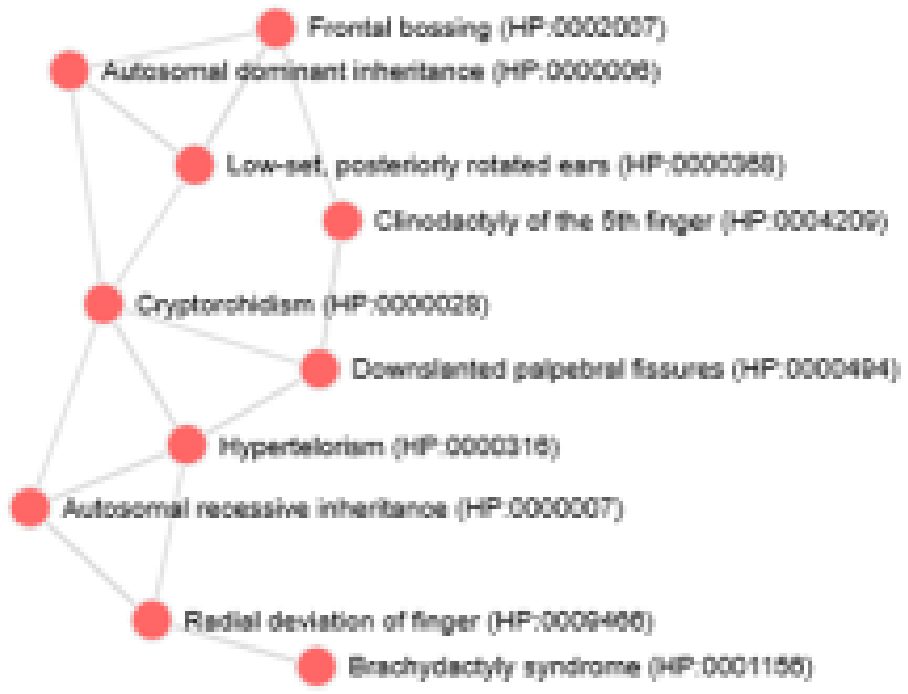

Human Phenotype Ontology: 8,405 genes

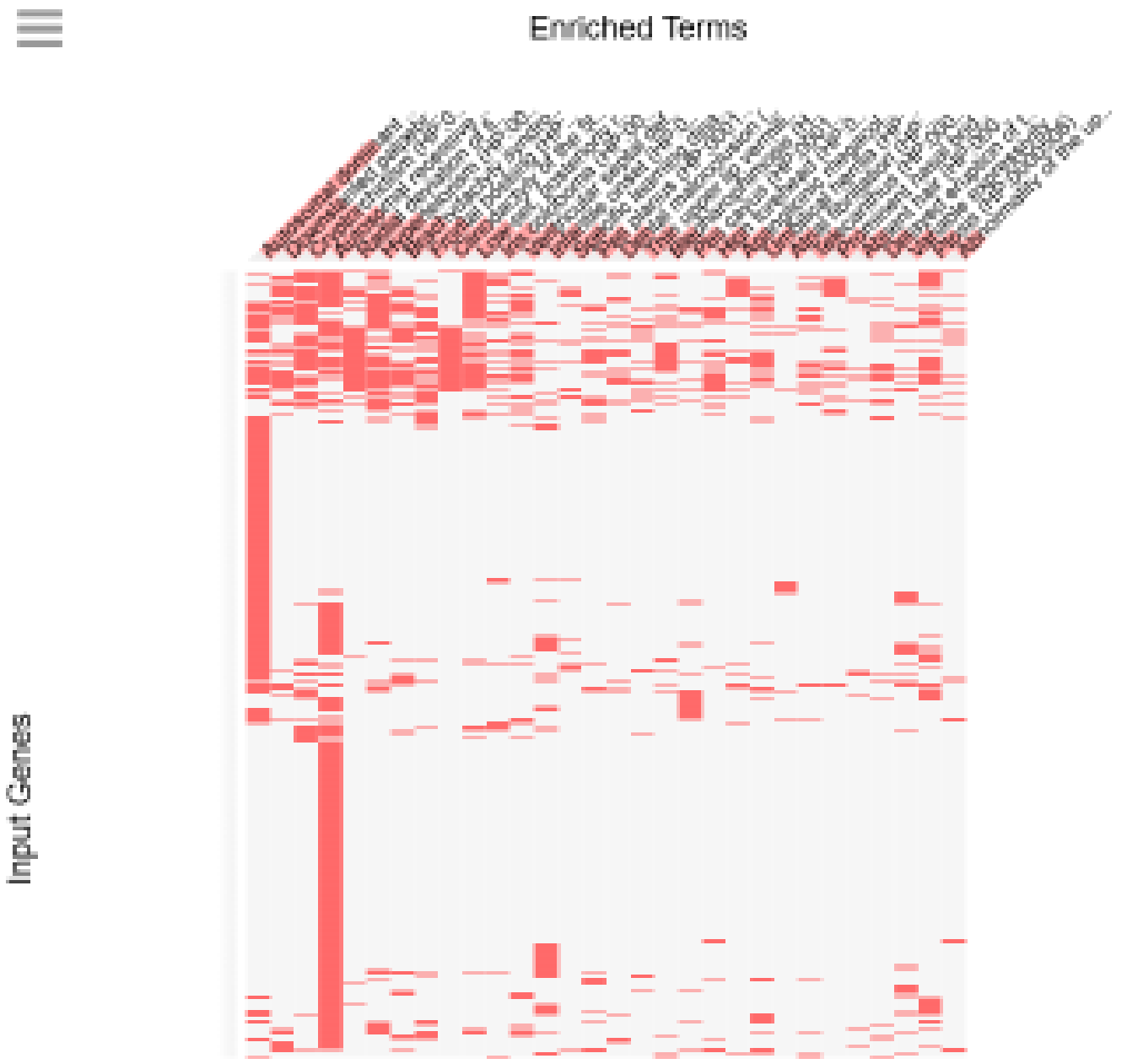

## Human Phenotype Ontology (8,405 genes): Top 35 of 298 significant records

| Term | Overlap | P-value | Adjusted P-value |
|---|---|---|---|
| Autosomal dominant inheritance (HP:0000006) | 645/1134 | 2.85E-25 | 5.07E-22 |
| Downslanted palpebral fissures (HP:0000494) | 127/188 | 1.17E-12 | 1.04E-09 |
| Cryptorchidism (HP:0000028) | 222/379 | 4.46E-11 | 2.64E-08 |
| Autosomal recessive inheritance (HP:0000007) | 849/1722 | 1.14E-10 | 5.07E-08 |
| Radial deviation of finger (HP:0009466) | 140/229 | 3.75E-09 | 1.11E-06 |
| Hypertelorism (HP:0000316) | 190/328 | 3.76E-09 | 1.11E-06 |
| Brachydactyly syndrome (HP:0001156) | 119/192 | 1.77E-08 | 4.5E-06 |
| Frontal bossing (HP:0002007) | 129/213 | 3.38E-08 | 7.51E-06 |
| Clinodactyly of the 5th finger (HP:0004209) | 118/192 | 4.03E-08 | 7.90E-06 |
| Low-set, posteriorly rotated ears (HP:0000368) | 158/276 | 2.14E-07 | 3.8E-05 |
| Iris coloboma (HP:0000612) | 72/110 | 5.77E-07 | 9.34E-05 |
| Ventricular septal defect (HP:0001629) | 106/176 | 7.99E-07 | 0.000113 |
| Infantile onset (HP:0003593) | 145/254 | 8.28E-07 | 0.000113 |
| Specific learning disability (HP:0001328) | 36/47 | 1.49E-06 | 0.000189 |
| Pes planus (HP:0001763) | 68/105 | 2.09E-06 | 0.000246 |
| Dental malocclusion (HP:0000689) | 55/81 | 2.21E-06 | 0.000246 |
| Thin upper lip vermilion (HP:0000219) | 38/51 | 2.5E-06 | 0.000261 |
| Blepharophimosis (HP:0000581) | 67/104 | 3.26E-06 | 0.000323 |
| Pes cavus (HP:0001761) | 80/129 | 3.55E-06 | 0.000323 |
| High forehead (HP:0000348) | 69/108 | 3.63E-06 | 0.000323 |
| Aganglionic megacolon (HP:0002251) | 63/97 | 4.26E-06 | 0.000361 |
| Umbilical hernia (HP:0001537) | 91/151 | 4.5E-06 | 0.000364 |
| Atrial fibrillation (HP:0005110) | 26/32 | 6.52E-06 | 0.000505 |
| Telecanthus (HP:0000506) | 55/83 | 6.96E-06 | 0.000516 |
| Prominent nasal bridge (HP:0000426) | 64/100 | 7.43E-06 | 0.000529 |
| Absent hair (HP:0002298) | 18/20 | 1.14E-05 | 0.000781 |
| Delayed eruption of teeth (HP:0000684) | 56/86 | 1.28E-05 | 0.000841 |
| Variable expressivity (HP:0003828) | 86/144 | 1.34E-05 | 0.00085 |
| Ptosis (HP:0000508) | 180/338 | 1.79E-05 | 0.001074 |
| Abnormality of the upper arm (HP:0001454) | 29/38 | 1.81E-05 | 0.001074 |
| Progressive disorder (HP:0003676) | 86/145 | 1.94E-05 | 0.001103 |
| Depressed nasal bridge (HP:0005280) | 127/228 | 1.99E-05 | 0.001103 |
| Sporadic (HP:0003745) | 42/61 | 2.05E-05 | 0.001103 |
| Left ventricular hypertrophy (HP:0001712) | 31/42 | 2.93E-05 | 0.0015 |
| Postaxial hand polydactyly (HP:0001162) | 53/82 | 2.95E-05 | 0.0015 |

## Database of Human Genotypes and Phenotypes (dbGaP): 8,405 genes

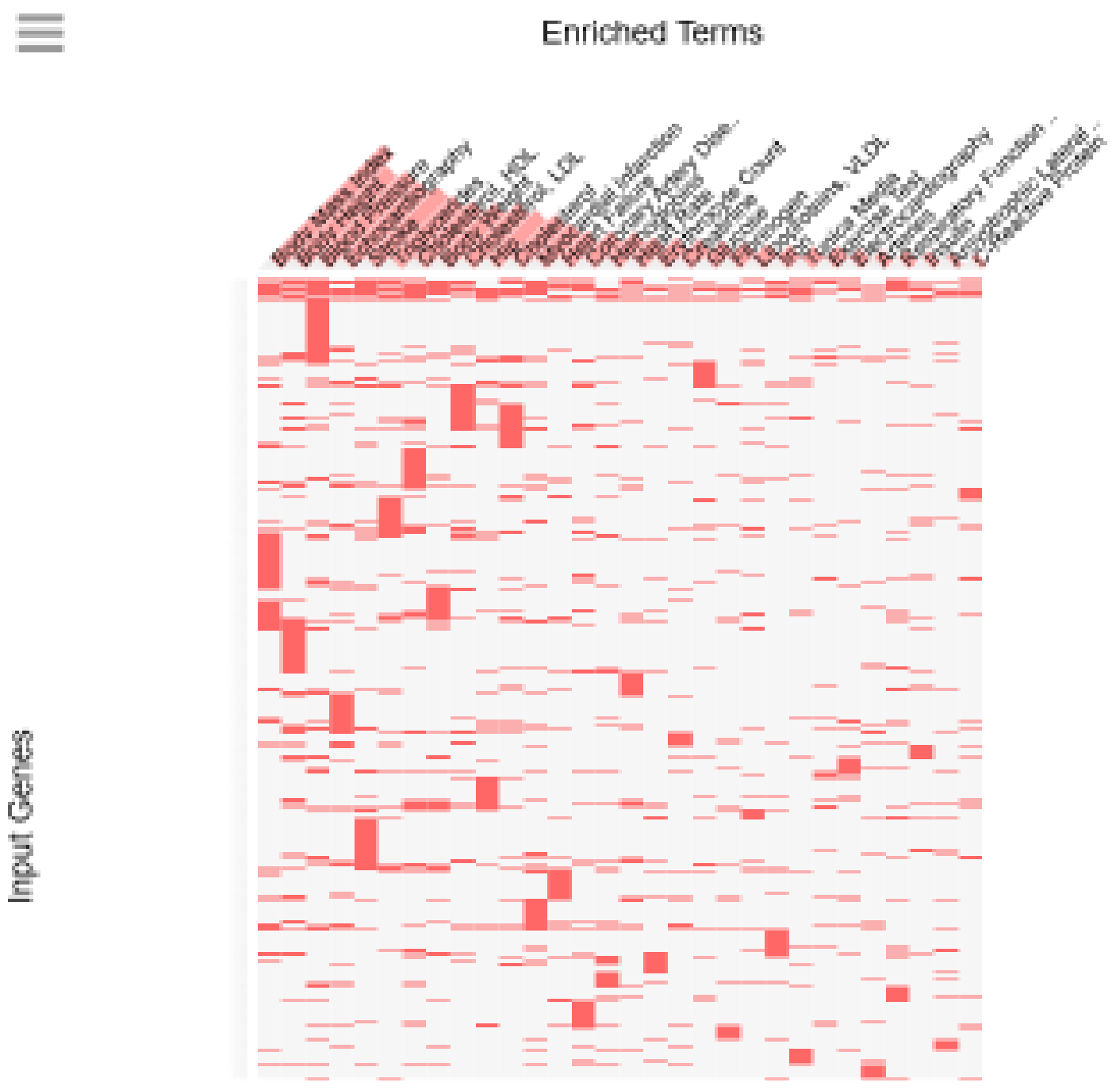

## Database of Human Genotypes and Phenotypes (dbGaP): 8,405 genes
## Top 20 of 136 significant records

| Term | Overlap | P-value | Adjusted P-value |
|---|---|---|---|
| Body Mass Index | 313/437 | 1.02E-36 | 3.46E-34 |
| Body Height | 276/385 | 1.32E-32 | 2.23E-30 |
| Blood Pressure | 310/454 | 3.19E-30 | 3.59E-28 |
| Echocardiography | 204/273 | 2.47E-28 | 2.08E-26 |
| Stroke | 211/289 | 6.02E-27 | 4.07E-25 |
| Triglycerides | 180/244 | 4.77E-24 | 2.68E-22 |
| Cholesterol, HDL | 242/357 | 3.57E-23 | 1.72E-21 |
| Body Weight | 161/216 | 1.91E-22 | 8.05E-21 |
| Cholesterol, LDL | 210/304 | 7.78E-22 | 2.92E-20 |
| Hip | 151/202 | 2.39E-21 | 8.06E-20 |
| Cholesterol | 183/268 | 2.28E-18 | 7.02E-17 |
| Myocardial Infarction | 157/229 | 3.32E-16 | 9.36E-15 |
| Coronary Artery Disease | 140/205 | 2.16E-14 | 5.31E-13 |
| Heart Failure | 130/187 | 2.2E-14 | 5.31E-13 |
| Hemoglobins | 113/157 | 2.38E-14 | 5.37E-13 |
| Erythrocyte Count | 87/115 | 2.09E-13 | 4.41E-12 |
| Heart Rate | 110/155 | 2.47E-13 | 4.86E-12 |
| Creatinine | 73/92 | 2.59E-13 | 4.86E-12 |
| Fibrinogen | 69/86 | 4.29E-13 | 7.05E-12 |
| Lipoproteins, VLDL | 69/86 | 4.29E-13 | 7.05E-12 |

## GWAS Catalog 2019: 8,405 genes

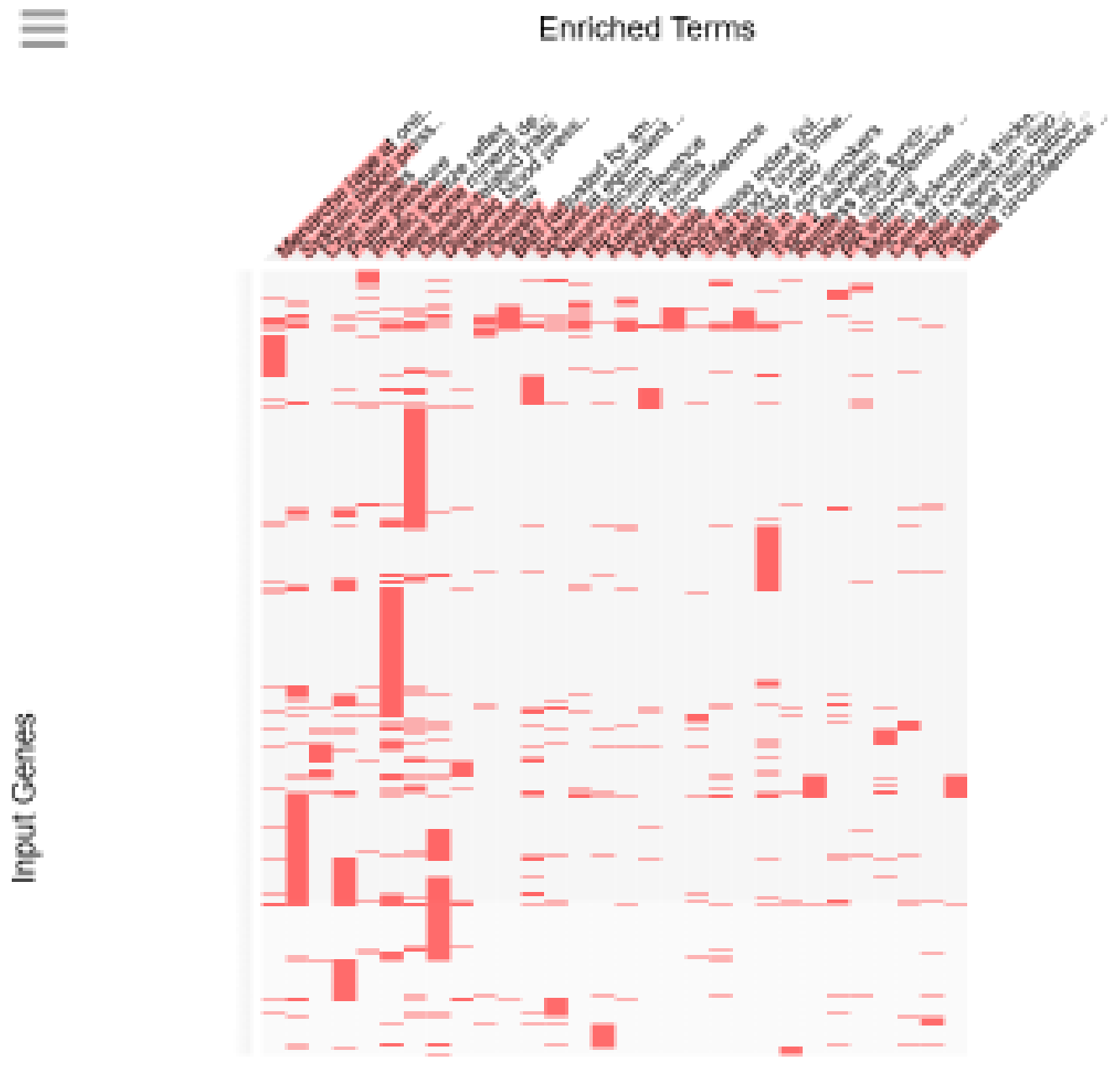

## GWAS Catalog 2019 (8,405 genes): Top 40 of 241 significant records

| Term | Overlap | P-value | Adjusted P-value |
|---|---|---|---|
| Menarche (age at onset) | 154/212 | 1.12E-19 | 1.4E-16 |
| Systolic blood pressure | 389/657 | 1.62E-19 | 1.4E-16 |
| Chin dimples | 62/74 | 1.51E-13 | 8.68E-11 |
| Pulse pressure | 321/567 | 9.66E-13 | 4.16E-10 |
| Photic sneeze reflex | 55/65 | 1.6E-12 | 5.51E-10 |
| Heel bone mineral density | 478/898 | 3.09E-12 | 8.88E-10 |
| Obesity-related traits | 427/804 | 6.98E-11 | 1.72E-08 |
| Diastolic blood pressure | 348/646 | 4.87E-10 | 1.05E-07 |
| Monobrow | 58/76 | 1.19E-09 | 2.28E-07 |
| Obesity | 45/55 | 1.61E-09 | 2.78E-07 |
| Height | 288/527 | 2.32E-09 | 3.63E-07 |
| Atrial fibrillation | 146/240 | 2.83E-09 | 3.78E-07 |
| BMI (adjusted for smoking behaviour) | 58/77 | 2.85E-09 | 3.78E-07 |
| Spherical equivalent or myopia (age of diagnosis) | 120/191 | 4.87E-09 | 5.99E-07 |
| Neuroticism | 69/97 | 5.87E-09 | 6.74E-07 |
| Hip circumference | 52/69 | 1.83E-08 | 1.97E-06 |
| Diverticular disease | 100/156 | 2.04E-08 | 2.06E-06 |
| Allergic rhinitis | 77/114 | 3.13E-08 | 3E-06 |
| Waist circumference | 55/75 | 3.67E-08 | 3.33E-06 |
| Myopia | 52/70 | 4.2E-08 | 3.62E-06 |
| Body mass index (joint analysis main effects and smoking interaction) | 56/77 | 4.5E-08 | 3.69E-06 |
| Hand grip strength | 99/156 | 5.05E-08 | 3.96E-06 |
| Total body bone mineral density | 31/36 | 5.77E-08 | 4.32E-06 |
| Waist-hip ratio | 44/58 | 1.65E-07 | 1.14E-05 |
| BMI in non-smokers | 44/58 | 1.65E-07 | 1.14E-05 |
| Type 2 diabetes | 217/397 | 2.08E-07 | 1.38E-05 |
| Restless legs syndrome | 32/39 | 3.48E-07 | 2.22E-05 |
| Blond vs. brown/black hair color | 98/160 | 6.9E-07 | 4.25E-05 |
| Waist circumference adjusted for BMI (adjusted for smoking behaviour) | 56/81 | 7.21E-07 | 4.29E-05 |
| Amyotrophic lateral sclerosis (sporadic) | 109/182 | 8.3E-07 | 4.77E-05 |
| PR interval | 48/67 | 8.6E-07 | 4.78E-05 |
| Motion sickness | 27/32 | 1.01E-06 | 5.45E-05 |
| Male-pattern baldness | 143/251 | 1.15E-06 | 6E-05 |
| Central corneal thickness | 47/66 | 1.48E-06 | 7.14E-05 |
| Paclitaxel disposition in epithelial ovarian cancer | 36/47 | 1.49E-06 | 7.14E-05 |
| Autism spectrum disorder, ADHD, bipolar disorder, MDD, and schizophrenia (combined) | 36/47 | 1.49E-06 | 7.14E-05 |
| Rosacea symptom severity | 87/141 | 1.84E-06 | 8.57E-05 |
| Waist circumference adjusted for BMI (joint analysis main effects and smoking interaction) | 54/79 | 1.98E-06 | 8.99E-05 |
| Waist circumference adjusted for body mass index | 80/128 | 2.29E-06 | 0.000101 |
| Glaucoma (primary open-angle) | 52/76 | 2.92E-06 | 0.000124 |

**Figure 4.** Identification of genes implicated in development and manifestations of hundreds physiological and pathological phenotypes and autosomal inheritance in Modern Humans.

## Disease Perturbations from GEO down: 8,405 genes

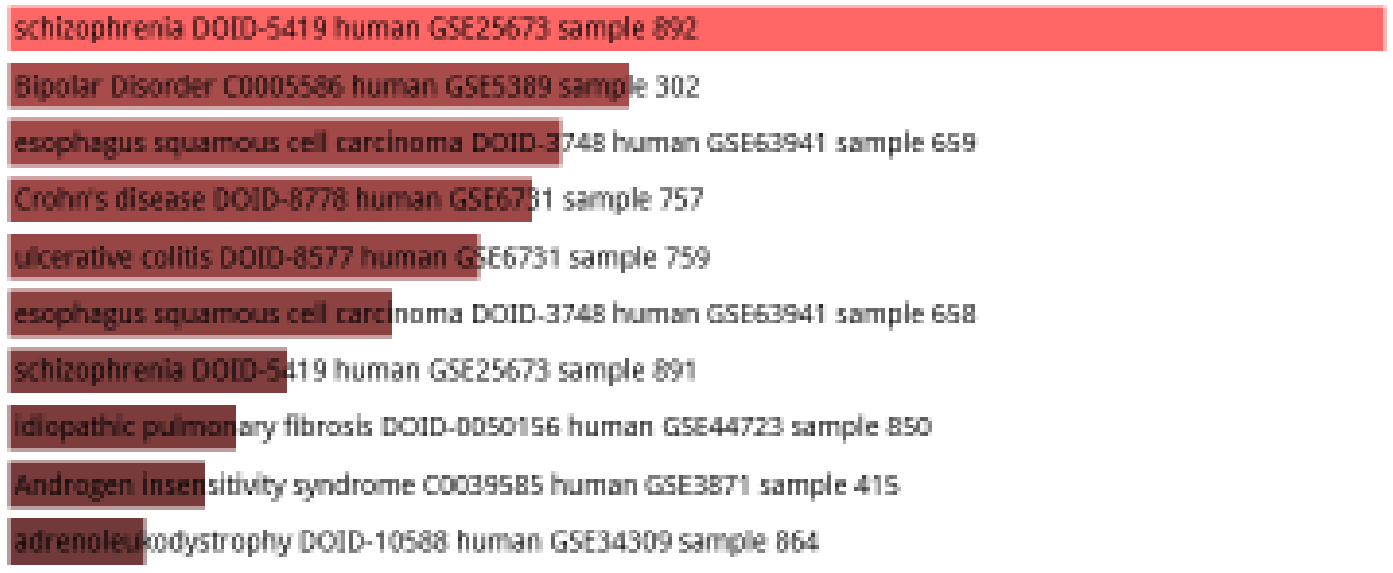

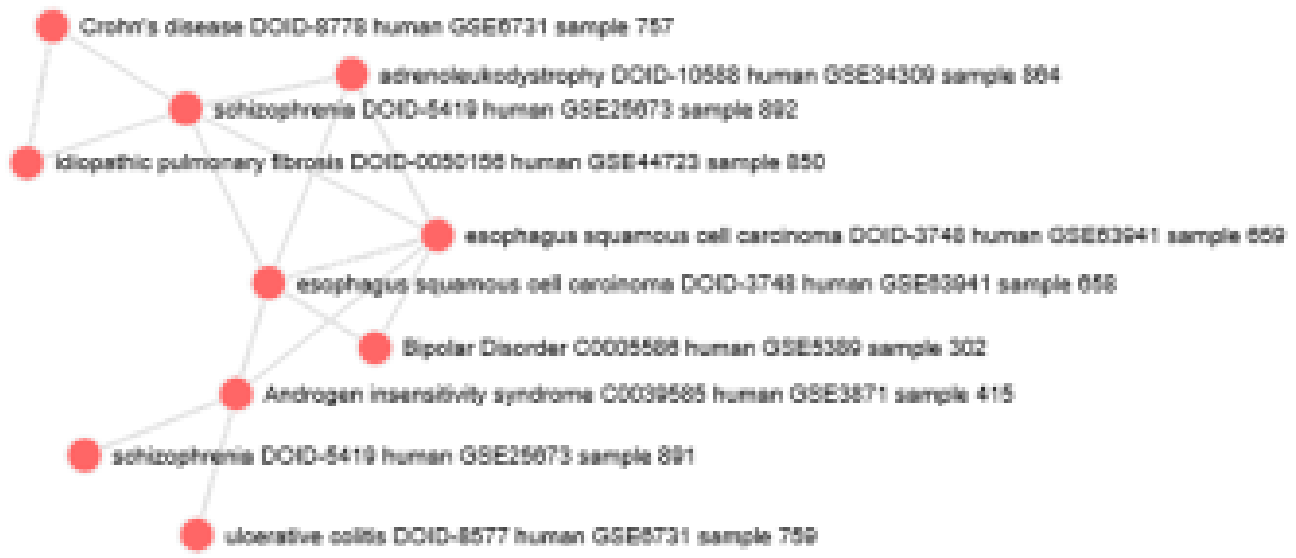

## Disease Perturbations from GEO down: 8,405 genes

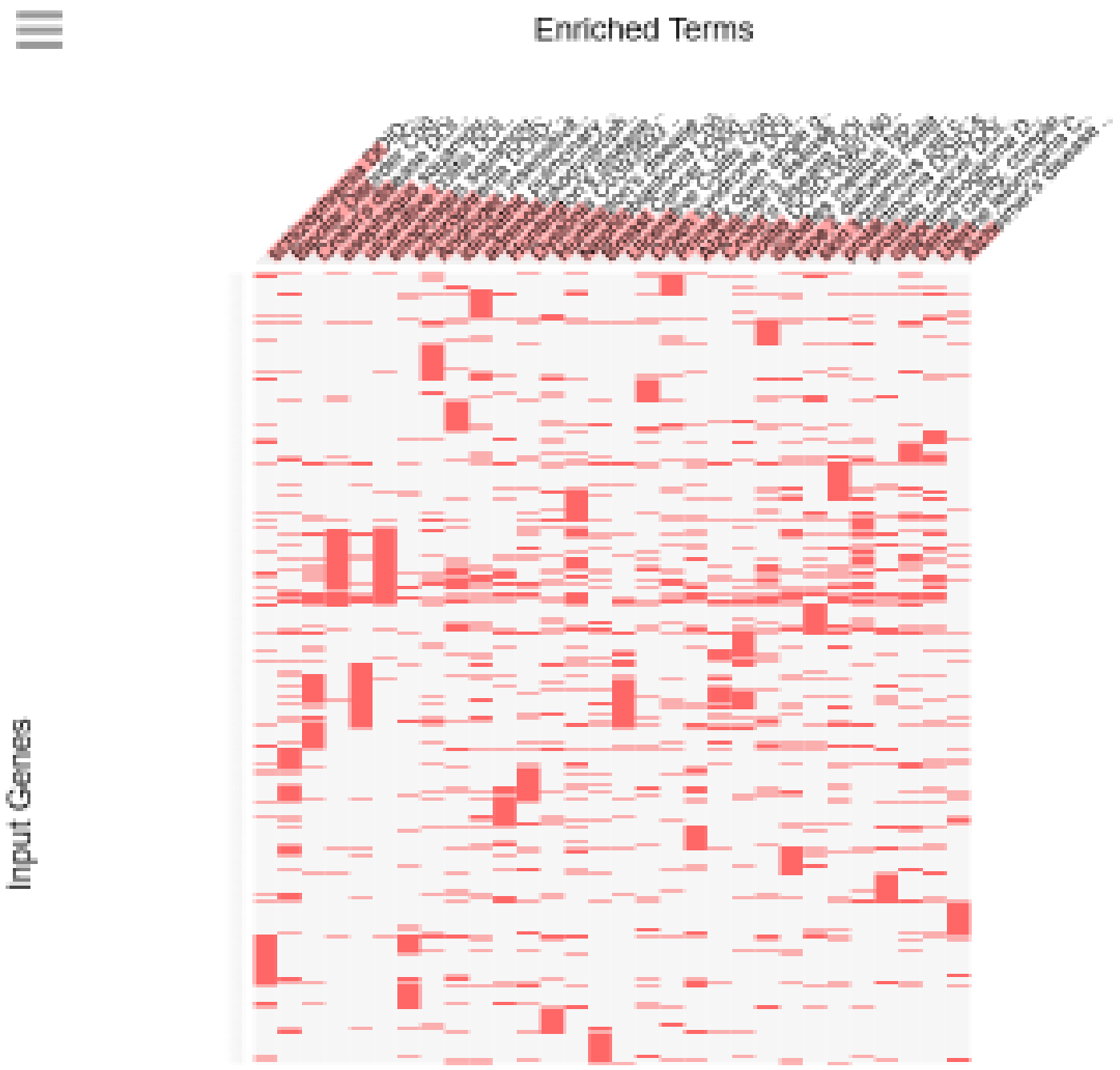

### Disease Perturbations from GEO down (8,405 genes): Top 30 of 240 significant records

| Term | Overlap | P-value | Adjusted P-value |
|---|---|---|---|
| schizophrenia DOID-5419 human GSE25673 sample 892 | 242/337 | 6.49E-29 | 5.44E-26 |
| Bipolar Disorder C0005586 human GSE5389 sample 302 | 256/395 | 2.78E-20 | 1.17E-17 |
| esophagus squamous cell carcinoma DOID-3748 human GSE63941 sample 659 | 252/391 | 1.65E-19 | 4.6E-17 |
| Crohn's disease DOID-8778 human GSE6731 sample 757 | 240/370 | 3.64E-19 | 7.63E-17 |
| ulcerative colitis DOID-8577 human GSE6731 sample 759 | 246/384 | 1.4E-18 | 2.34E-16 |
| esophagus squamous cell carcinoma DOID-3748 human GSE63941 sample 658 | 256/408 | 1.43E-17 | 2E-15 |
| schizophrenia DOID-5419 human GSE25673 sample 891 | 182/274 | 2.18E-16 | 2.61E-14 |
| idiopathic pulmonary fibrosis DOID-0050156 human GSE44723 sample 850 | 201/312 | 8.28E-16 | 8.68E-14 |
| Androgen insensitivity syndrome C0039585 human GSE3871 sample 415 | 208/327 | 1.94E-15 | 1.81E-13 |
| adrenoleukodystrophy DOID-10588 human GSE34309 sample 864 | 212/338 | 9.23E-15 | 7.75E-13 |
| Huntington's disease DOID-12858 mouse GSE3621 sample 704 | 219/356 | 6.55E-14 | 5E-12 |
| Dystonia C0393593 human GSE3064 sample 329 | 198/317 | 1.27E-13 | 8.89E-12 |
| Nephroblastoma C0027708 human GSE2712 sample 418 | 248/419 | 6.95E-13 | 4.48E-11 |
| Huntington's disease DOID-12858 mouse GSE3583 sample 929 | 183/293 | 1.08E-12 | 6.44E-11 |
| Ulcerative Colitis C0009324 human GSE6731 sample 249 | 213/354 | 3.26E-12 | 1.82E-10 |
| Breast Cancer C0006142 human GSE1378 sample 52 | 184/299 | 6.23E-12 | 3.27E-10 |
| Down syndrome DOID-14250 human GSE42956 sample 1060 | 156/247 | 1.41E-11 | 6.95E-10 |
| Primary open angle glaucoma C0339573 human GSE2705 sample 257 | 156/249 | 3.5E-11 | 1.63E-09 |
| colitis DOID-0060180 human GSE6731 sample 761 | 211/359 | 8.78E-11 | 3.71E-09 |
| Alzheimer's disease DOID-10652 human GSE4757 sample 592 | 216/369 | 8.85E-11 | 3.71E-09 |
| Crohn's disease DOID-8778 human GSE6731 sample 758 | 162/263 | 1.01E-10 | 4.01E-09 |
| prolactinoma DOID-5394 human GSE36314 sample 636 | 251/440 | 1.05E-10 | 4.01E-09 |
| diabetes mellitus type 2 DOID-9352 human GSE12643 sample 766 | 204/346 | 1.22E-10 | 4.34E-09 |
| type 2 diabetes mellitus DOID-9352 human GSE13760 sample 882 | 221/380 | 1.24E-10 | 4.34E-09 |
| skin squamous cell carcinoma DOID-3151 human GSE45164 sample 657 | 177/295 | 2.98E-10 | 1E-08 |
| prostate cancer DOID-10283 human GSE3868 sample 638 | 201/344 | 4.9E-10 | 1.58E-08 |
| Dental cavity, complex C0399396 human GSE1629 sample 175 | 228/401 | 1.12E-09 | 3.49E-08 |
| Alzheimer's disease DOID-10652 human GSE36980 sample 520 | 190/325 | 1.37E-09 | 4.1E-08 |
| oligodendroglioma DOID-3181 human GSE15824 sample 858 | 183/313 | 2.71E-09 | 7.83E-08 |
| breast cancer DOID-1612 human GSE3744 sample 978 | 247/443 | 2.84E-09 | 7.95E-08 |

### Disease Perturbations from GEO up: 8,405 genes

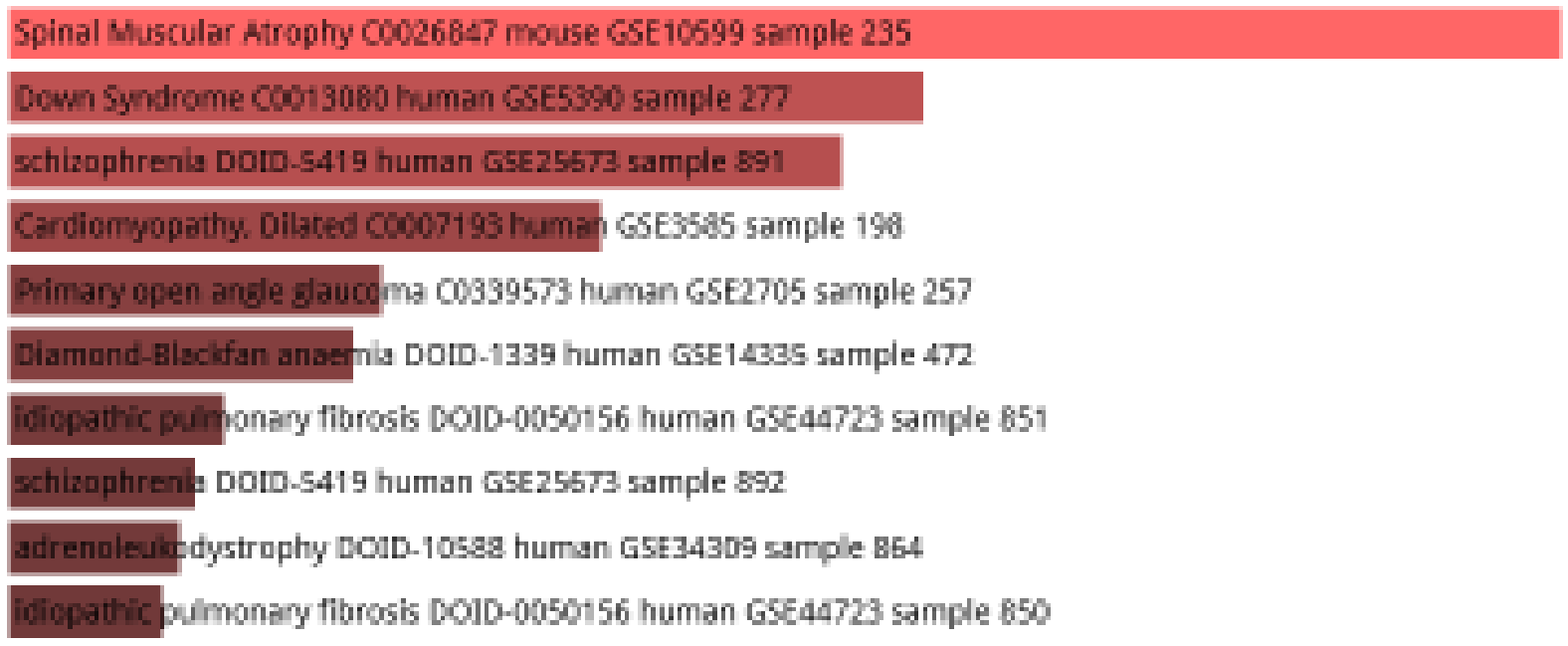

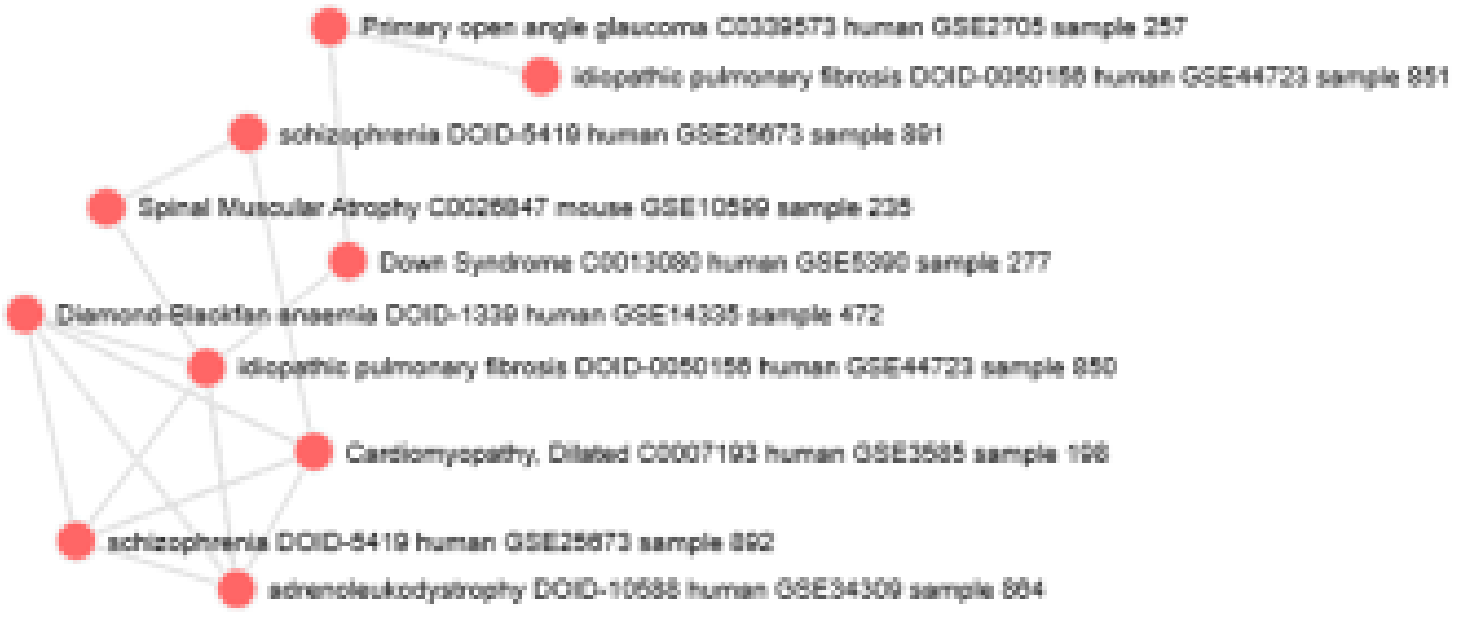

**Disease Perturbations from GEO up: 8,405 genes**

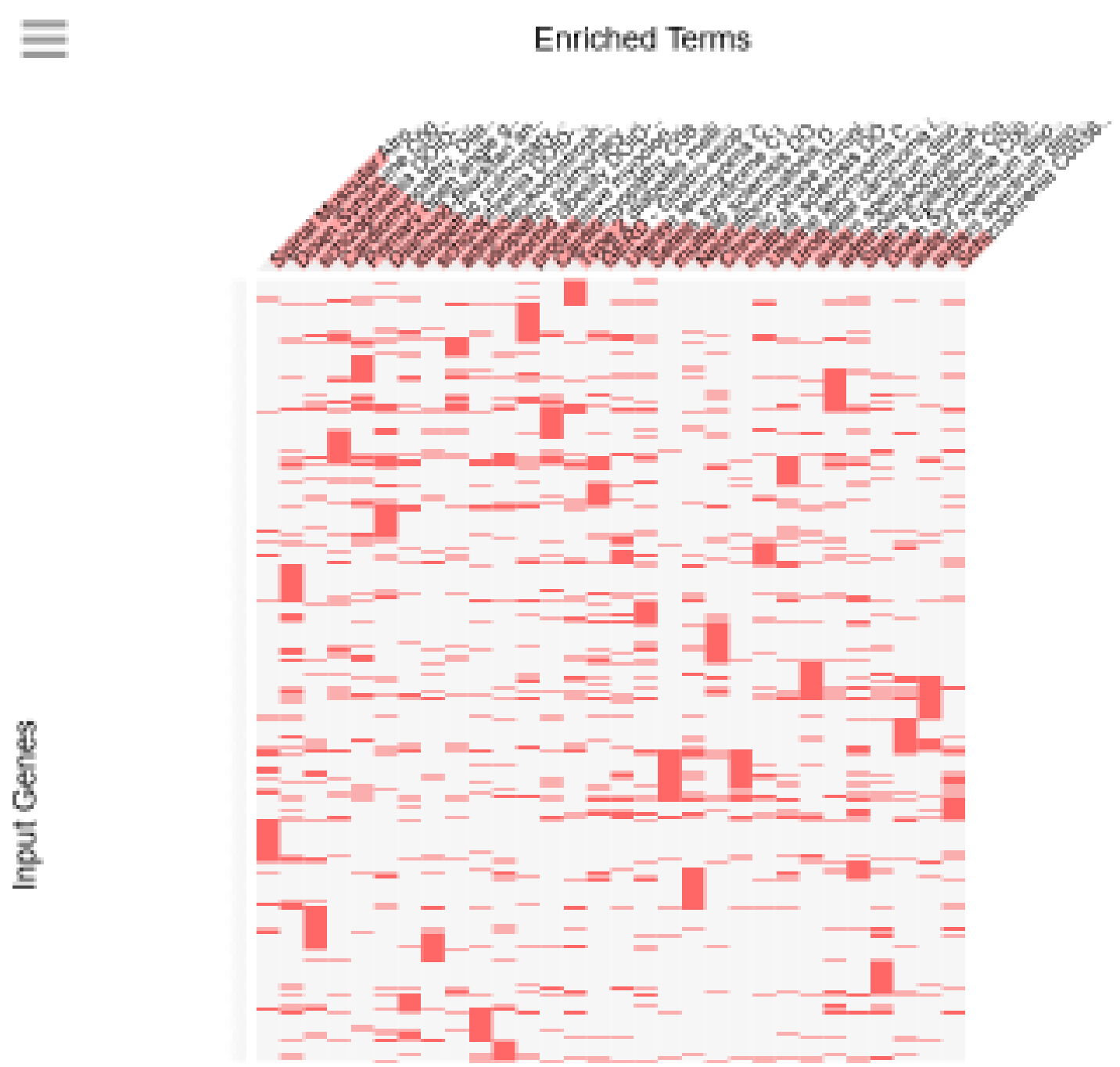

**Disease Perturbations from GEO up (8,405 genes): Top 30 of 204 significant records**

| Term | Overlap | P-value | Adjusted P-value |
|---|---|---|---|
| Spinal Muscular Atrophy C0026847 mouse GSE10599 sample 235 | 254/368 | 4.69E-26 | 3.93E-23 |
| Down Syndrome C0013080 human GSE5390 sample 277 | 288/460 | 2.08E-19 | 8.72E-17 |
| schizophrenia DOID-5419 human GSE25673 sample 891 | 215/326 | 1.48E-18 | 4.14E-16 |
| Cardiomyopathy, Dilated C0007193 human GSE3585 sample 198 | 209/326 | 4.82E-16 | 1.01E-13 |
| Primary open angle glaucoma C0339573 human GSE2705 sample 257 | 216/351 | 9.23E-14 | 1.55E-11 |
| Diamond-Blackfan anaemia DOID-1339 human GSE14335 sample 472 | 231/382 | 1.88E-13 | 2.64E-11 |
| idiopathic pulmonary fibrosis DOID-0050156 human GSE44723 sample 851 | 185/300 | 4.2E-12 | 5.03E-10 |
| schizophrenia DOID-5419 human GSE25673 sample 892 | 165/263 | 8.13E-12 | 8.53E-10 |
| adrenoleukodystrophy DOID-10588 human GSE34309 sample 864 | 164/262 | 1.23E-11 | 1.14E-09 |
| idiopathic pulmonary fibrosis DOID-0050156 human GSE44723 sample 850 | 177/288 | 1.83E-11 | 1.5E-09 |
| morbid obesity DOID-11981 human GSE48964 sample 583 | 180/294 | 1.96E-11 | 1.5E-09 |
| Cardiomyopathy C0878544 human GSE1869 sample 79 | 197/330 | 5.63E-11 | 3.87E-09 |
| Type 2 diabetes mellitus C0011860 human GSE12643 sample 274 | 205/346 | 5.99E-11 | 3.87E-09 |
| GERD - Gastro-esophageal reflux disease C0017168 human GSE2144 sample 27 | 203/346 | 2.46E-10 | 1.48E-08 |
| Uterine leiomyoma C0042133 human GSE2725 sample 399 | 158/258 | 3.19E-10 | 1.78E-08 |
| multiple sclerosis DOID-2377 human GSE38010 sample 737 | 183/310 | 9.47E-10 | 4.96E-08 |
| Chronic phase chronic myelogenous leukemia DOID-8552 human GSE5550 sample 456 | 192/330 | 1.92E-09 | 9.47E-08 |
| autism spectrum disorder DOID-0060041 human GSE28521 sample 1041 | 185/317 | 2.71E-09 | 1.26E-07 |
| Setleis syndrome C1744559 human GSE16524 sample 285 | 222/393 | 4E-09 | 1.77E-07 |
| Neurofibromatosis DOID-8712 mouse GSE1482 sample 667 | 210/369 | 4.54E-09 | 1.9E-07 |
| multiple sclerosis DOID-2377 human GSE38010 sample 738 | 180/309 | 5.36E-09 | 2.14E-07 |
| Uterine leiomyoma C0042133 human GSE593 sample 16 | 160/270 | 7.23E-09 | 2.76E-07 |
| autism spectrum disorder DOID-0060041 human GSE28521 sample 1040 | 194/341 | 1.8E-08 | 6.56E-07 |
| Urothelial carcinoma in situ C0334267 human GSE3167 sample 229 | 213/380 | 1.94E-08 | 6.77E-07 |
| Status Epilepticus C0038220 rat GSE4236 sample 391 | 191/336 | 2.51E-08 | 8.11E-07 |
| autism spectrum disorder DOID-0060041 human GSE28521 sample 1039 | 191/336 | 2.51E-08 | 8.11E-07 |
| adrenoleukodystrophy DOID-10588 human GSE34308 sample 709 | 205/365 | 2.88E-08 | 8.95E-07 |
| Down Syndrome C0013080 human GSE10758 sample 310 | 210/376 | 3.57E-08 | 1.07E-06 |
| chronic lymphocytic leukemia DOID-1040 human GSE6691 sample 786 | 180/315 | 3.77E-08 | 1.09E-06 |
| Oligodendroglioma C0028945 human GSE2223 sample 116 | 167/289 | 4.08E-08 | 1.14E-06 |

**Figure 5.** Identification of genes expression of which is altered in several hundred common human disorders.

## Rare Diseases GeneRIF Gene Lists: 8,405 genes

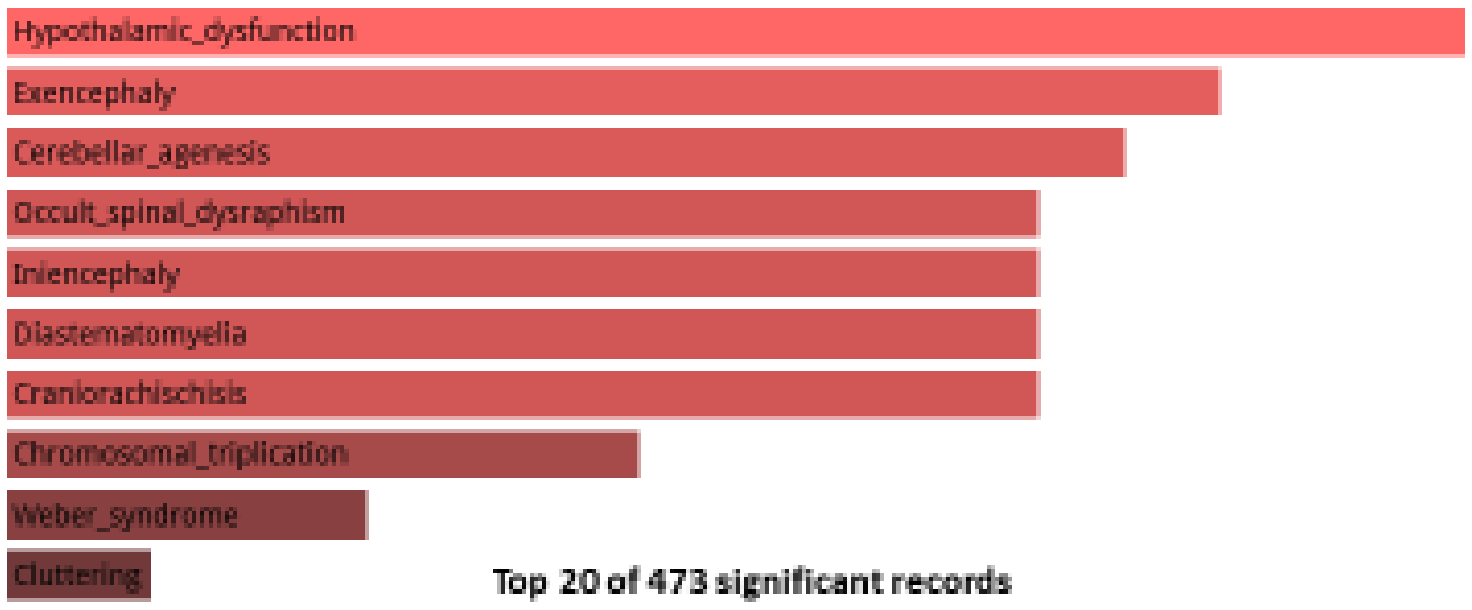

Top 20 of 473 significant records

| Term | Overlap | P-value | Adjusted P-value |
|---|---|---|---|
| Hypothalamic_dysfunction | 90/128 | 8.12E-11 | 6.55E-08 |
| Exencephaly | 158/256 | 1.38E-10 | 6.55E-08 |
| Cerebellar_agenesis | 198/335 | 1.7E-10 | 6.55E-08 |
| Occult_spinal_dysraphism | 157/255 | 2.05E-10 | 6.55E-08 |
| Diastematomyelia | 157/255 | 2.05E-10 | 6.55E-08 |
| Craniorachischisis | 157/255 | 2.05E-10 | 6.55E-08 |
| Iniencephaly | 157/255 | 2.05E-10 | 6.55E-08 |
| Chromosomal_triplication | 201/344 | 4.9E-10 | 1.37E-07 |
| Weber_syndrome | 94/139 | 8.85E-10 | 2.2E-07 |
| Cluttering | 106/162 | 1.41E-09 | 3.16E-07 |
| Mental_retardation_epilepsy | 190/326 | 1.92E-09 | 3.9E-07 |
| Single_ventricular_heart | 73/104 | 5.49E-09 | 1.02E-06 |
| Sydenham's_chorea | 209/369 | 8.42E-09 | 1.34E-06 |
| Chorea_minor | 209/369 | 8.42E-09 | 1.34E-06 |
| Antisocial_personality_disorder | 46/59 | 1.93E-08 | 2.87E-06 |
| Sudden_infant_death_syndrome | 103/162 | 2.37E-08 | 3.31E-06 |
| Stress_cardiomyopathy | 110/176 | 3.16E-08 | 4.16E-06 |
| N_syndrome | 222/401 | 3.89E-08 | 4.83E-06 |
| Basilar_migraine | 127/210 | 4.85E-08 | 5.7E-06 |
| Corpus_callosum_agenesis | 89/138 | 7.92E-08 | 8.85E-06 |

## Rare Diseases GeneRIF Gene Lists: 8,405 genes

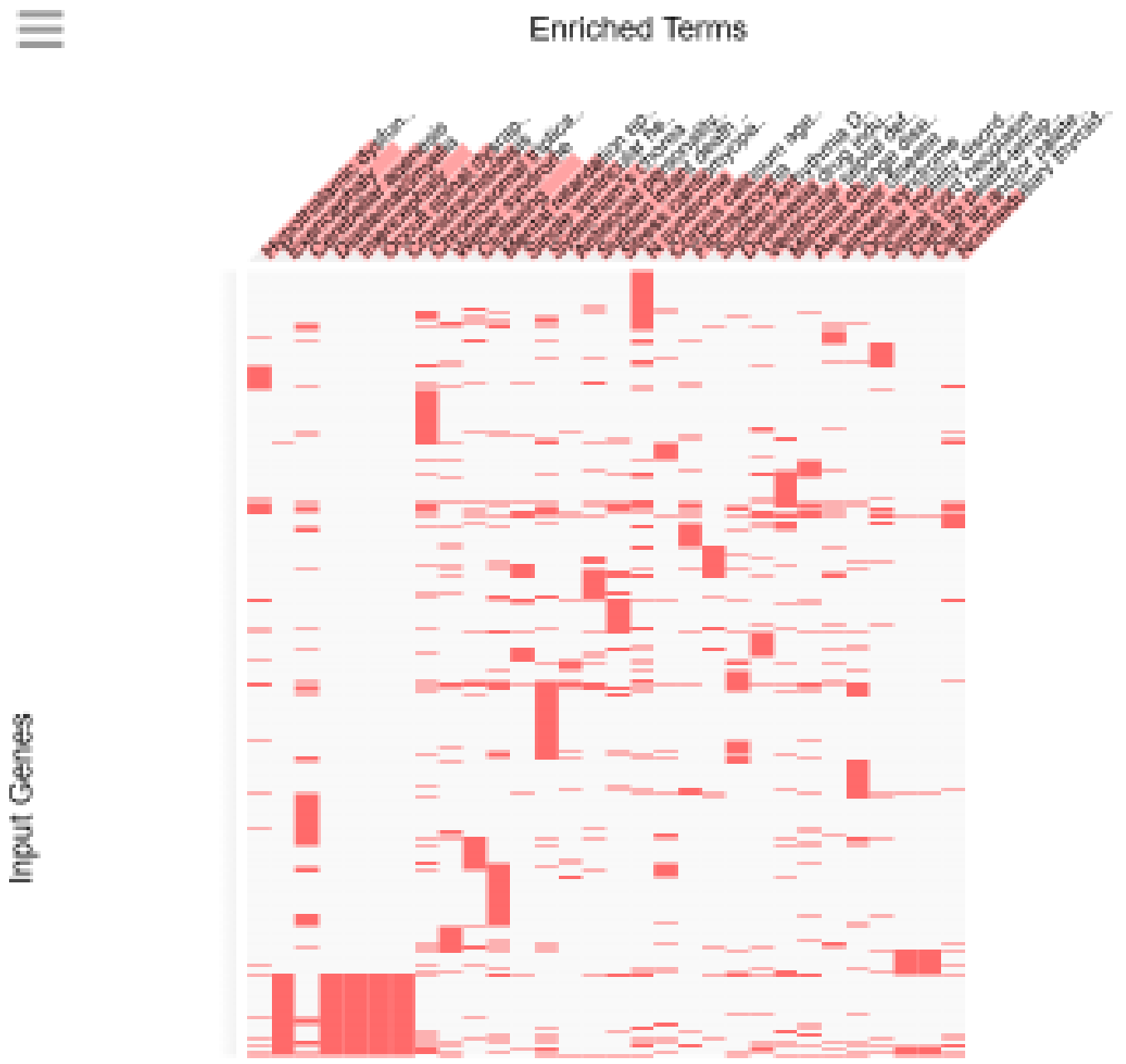

## Rare Diseases GeneRIF ARCHS4 Predictions: 8,405 genes

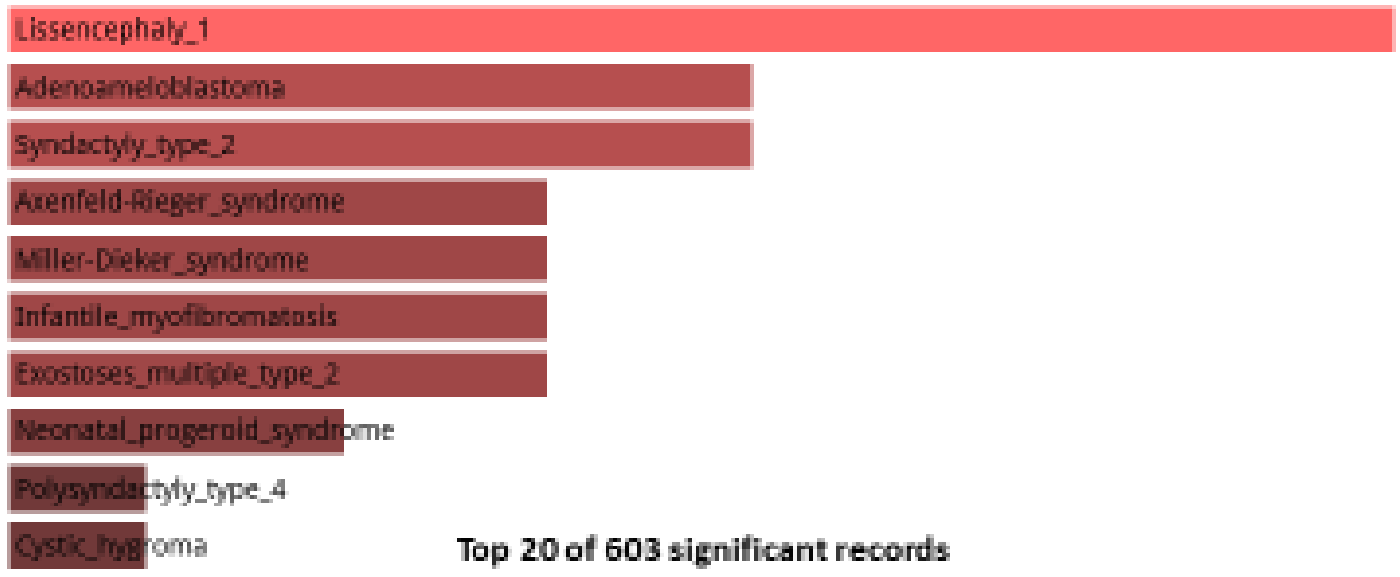

**Top 20 of 603 significant records**

| Term | Overlap | P-value | Adjusted P-value |
|---|---|---|---|
| Lissencephaly_1 | 154/200 | 4.79E-24 | 1.08E-20 |
| Adenoameloblastoma | 151/200 | 4.35E-22 | 3.26E-19 |
| Syndactyly_type_2 | 151/200 | 4.35E-22 | 3.26E-19 |
| Axenfeld-Rieger_syndrome | 150/200 | 1.85E-21 | 5.94E-19 |
| Miller-Dieker_syndrome | 150/200 | 1.85E-21 | 5.94E-19 |
| Exostoses_multiple_type_2 | 150/200 | 1.85E-21 | 5.94E-19 |
| Infantile_myofibromatosis | 150/200 | 1.85E-21 | 5.94E-19 |
| Neonatal_progeroid_syndrome | 149/200 | 7.69E-21 | 2.16E-18 |
| Polysyndactyly_type_4 | 148/200 | 3.11E-20 | 6.97E-18 |
| Cystic_hygroma | 148/200 | 3.11E-20 | 6.97E-18 |
| Acromesomelic_dysplasia_Hunter_Thompson_type | 147/200 | 1.22E-19 | 2.11E-17 |
| Acromesomelic_dysplasia | 147/200 | 1.22E-19 | 2.11E-17 |
| Exostoses_multiple_type_1 | 147/200 | 1.22E-19 | 2.11E-17 |
| Scholte_syndrome | 146/200 | 4.69E-19 | 6.58E-17 |
| Congenital_diaphragmatic_hernia | 146/200 | 4.69E-19 | 6.58E-17 |
| Hereditary_multiple_osteochondromas | 146/200 | 4.69E-19 | 6.58E-17 |
| Aortopulmonary_window | 145/200 | 1.76E-18 | 1.88E-16 |
| Corpus_callosum_agenesis | 145/200 | 1.76E-18 | 1.88E-16 |
| Apert_syndrome | 145/200 | 1.76E-18 | 1.88E-16 |
| Subependymoma | 145/200 | 1.76E-18 | 1.88E-16 |

## Rare Diseases GeneRIF ARCHS4 Predictions: 8,405 genes

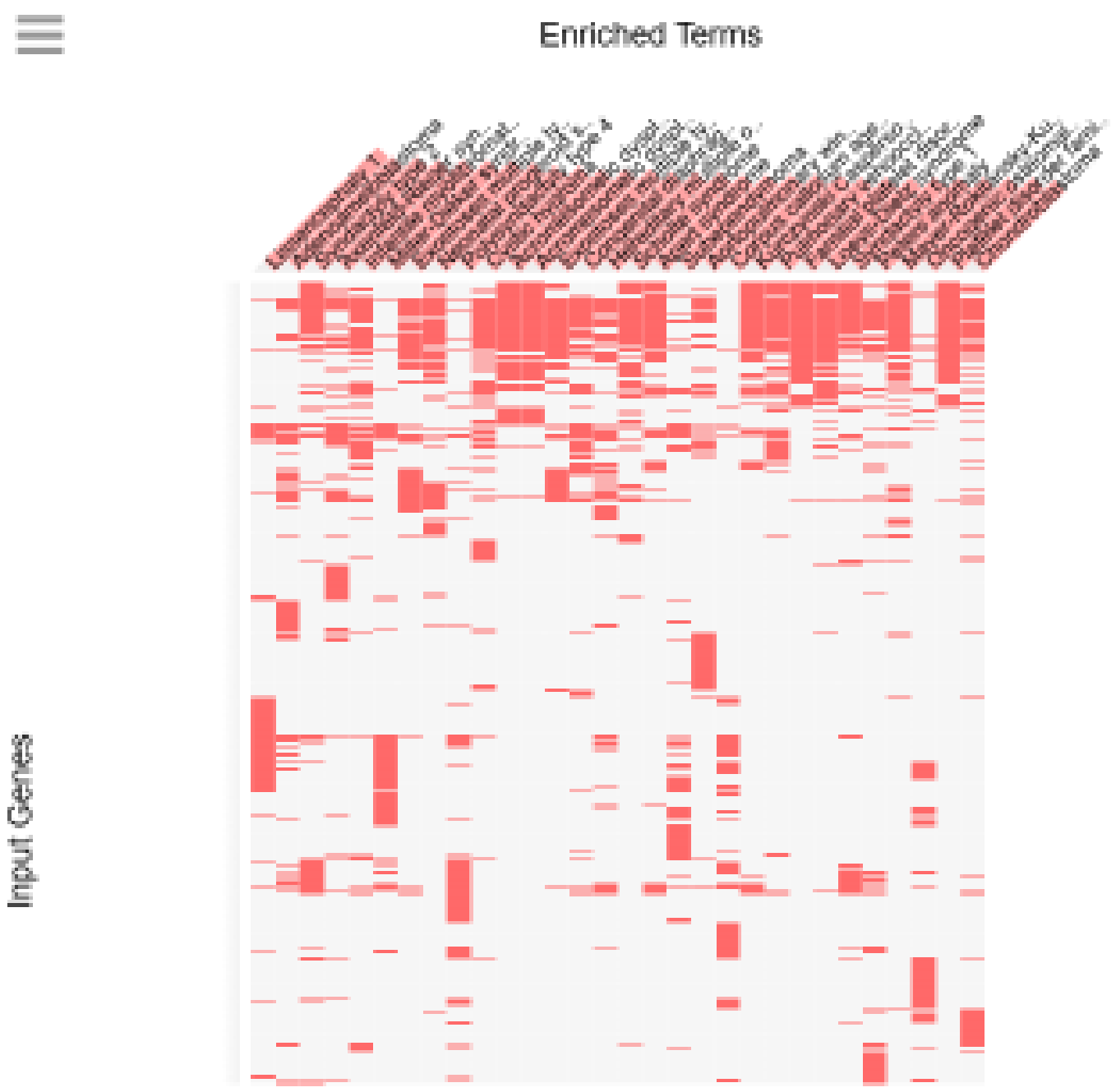

**Rare Diseases AutoRIF ARCHS4 Predictions: 8,405 genes**

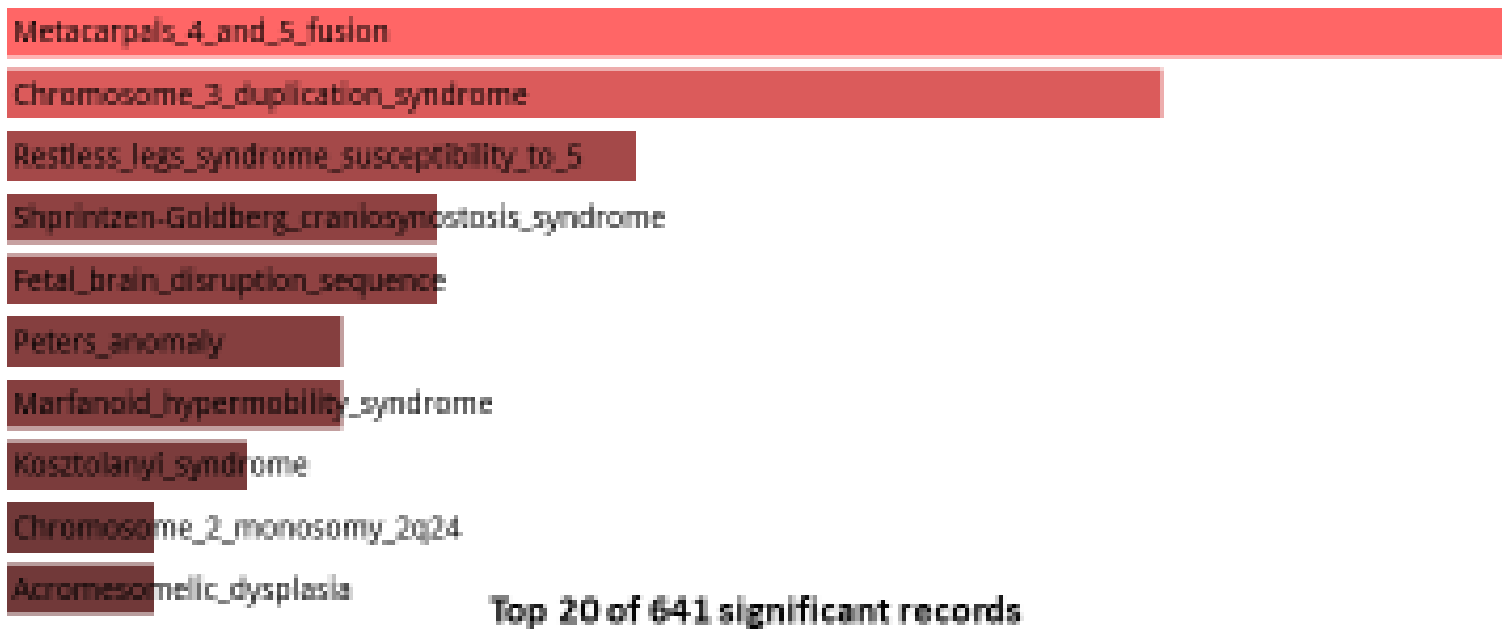

**Top 20 of 641 significant records**

| Term | Overlap | P-value | Adjusted P-value |
|---|---|---|---|
| Metacarpals_4_and_5_fusion | 163/200 | 1.33E-30 | 4.97E-27 |
| Chromosome_3_duplication_syndrome | 160/200 | 2.69E-28 | 5.01E-25 |
| Restless_legs_syndrome_susceptibility_to_5 | 155/200 | 1.01E-24 | 1.25E-21 |
| Shprintzen-Goldberg_craniosynostosis_syndrome | 153/200 | 2.21E-23 | 1.65E-20 |
| Fetal_brain_disruption_sequence | 153/200 | 2.21E-23 | 1.65E-20 |
| Peters_anomaly | 152/200 | 9.95E-23 | 5.3E-20 |
| Marfanoid_hypermobility_syndrome | 152/200 | 9.95E-23 | 5.3E-20 |
| Kosztolanyi_syndrome | 151/200 | 4.35E-22 | 2.03E-19 |
| Chromosome_2_monosomy_2q24 | 150/200 | 1.85E-21 | 6.28E-19 |
| Acromesomelic_dysplasia_Hunter_Thompson_type | 150/200 | 1.85E-21 | 6.28E-19 |
| Acromesomelic_dysplasia | 150/200 | 1.85E-21 | 6.28E-19 |
| Chromosome_3_trisomy_3p | 149/200 | 7.69E-21 | 2.2E-18 |
| Buschke_Ollendorff_syndrome | 149/200 | 7.69E-21 | 2.2E-18 |
| Fraser_like_syndrome | 148/200 | 3.11E-20 | 6.81E-18 |
| Crandall_syndrome | 148/200 | 3.11E-20 | 6.81E-18 |
| Orofaciodigital_syndrome_11 | 148/200 | 3.11E-20 | 6.81E-18 |
| Postaxial_polydactyly_mental_retardation | 148/200 | 3.11E-20 | 6.81E-18 |
| Short_rib-polydactyly_syndrome_type_4 | 147/200 | 1.22E-19 | 2.28E-17 |
| Trichorhinophalangeal_syndrome_type_3 | 147/200 | 1.22E-19 | 2.28E-17 |
| Duane_syndrome | 147/200 | 1.22E-19 | 2.28E-17 |

**Rare Diseases AutoRIF ARCHS4 Predictions: 8,405 genes**

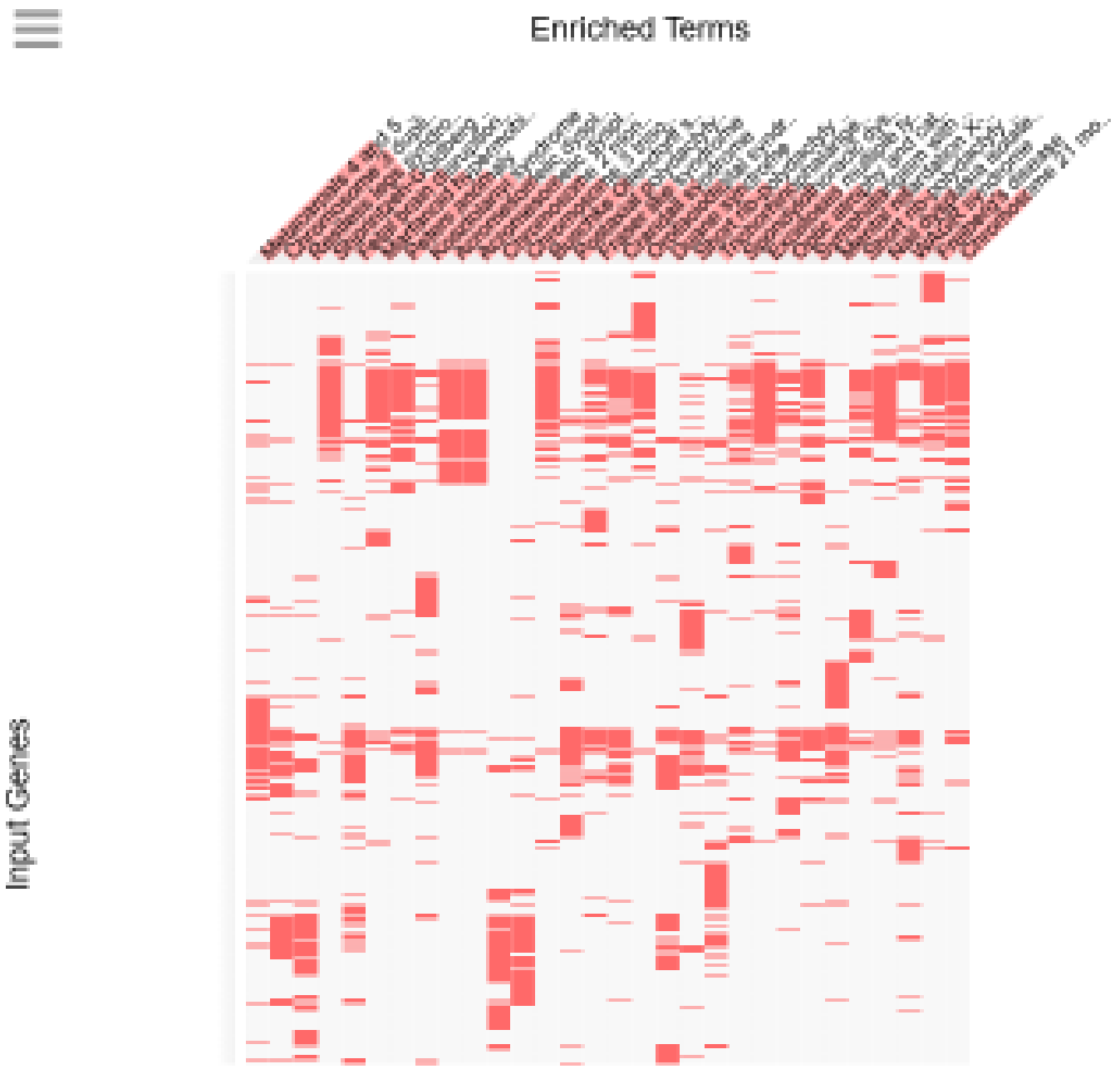

## Rare Diseases AutoRIF Gene Lists: 8,405 genes

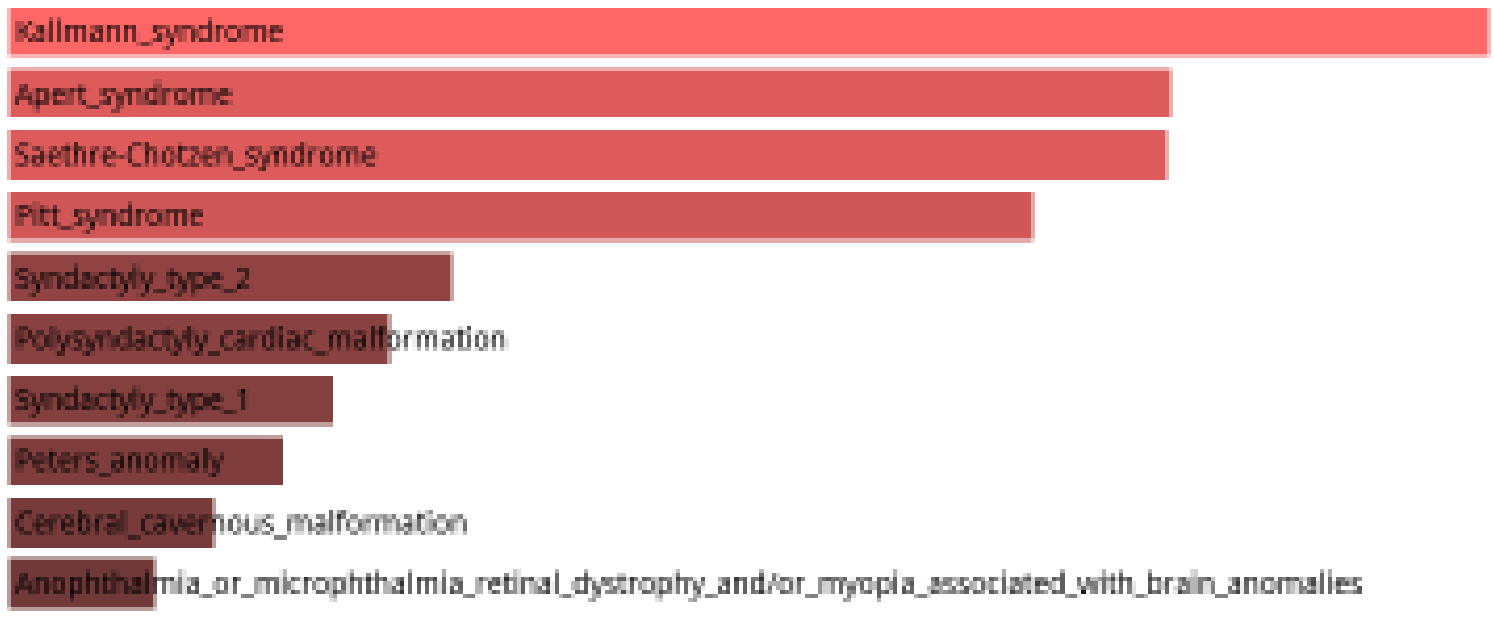

**Top 20 of 1,116 significant records**

| Term | Overlap | P-value | Adjusted P-value |
|---|---|---|---|
| Kallmann_syndrome | 159/247 | 9.25E-13 | 3.25E-09 |
| Apert_syndrome | 145/219 | 2.58E-12 | 3.25E-09 |
| Saethre-Chotzen_syndrome | 136/206 | 2.62E-12 | 3.25E-09 |
| Pitt_syndrome | 241/410 | 4.01E-12 | 3.73E-09 |
| Syndactyly_type_2 | 124/188 | 2.62E-11 | 1.95E-08 |
| Polysyndactyly_cardiac_malformation | 77/104 | 3.2E-11 | 1.98E-08 |
| Syndactyly_type_1 | 122/185 | 3.85E-11 | 2.04E-08 |
| Peters_anomaly | 75/101 | 4.53E-11 | 2.1E-08 |
| Cerebral_cavernous_malformation | 157/252 | 5.65E-11 | 2.33E-08 |
| Anophthalmia_or_microphthalmia_retinal_dystrophy_and/or_myopia_associated_with_brain_anomalies | 182/301 | 6.82E-11 | 2.36E-08 |
| X-linked_periventricular_heterotopia | 89/126 | 6.98E-11 | 2.36E-08 |
| Aniridia | 171/282 | 1.85E-10 | 5.73E-08 |
| Dominant_cleft_palate | 163/267 | 2.37E-10 | 6.76E-08 |
| Craniofacial_and_skeletal_defects | 94/137 | 2.64E-10 | 7.01E-08 |
| Anodontia | 140/223 | 2.84E-10 | 7.03E-08 |
| Glaucoma_congenital | 202/345 | 3.48E-10 | 8.08E-08 |
| Childhood-Onset_Schizophrenia | 168/278 | 3.87E-10 | 8.46E-08 |
| Hennekam_syndrome | 139/223 | 6.68E-10 | 1.38E-07 |
| Kurczynski_Casperson_syndrome | 114/176 | 8.57E-10 | 1.68E-07 |
| Osteochondroma | 177/299 | 1.31E-09 | 2.44E-07 |

## Rare Diseases AutoRIF Gene Lists: 8,405 genes

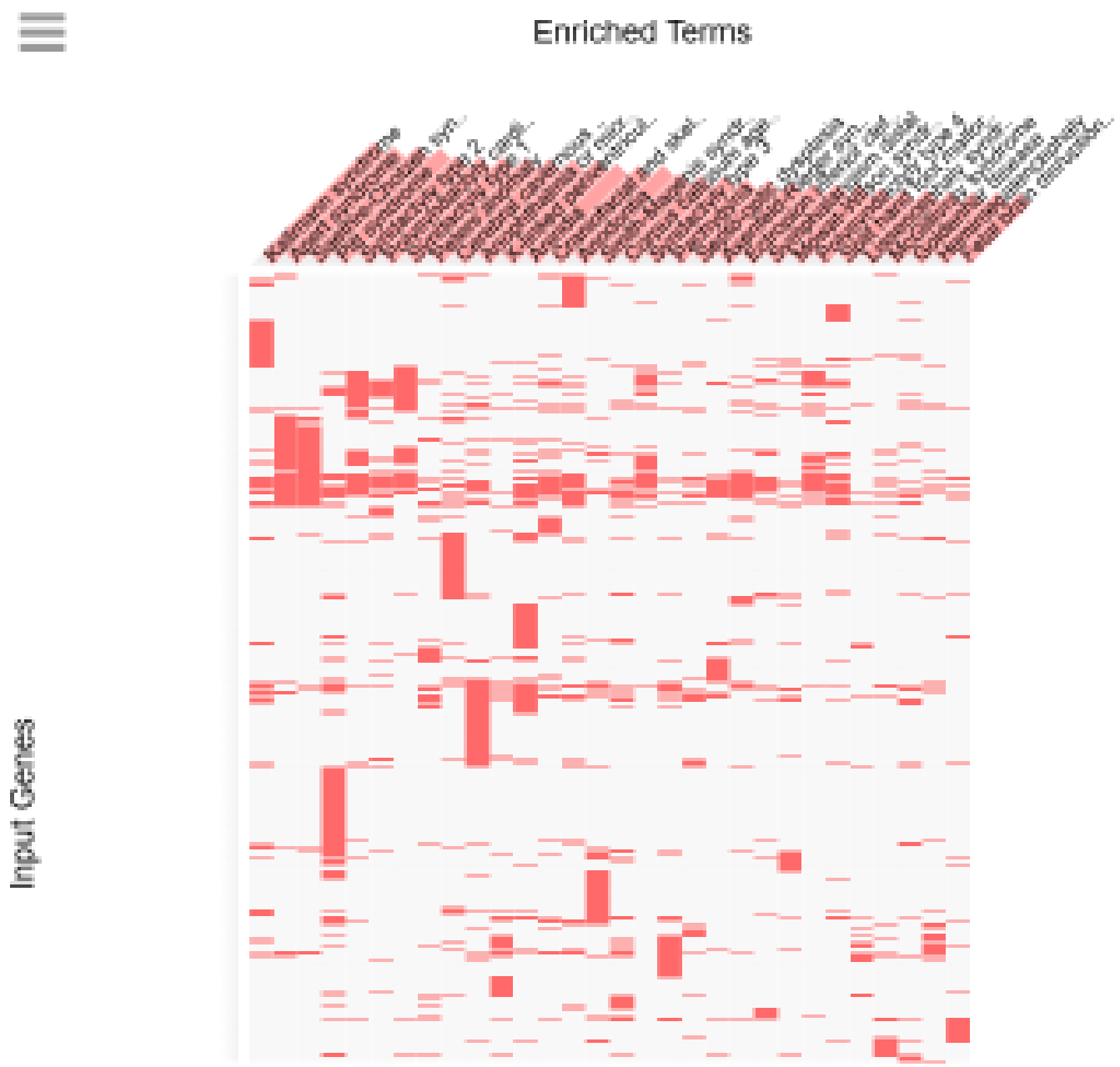

**Figure 6.** Identification of genes implicated in more than 1,000 records classified as human rare diseases.

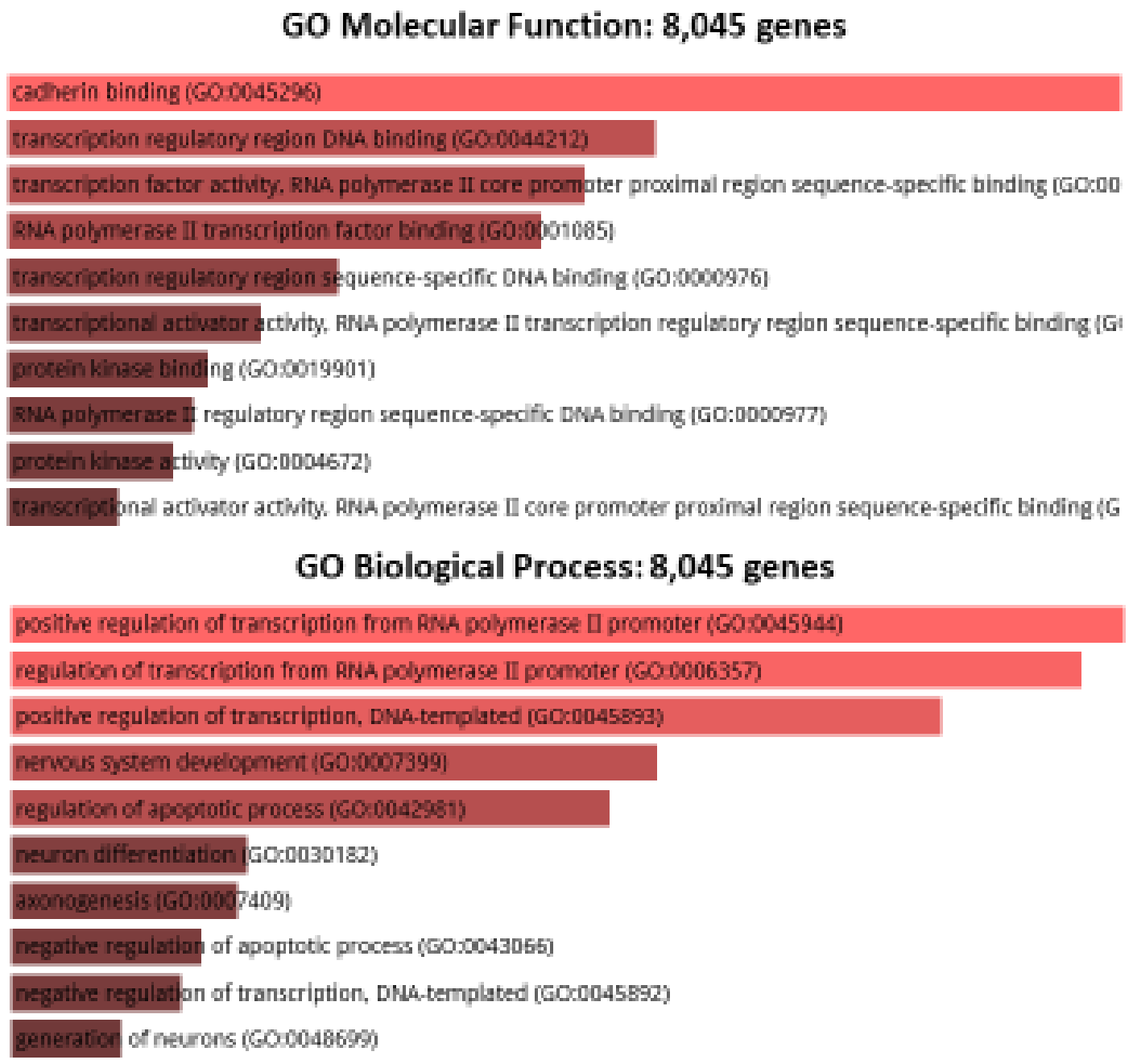
GO Molecular Function: 8,045 genes
cadherin binding (GO:0045296)
transcription regulatory region DNA binding (GO:0044212)
transcription factor activity, RNA polymerase II core promoter proximal region sequence-specific binding (GO:00
RNA polymerase II transcription factor binding (GO:0001085)
transcription regulatory region sequence-specific DNA binding (GO:0000976)
transcriptional activator activity, RNA polymerase II transcription regulatory region sequence-specific binding (G
protein kinase binding (GO:0019901)
RNA polymerase II regulatory region sequence-specific DNA binding (GO:0000977)
protein kinase activity (GO:0004672)
transcriptional activator activity, RNA polymerase II core promoter proximal region sequence-specific binding (G
GO Biological Process: 8,045 genes
positive regulation of transcription from RNA polymerase II promoter (GO:0045944)
regulation of transcription from RNA polymerase II promoter (GO:0006357)
positive regulation of transcription, DNA-templated (GO:0045893)
nervous system development (GO:0007399)
regulation of apoptotic process (GO:0042981)
neuron differentiation (GO:0030182)
axonogenesis (GO:0007409)
negative regulation of apoptotic process (GO:0043066)
negative regulation of transcription, DNA-templated (GO:0045892)
generation of neurons (GO:0048699)

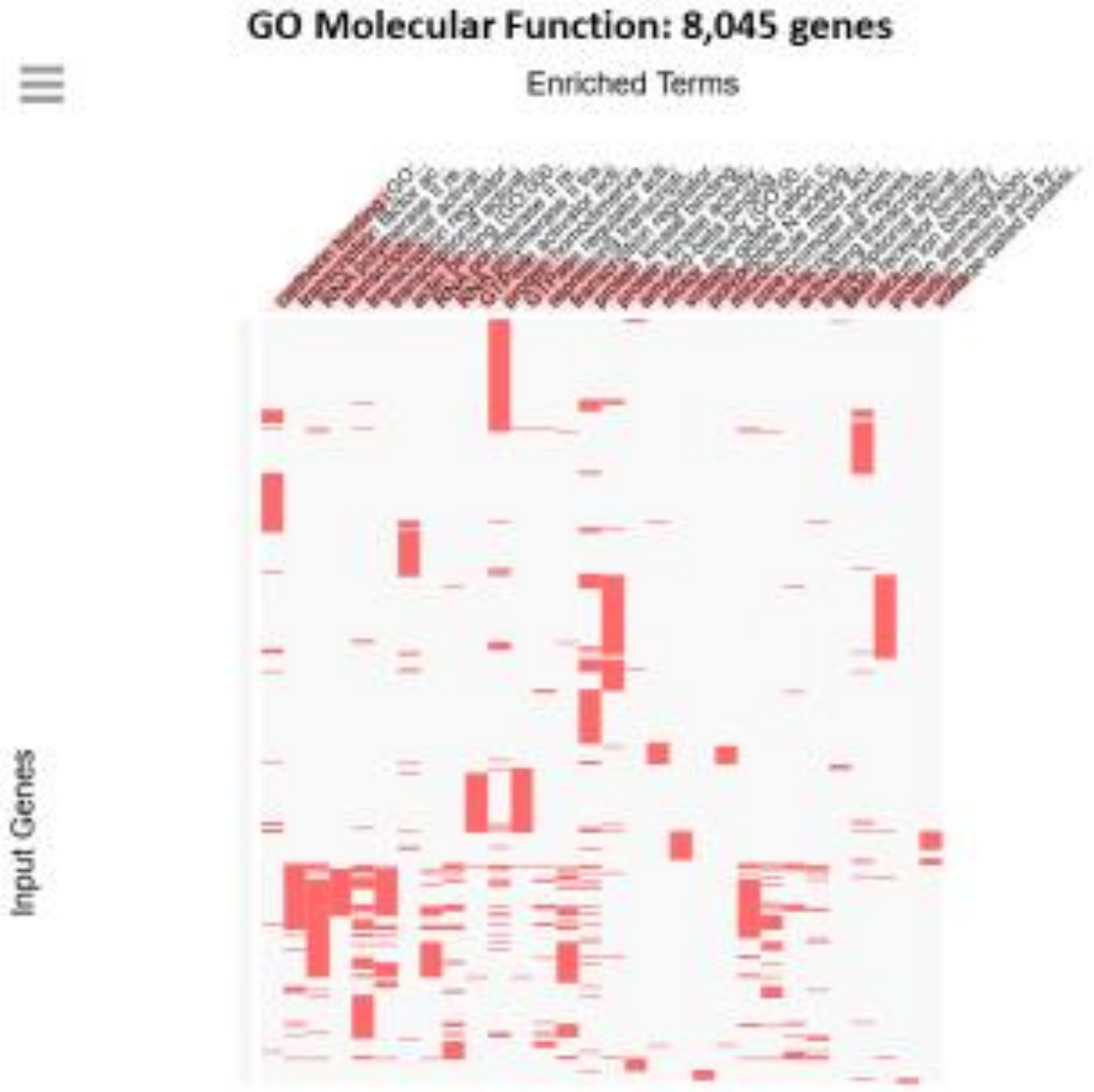
GO Molecular Function: 8,045 genes
Enriched Terms
Input Genes

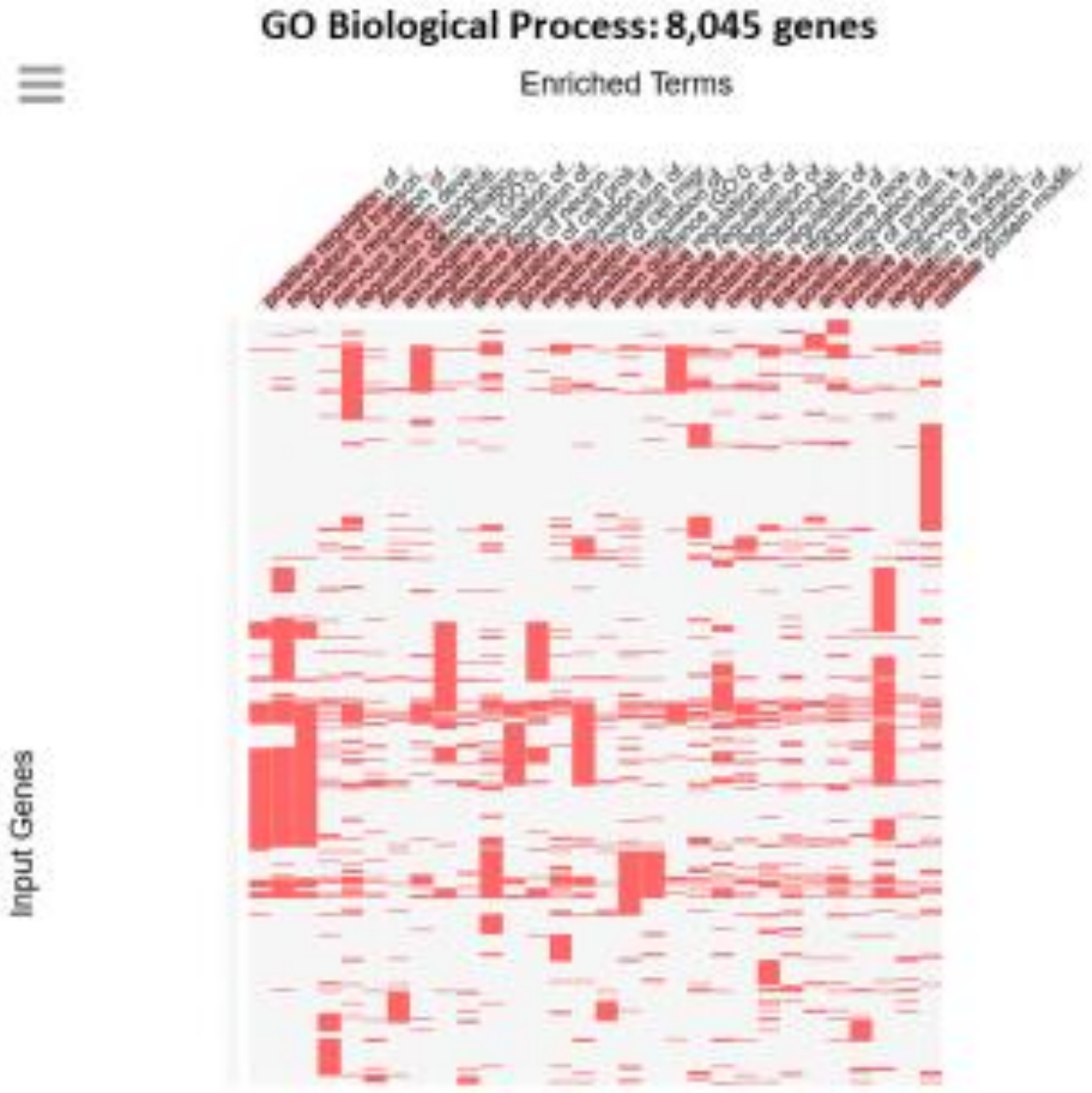
GO Biological Process: 8,045 genes
Enriched Terms
Input Genes

### GO Biological Process (8,045 genes): Top 35 of 308 significant records

| Term | Overlap | P-value | Adjusted P-value |
|---|---|---|---|
| positive regulation of transcription from RNA polymerase II promoter (GO:0045944) | 485/849 | 1.12E-19 | 5.67E-16 |
| regulation of transcription from RNA polymerase II promoter (GO:0006357) | 786/1479 | 2.51E-19 | 6.37E-16 |
| positive regulation of transcription, DNA-templated (GO:0045893) | 611/1121 | 3.46E-18 | 5.85E-15 |
| nervous system development (GO:0007399) | 276/456 | 7.07E-16 | 8.98E-13 |
| regulation of apoptotic process (GO:0042981) | 453/816 | 1.74E-15 | 1.77E-12 |
| neuron differentiation (GO:0030182) | 100/140 | 1.57E-12 | 1.33E-09 |
| axonogenesis (GO:0007409) | 146/224 | 1.88E-12 | 1.37E-09 |
| negative regulation of apoptotic process (GO:0043066) | 279/486 | 3.63E-12 | 2.31E-09 |
| negative regulation of transcription, DNA-templated (GO:0045892) | 437/814 | 5.44E-12 | 3.07E-09 |
| generation of neurons (GO:0048699) | 93/131 | 1.66E-11 | 7.69E-09 |
| regulation of cell proliferation (GO:0042127) | 400/741 | 1.67E-11 | 7.69E-09 |
| positive regulation of nucleic acid-templated transcription (GO:1903508) | 284/503 | 2.98E-11 | 1.26E-08 |
| negative regulation of transcription from RNA polymerase II promoter (GO:0000122) | 314/566 | 4.51E-11 | 1.76E-08 |
| regulation of cell migration (GO:0030334) | 190/317 | 7.61E-11 | 2.76E-08 |
| positive regulation of gene expression (GO:0010628) | 410/772 | 1.69E-10 | 5.73E-08 |
| axon guidance (GO:0007411) | 106/159 | 2.78E-10 | 8.82E-08 |
| negative regulation of cellular process (GO:0048523) | 295/535 | 4.21E-10 | 1.26E-07 |
| negative regulation of cell proliferation (GO:0008285) | 210/364 | 9.35E-10 | 2.64E-07 |
| negative regulation of programmed cell death (GO:0043069) | 232/409 | 1.08E-09 | 2.88E-07 |
| protein phosphorylation (GO:0006468) | 261/471 | 2.26E-09 | 5.62E-07 |
| negative regulation of gene expression (GO:0010629) | 332/619 | 2.32E-09 | 5.62E-07 |
| positive regulation of macromolecule metabolic process (GO:0010604) | 165/277 | 2.5E-09 | 5.78E-07 |
| transmembrane receptor protein tyrosine kinase signaling pathway (GO:0007169) | 224/397 | 3.91E-09 | 8.63E-07 |
| positive regulation of cell differentiation (GO:0045597) | 122/195 | 5.25E-09 | 1.11E-06 |
| activation of protein kinase activity (GO:0032147) | 142/234 | 5.85E-09 | 1.19E-06 |
| positive regulation of protein phosphorylation (GO:0001934) | 231/413 | 6.36E-09 | 1.24E-06 |
| central nervous system development (GO:0007417) | 133/218 | 1.1E-08 | 2.06E-06 |
| negative regulation of cellular macromolecule biosynthetic process (GO:2000113) | 278/513 | 1.29E-08 | 2.34E-06 |
| regulation of transcription, DNA-templated (GO:0006355) | 776/1599 | 2.65E-08 | 4.64E-06 |
| positive regulation of epithelial cell migration (GO:0010634) | 55/75 | 3.67E-08 | 6.13E-06 |
| cellular protein modification process (GO:0006464) | 504/1002 | 3.74E-08 | 6.13E-06 |
| positive regulation of cell migration (GO:0030335) | 133/222 | 5.27E-08 | 8.35E-06 |
| positive regulation of cell proliferation (GO:0008284) | 233/425 | 5.42E-08 | 8.35E-06 |
| neuron projection morphogenesis (GO:0048812) | 103/164 | 5.98E-08 | 8.94E-06 |
| positive regulation of multicellular organismal process (GO:0051240) | 123/203 | 6.72E-08 | 9.66E-06 |

### GO Molecular Function (8,045 genes): Top 30 of 81 significant records

| Term | Overlap | P-value | Adjusted P-value |
|---|---|---|---|
| cadherin binding (GO:0045296) | 191/314 | 1.1E-11 | 1.26E-08 |
| transcription regulatory region DNA binding (GO:0044212) | 216/375 | 6.56E-10 | 3.76E-07 |
| transcription factor activity, RNA polymerase II core promoter proximal region sequence-specific binding (GO:0000982) | 168/281 | 1.22E-09 | 4.66E-07 |
| RNA polymerase II transcription factor binding (GO:0001085) | 84/122 | 1.79E-09 | 5.11E-07 |
| transcription regulatory region sequence-specific DNA binding (GO:0000976) | 171/293 | 1.06E-08 | 2.43E-06 |
| transcriptional activator activity, RNA polymerase II transcription regulatory region sequence-specific binding (GO:0001228) | 166/285 | 2.1E-08 | 4.01E-06 |
| protein kinase binding (GO:0019901) | 268/496 | 3.35E-08 | 5.4E-06 |
| RNA polymerase II regulatory region sequence-specific DNA binding (GO:0000977) | 251/461 | 3.78E-08 | 5.4E-06 |
| protein kinase activity (GO:0004672) | 276/514 | 4.5E-08 | 5.72E-06 |
| transcriptional activator activity, RNA polymerase II core promoter proximal region sequence-specific binding (GO:0001077) | 109/176 | 7.35E-08 | 8.4E-06 |
| GTPase regulator activity (GO:0030695) | 154/276 | 2.44E-06 | 0.000254 |
| amyloid-beta binding (GO:0001540) | 37/50 | 4.47E-06 | 0.000426 |
| protein serine/threonine kinase activity (GO:0004674) | 197/369 | 5.99E-06 | 0.000518 |
| repressing transcription factor binding (GO:0070491) | 39/54 | 6.67E-06 | 0.000518 |
| RNA polymerase II regulatory region DNA binding (GO:0001012) | 116/202 | 6.8E-06 | 0.000518 |
| GTPase activator activity (GO:0005096) | 139/250 | 9.43E-06 | 0.000675 |
| voltage-gated cation channel activity (GO:0022843) | 64/101 | 1.2E-05 | 0.000807 |
| motor activity (GO:0003774) | 56/86 | 1.28E-05 | 0.000811 |
| regulatory region DNA binding (GO:0000975) | 126/225 | 1.52E-05 | 0.000916 |
| core promoter proximal region sequence-specific DNA binding (GO:0000987) | 152/279 | 1.65E-05 | 0.000943 |
| protein homodimerization activity (GO:0042803) | 332/665 | 1.79E-05 | 0.000973 |
| actin binding (GO:0003779) | 140/255 | 2.09E-05 | 0.001086 |
| PDZ domain binding (GO:0030165) | 43/63 | 2.29E-05 | 0.001141 |
| RNA polymerase II core promoter proximal region sequence-specific DNA binding (GO:0000978) | 143/263 | 3.29E-05 | 0.001569 |
| transcriptional repressor activity, RNA polymerase II transcription regulatory region sequence-specific binding (GO:0001227) | 92/160 | 5.43E-05 | 0.002452 |
| protein tyrosine kinase activity (GO:0004713) | 86/148 | 5.57E-05 | 0.002452 |
| microtubule motor activity (GO:0003777) | 41/61 | 6.16E-05 | 0.002608 |
| tubulin binding (GO:0015631) | 138/256 | 7.7E-05 | 0.003145 |
| microtubule binding (GO:0008017) | 109/196 | 8.13E-05 | 0.003208 |
| acetylgalactosaminyltransferase activity (GO:0008376) | 34/49 | 9.81E-05 | 0.003742 |

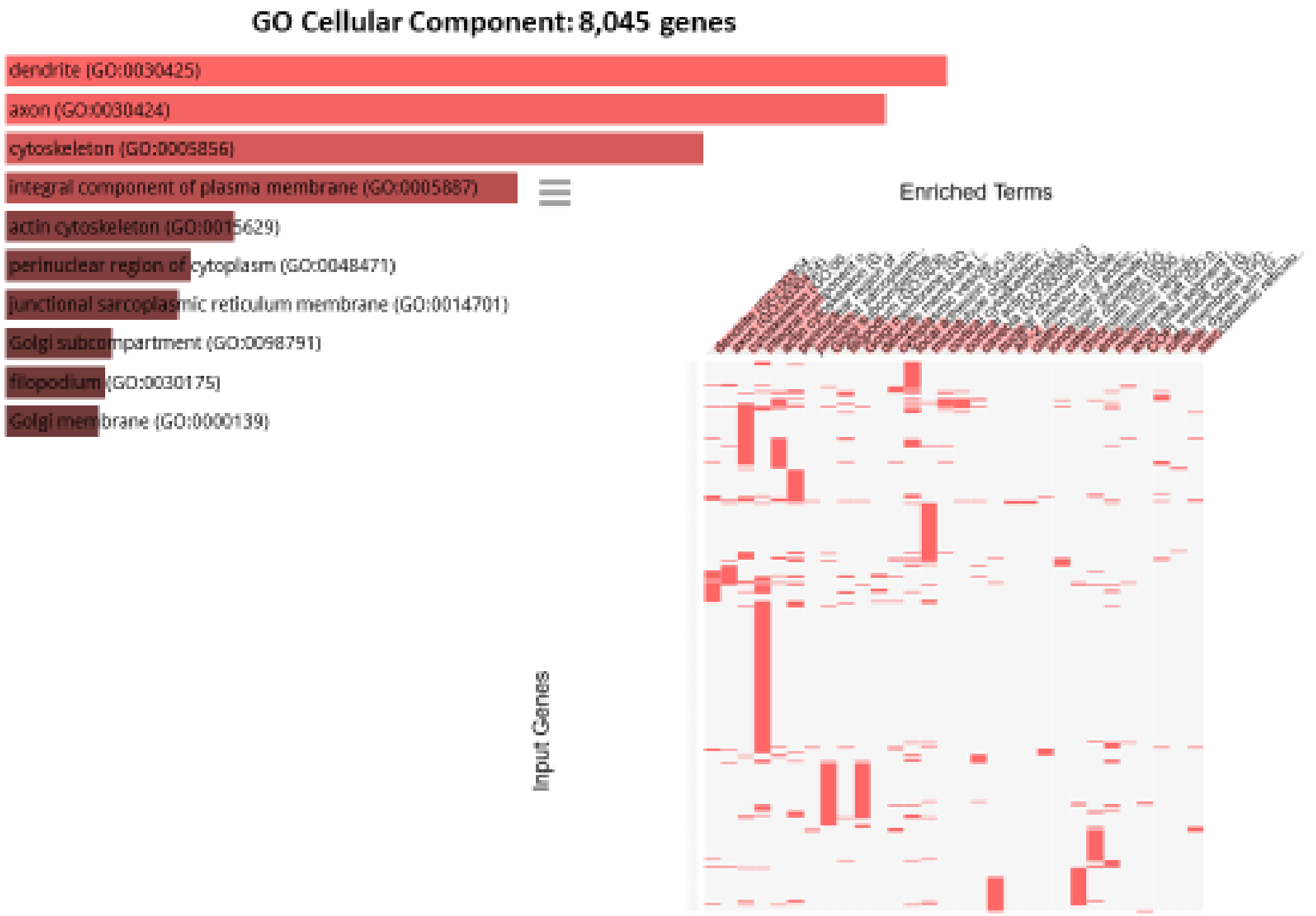

GO Cellular Component (8,045 genes): 29 significant records

| Term | Overlap | P-value | Adjusted P-value |
|---|---|---|---|
| cytoskeleton (GO:0005856) | 284/521 | 4.18E-09 | 1.85E-06 |
| dendrite (GO:0030425) | 131/216 | 2.35E-08 | 5.12E-06 |
| axon (GO:0030424) | 92/142 | 3.47E-08 | 5.12E-06 |
| actin cytoskeleton (GO:0015629) | 169/295 | 7.67E-08 | 8.23E-06 |
| integral component of plasma membrane (GO:0005887) | 711/1464 | 9.29E-08 | 8.23E-06 |
| focal adhesion (GO:0005925) | 194/357 | 1.52E-06 | 0.000112 |
| Golgi subcompartment (GO:0098791) | 250/480 | 4.46E-06 | 0.000282 |
| perinuclear region of cytoplasm (GO:0048471) | 198/379 | 3.32E-05 | 0.001839 |
| cytoplasmic vesicle membrane (GO:0030659) | 38/55 | 4.48E-05 | 0.002203 |
| Golgi membrane (GO:0000139) | 225/443 | 0.000104 | 0.004546 |
| cortical cytoskeleton (GO:0030863) | 36/53 | 0.000123 | 0.004546 |
| cortical actin cytoskeleton (GO:0030864) | 36/53 | 0.000123 | 0.004546 |
| caveola (GO:0005901) | 38/57 | 0.000148 | 0.00505 |
| filopodium (GO:0030175) | 40/61 | 0.000173 | 0.005461 |
| membrane raft (GO:0045121) | 70/120 | 0.000224 | 0.006622 |
| endoplasmic reticulum lumen (GO:0005788) | 142/271 | 0.000339 | 0.009382 |
| microtubule organizing center (GO:0005815) | 251/508 | 0.000401 | 0.010456 |
| cytoplasmic vesicle (GO:0031410) | 115/216 | 0.000546 | 0.013065 |
| chromatin (GO:0000785) | 153/297 | 0.00056 | 0.013065 |
| nuclear chromatin (GO:0000790) | 132/254 | 0.000826 | 0.018285 |
| catenin complex (GO:0016342) | 21/29 | 0.0009 | 0.018978 |
| ionotropic glutamate receptor complex (GO:0008328) | 27/40 | 0.001001 | 0.020165 |
| dendrite membrane (GO:0032590) | 16/21 | 0.001571 | 0.030261 |
| cation channel complex (GO:0034703) | 40/66 | 0.00177 | 0.032667 |
| centrosome (GO:0005813) | 225/462 | 0.00199 | 0.035259 |
| main axon (GO:0044304) | 23/34 | 0.002257 | 0.038461 |
| junctional sarcoplasmic reticulum membrane (GO:0014701) | 10-Sep | 0.002537 | 0.040494 |
| microtubule cytoskeleton (GO:0015630) | 191/389 | 0.002642 | 0.040494 |
| actin-based cell projection (GO:0098858) | 42/71 | 0.002651 | 0.040494 |
| ruffle membrane (GO:0032587) | 33/54 | 0.003598 | 0.053124 |

**Figure 7.** Gene ontology analyses of putative regulatory targets of genetic loci harboring human-specific SNCs.

**KEGG 2019 Human: 8,405 genes**

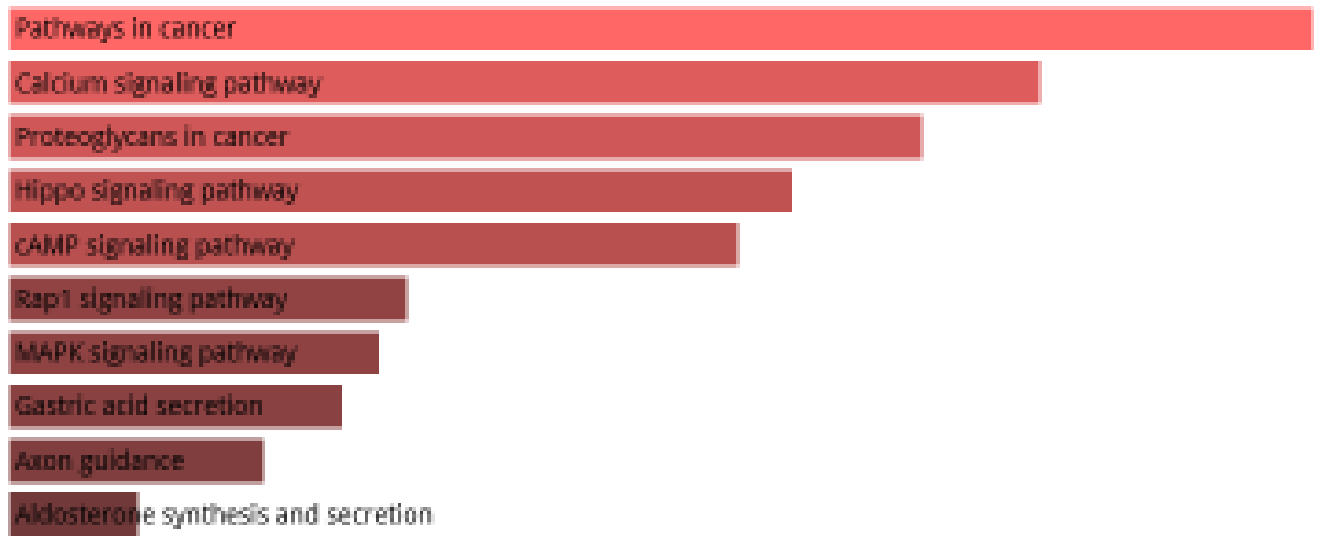

**KEGG 2019 Mouse: 8,405 genes**

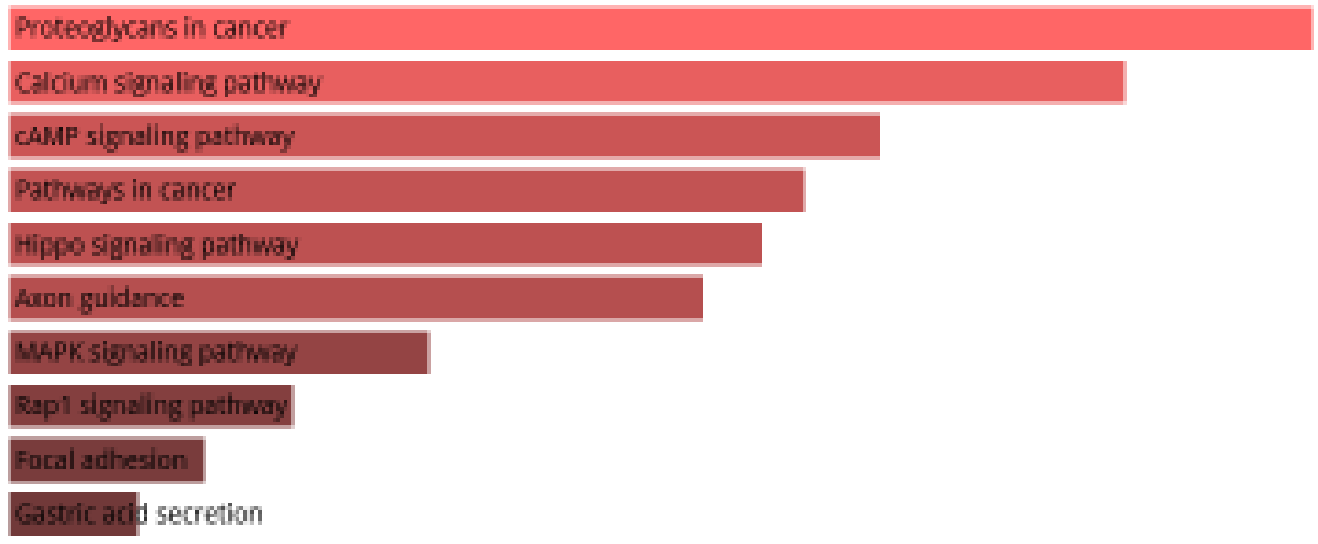

**KEGG 2019 Human: 8,405 genes**

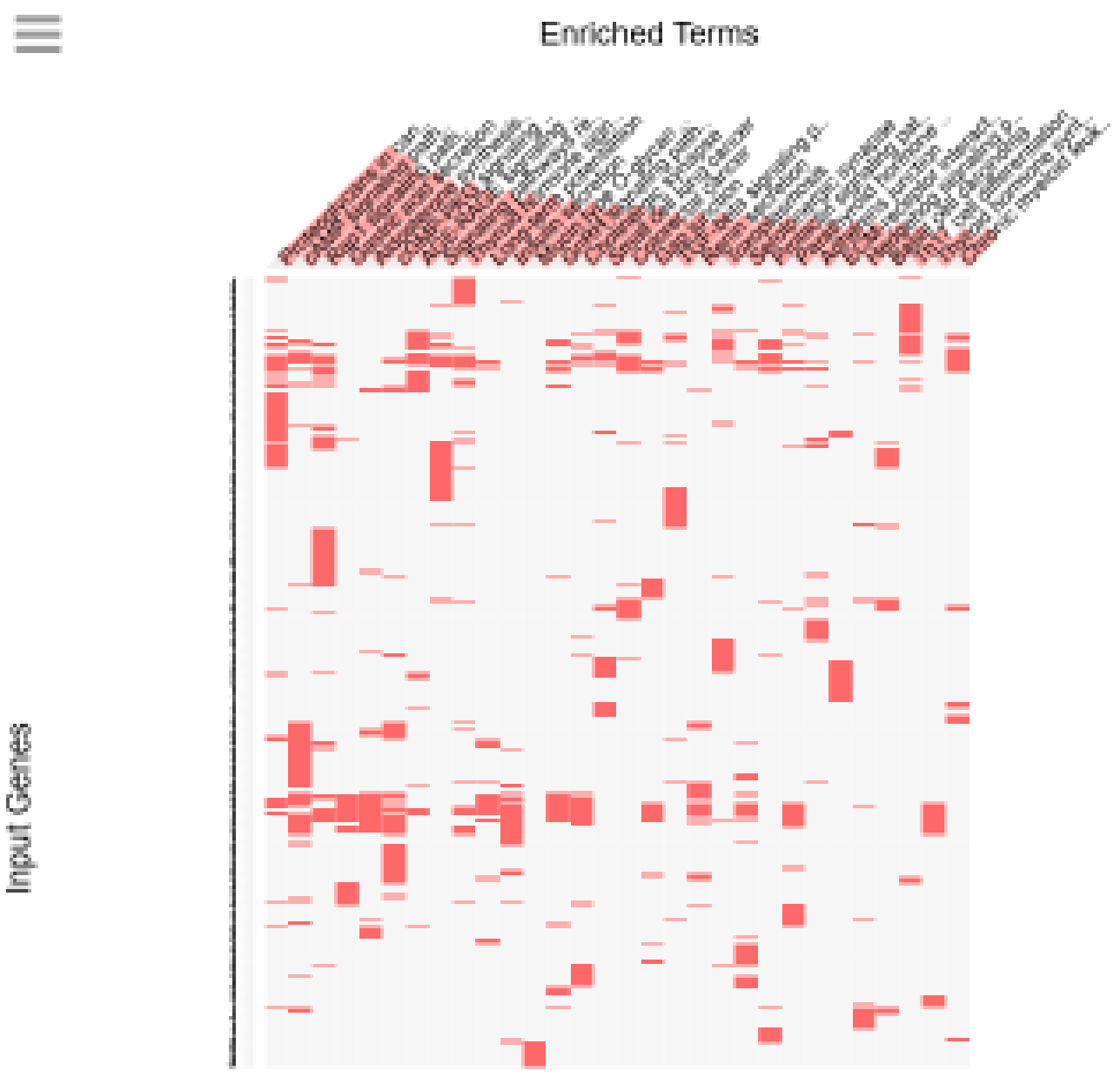

**KEGG 2019 Human (8,405 genes): Top 40 of 129 significant records**

| Term | Overlap | P-value | Adjusted P-value |
|---|---|---|---|
| Pathways in cancer | 312/530 | 2.04E-15 | 6.28E-13 |
| Calcium signaling pathway | 130/188 | 4.19E-14 | 6.45E-12 |
| Proteoglycans in cancer | 136/201 | 1.55E-13 | 1.59E-11 |
| Hippo signaling pathway | 112/160 | 6.61E-13 | 5.09E-11 |
| cAMP signaling pathway | 140/212 | 1.21E-12 | 7.48E-11 |
| Rap1 signaling pathway | 133/206 | 4.85E-11 | 2.49E-09 |
| MAPK signaling pathway | 179/295 | 6.48E-11 | 2.85E-09 |
| Gastric acid secretion | 59/75 | 9.75E-11 | 3.75E-09 |
| Axon guidance | 118/181 | 2.39E-10 | 8.17E-09 |
| Aldosterone synthesis and secretion | 71/98 | 9.54E-10 | 2.94E-08 |
| cGMP-PKG signaling pathway | 108/166 | 1.62E-09 | 4.54E-08 |
| Focal adhesion | 125/199 | 2.39E-09 | 6.14E-08 |
| AGE-RAGE signaling pathway in diabetic complications | 70/100 | 1.35E-08 | 3.13E-07 |
| Dopaminergic synapse | 87/131 | 1.42E-08 | 3.13E-07 |
| Signaling pathways regulating pluripotency of stem cells | 91/139 | 1.94E-08 | 3.98E-07 |
| Wnt signaling pathway | 101/158 | 2.15E-08 | 4.13E-07 |
| Adrenergic signaling in cardiomyocytes | 94/145 | 2.36E-08 | 4.28E-07 |
| Gastric cancer | 96/149 | 2.67E-08 | 4.56E-07 |
| Amoebiasis | 67/96 | 3.33E-08 | 5.32E-07 |
| Thyroid hormone signaling pathway | 78/116 | 3.45E-08 | 5.32E-07 |
| PI3K-Akt signaling pathway | 199/354 | 4.17E-08 | 6.11E-07 |
| Long-term potentiation | 50/67 | 5.95E-08 | 8.32E-07 |
| Melanogenesis | 69/101 | 8.04E-08 | 1.08E-06 |
| Breast cancer | 93/147 | 1.57E-07 | 2.02E-06 |
| Insulin secretion | 60/86 | 1.75E-07 | 2.16E-06 |
| Cushing syndrome | 97/155 | 1.84E-07 | 2.18E-06 |
| Circadian entrainment | 66/97 | 1.94E-07 | 2.22E-06 |
| Glutamatergic synapse | 75/114 | 2.48E-07 | 2.73E-06 |
| Mucin type O-glycan biosynthesis | 27/31 | 2.65E-07 | 2.81E-06 |
| Relaxin signaling pathway | 83/130 | 4.06E-07 | 4.17E-06 |
| GnRH signaling pathway | 63/93 | 4.68E-07 | 4.65E-06 |
| Cholinergic synapse | 73/112 | 6.18E-07 | 5.95E-06 |
| Oxytocin signaling pathway | 94/153 | 9.5E-07 | 8.87E-06 |
| Inflammatory mediator regulation of TRP channels | 66/100 | 1.08E-06 | 9.78E-06 |
| Small cell lung cancer | 62/93 | 1.35E-06 | 1.19E-05 |
| Human papillomavirus infection | 181/330 | 1.55E-06 | 1.31E-05 |
| Neuroactive ligand-receptor interaction | 184/338 | 2.41E-06 | 2.01E-05 |
| Colorectal cancer | 57/86 | 4.72E-06 | 3.82E-05 |
| Gap junction | 58/88 | 5.1E-06 | 4.03E-05 |
| TGF-beta signaling pathway | 59/90 | 5.49E-06 | 4.22E-05 |

**KEGG 2019 Mouse: 8,405 genes**

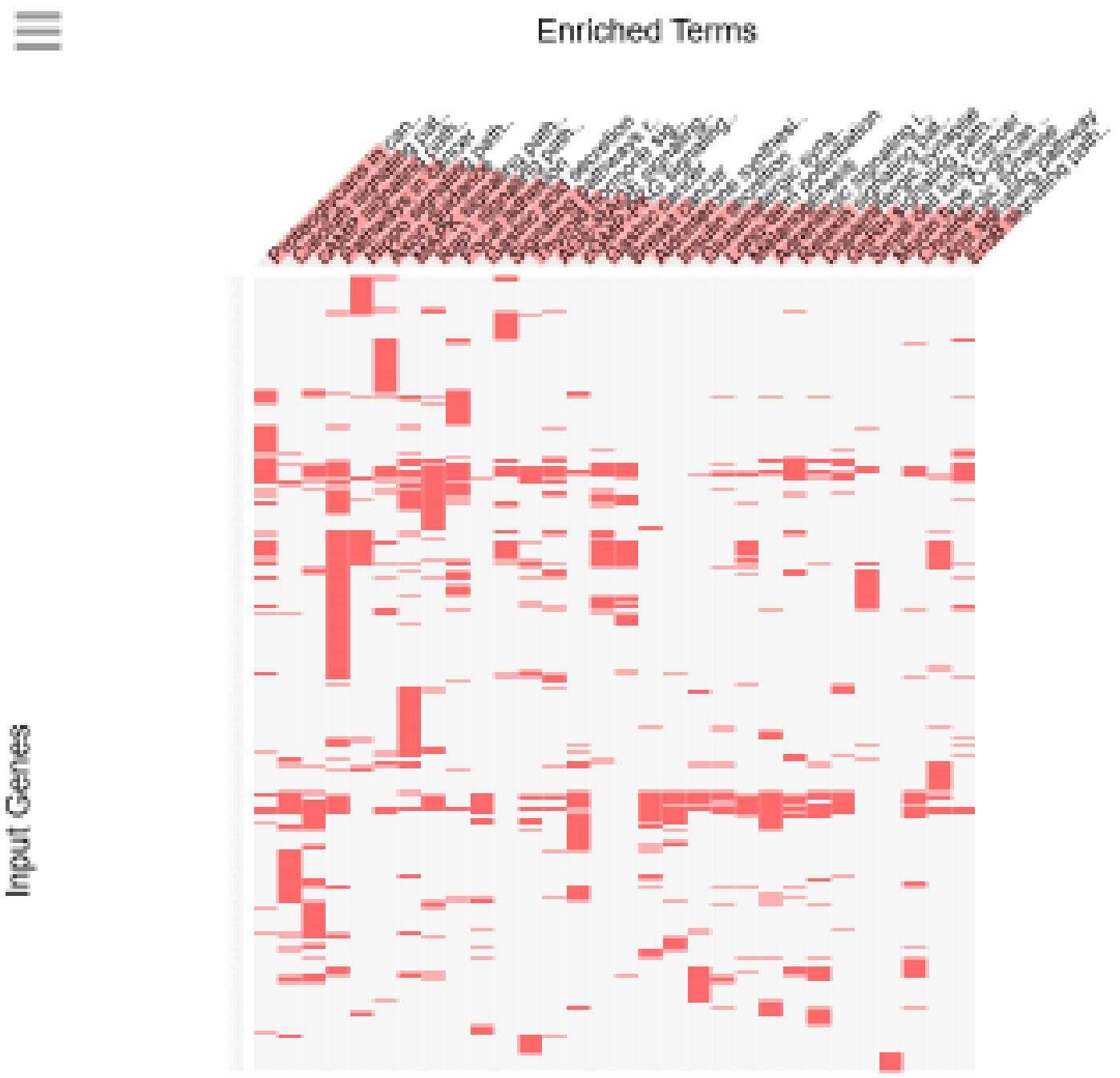

**KEGG 2019 Mouse (8,405 genes): Top 35 of 106 significant records**

| Term | Overlap | P-value | Adjusted P-value |
|---|---|---|---|
| Proteoglycans in cancer | 136/203 | 4.94E-13 | 1.5E-10 |
| Calcium signaling pathway | 127/189 | 2.09E-12 | 3.16E-10 |
| cAMP signaling pathway | 137/211 | 1.36E-11 | 1.38E-09 |
| Pathways in cancer | 300/535 | 2.41E-11 | 1.83E-09 |
| Hippo signaling pathway | 108/159 | 3.32E-11 | 2.01E-09 |
| Axon guidance | 119/180 | 5.22E-11 | 2.64E-09 |
| MAPK signaling pathway | 176/294 | 4.31E-10 | 1.86E-08 |
| Rap1 signaling pathway | 131/209 | 1.22E-09 | 4.62E-08 |
| Focal adhesion | 125/199 | 2.39E-09 | 8.05E-08 |
| Gastric acid secretion | 56/74 | 3.97E-09 | 1.2E-07 |
| Signaling pathways regulating pluripotency of stem cells | 90/137 | 1.8E-08 | 4.96E-07 |
| Thyroid hormone signaling pathway | 77/115 | 5.58E-08 | 1.41E-06 |
| AGE-RAGE signaling pathway in diabetic complications | 69/101 | 8.04E-08 | 1.88E-06 |
| cGMP-PKG signaling pathway | 106/172 | 1.58E-07 | 3.43E-06 |
| Gastric cancer | 94/150 | 2.59E-07 | 5.22E-06 |
| Breast cancer | 92/147 | 3.76E-07 | 7.12E-06 |
| Aldosterone synthesis and secretion | 68/102 | 4.22E-07 | 7.52E-06 |
| Insulin secretion | 59/86 | 5.53E-07 | 9.32E-06 |
| Glutamatergic synapse | 74/114 | 6.6E-07 | 1.05E-05 |
| Long-term potentiation | 48/67 | 8.6E-07 | 1.3E-05 |
| Melanogenesis | 66/100 | 1.08E-06 | 1.56E-05 |
| Adrenergic signaling in cardiomyocytes | 91/148 | 1.34E-06 | 1.84E-05 |
| PI3K-Akt signaling pathway | 194/357 | 1.52E-06 | 1.98E-05 |
| Relaxin signaling pathway | 82/131 | 1.57E-06 | 1.98E-05 |
| Dopaminergic synapse | 84/135 | 1.68E-06 | 2.04E-05 |
| GnRH signaling pathway | 60/90 | 1.99E-06 | 2.32E-05 |
| Small cell lung cancer | 61/92 | 2.17E-06 | 2.43E-05 |
| Mucin type O-glycan biosynthesis | 24/28 | 2.36E-06 | 2.56E-05 |
| Cholinergic synapse | 72/113 | 2.6E-06 | 2.72E-05 |
| Wnt signaling pathway | 96/160 | 3.27E-06 | 3.3E-05 |
| ErbB signaling pathway | 56/84 | 4.34E-06 | 4.24E-05 |
| Circadian entrainment | 64/99 | 4.53E-06 | 4.29E-05 |
| Oxytocin signaling pathway | 92/154 | 6.68E-06 | 6.13E-05 |
| Cushing syndrome | 94/159 | 9.8E-06 | 8.74E-05 |
| Gap junction | 56/86 | 1.28E-05 | 0.000111 |

**Figure 8.** KEGG analyses of putative regulatory targets of genetic loci harboring human-specific SNCs.

**Association with networks of human-specific regulatory sequences (HSGRS) and stem cell-associated retroviral sequences (SCARS) of 8,405 genes associated with 35,074 fixed human-specific single nucleotide changes located in differentially-accessible chromatin regions during human neurogenesis in cerebral organoids**

| Classification category | Number of genes | Perent |
|---|---|---|
| Unique genes | 8405 | 100.00 |
| In network of human-specific genomic regulatory sequences (HSGRS) | 7406 | 88.11 |
| LTR5_Hs/SVA_D enhancers-regulated genes | 1387 | 16.50 |
| HERVH lncRNA-regulated genes | 3191 | 37.97 |
| LTR7Y/B enhancers-regulated genes | 3306 | 39.33 |
| In network of stem cell-associated retroviral sequences (SCARS) | 4029 | 47.94 |
| Both HSRGS & SCARS-regulated genes | 3602 | 42.86 |
| All HSGRS & SCARS-regulated genes | 7833 | 93.19 |

**Effect of CRISPR-guided epigentic silencing of LRT5_Hs/SVA_D enhancers on expression of 1,387 genes associated with human-specific SNC in DA chromatin regions**

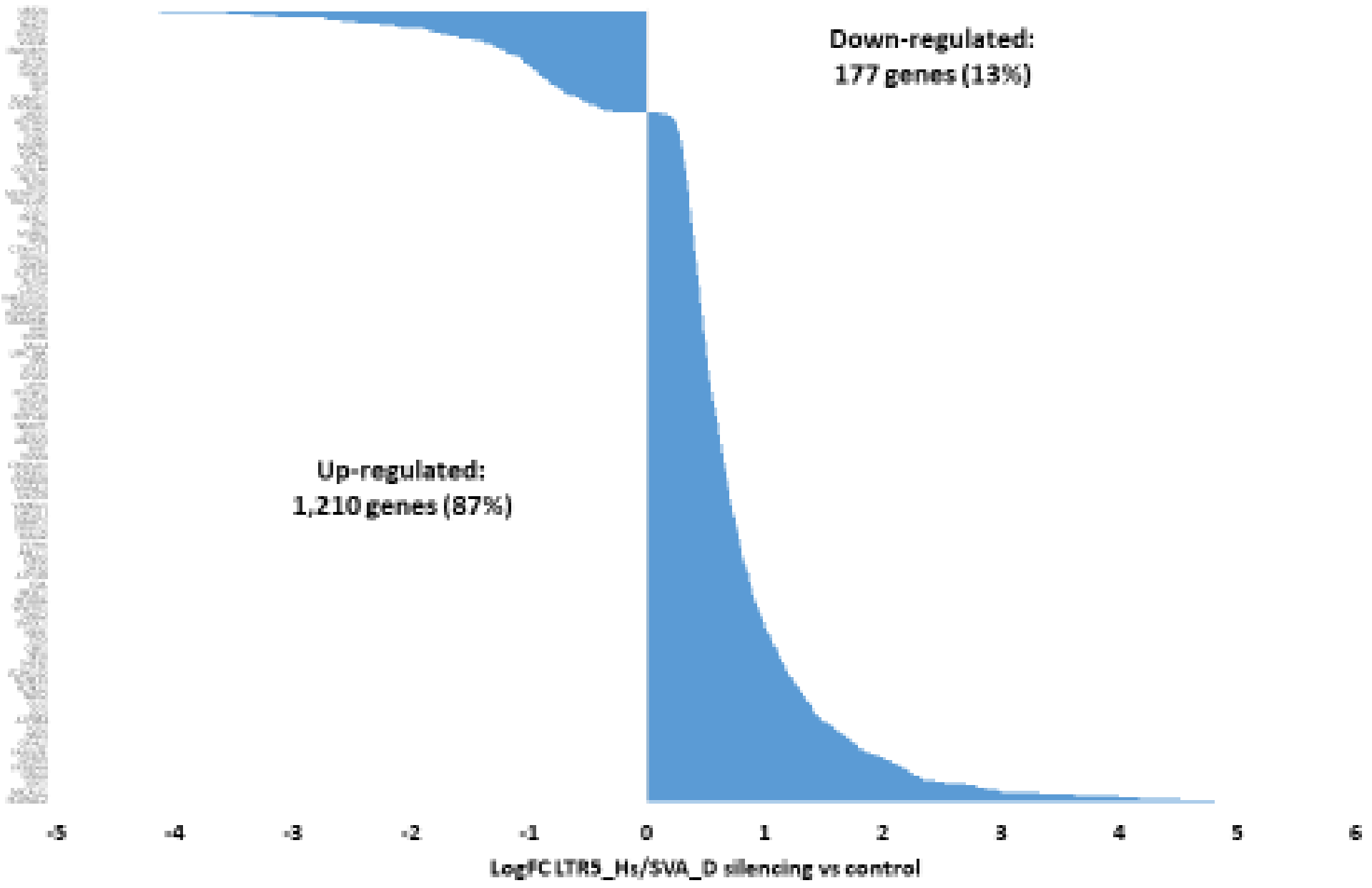

Effects of stem cell-associated regulatory sequences (SCARS) on expression of 4,029 genes associated with human-specific SNCs located in DA chromatin regions

| Classification category | Number of genes | Down-regulated | Percent | Up-regulated | Percent |
|---|---|---|---|---|---|
| LTR5_Hs/SVA_D enhancers-regulated genes | 1387 | 1210 | 87.24 | 177 | 12.76 |
| HERVH lncRNA-regulated genes | 3191 | 1733 | 54.31 | 1458 | 45.69 |
| LTR7Y/B enhancers-regulated genes | 3306 | 2494 | 75.44 | 812 | 24.56 |

**Figure 9.** Structurally, functionally, and evolutionary distinct classes of HSRS share the relatively restricted elite set of common genetic targets.

MGI Mammalian Phenotype 2017: 8,405 genes

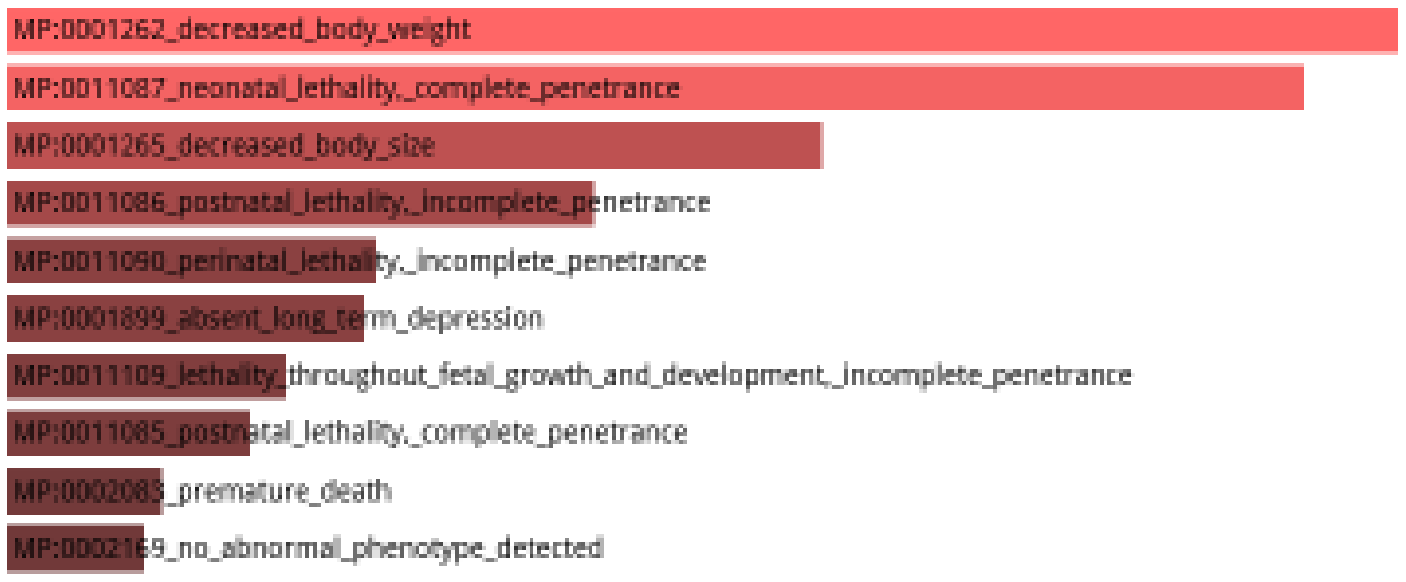

MGI Mammalian Phenotype 2017: 8,405 genes

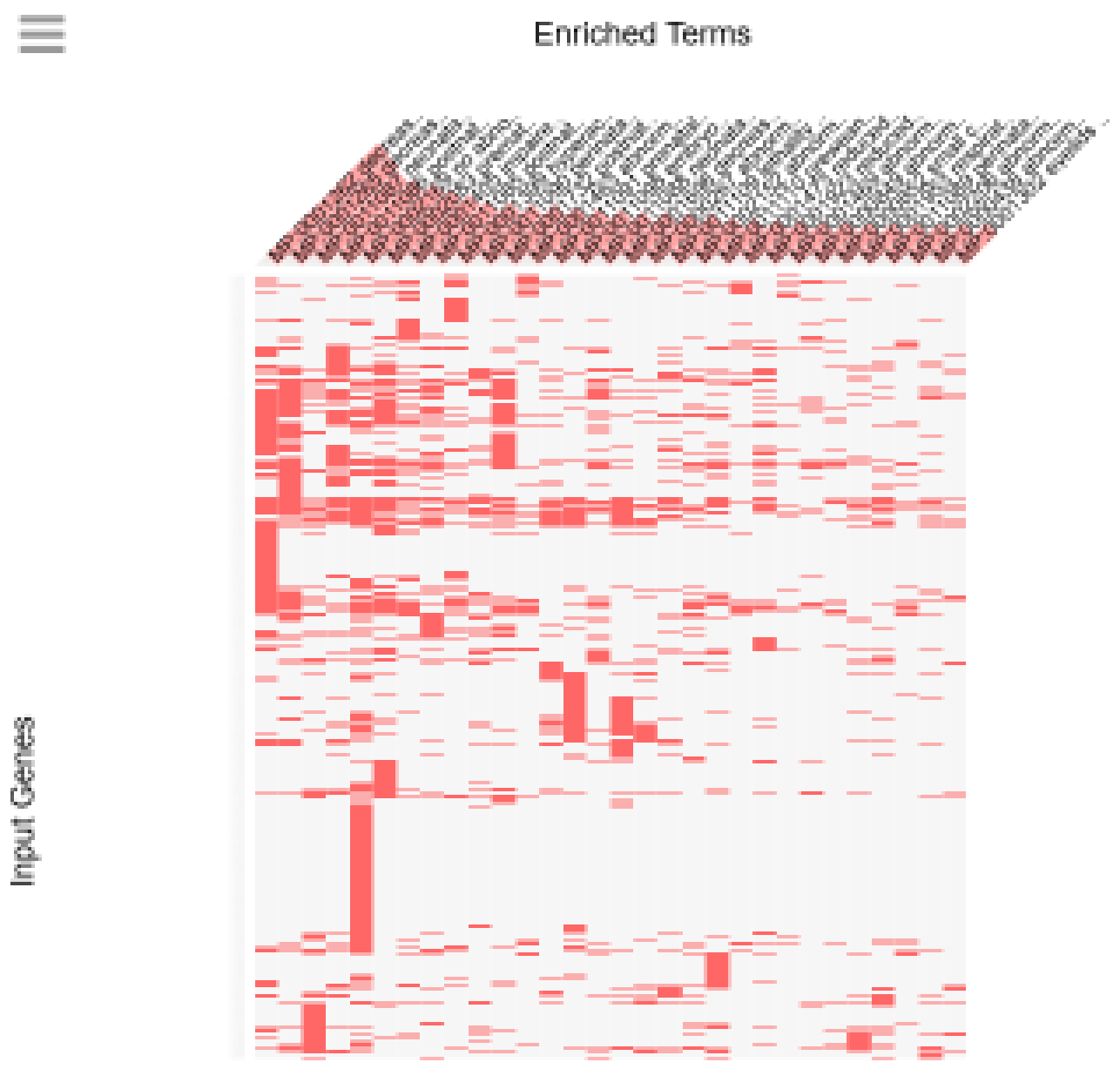

## MGI Mammalian Phenotype 2017 (8,405 genes): Top 40 of 749 significant records

| Term | Overlap | P-value | Adjusted P-value |
|---|---|---|---|
| MP:0001262_decreased_body_weight | 692/1189 | 4.71E-31 | 2.44E-27 |
| MP:0001265_decreased_body_size | 472/774 | 2.25E-27 | 5.82E-24 |
| MP:0011087_neonatal_lethality,_complete_penetrance | 299/462 | 2.47E-23 | 4.25E-20 |
| MP:0011086_postnatal_lethality,_incomplete_penetrance | 346/563 | 4.06E-21 | 5.26E-18 |
| MP:0002169_no_abnormal_phenotype_detected | 882/1674 | 2.98E-20 | 3.08E-17 |
| MP:0002083_premature_death | 474/834 | 1.12E-18 | 9.65E-16 |
| MP:0001405_impaired_coordination | 215/332 | 3.43E-17 | 2.54E-14 |
| MP:0011085_postnatal_lethality,_complete_penetrance | 238/384 | 1.57E-15 | 1.02E-12 |
| MP:0001399_hyperactivity | 216/344 | 4.36E-15 | 2.51E-12 |
| MP:0011090_perinatal_lethality,_incomplete_penetrance | 152/226 | 1.28E-14 | 6.65E-12 |
| MP:0001732_postnatal_growth_retardation | 359/590 | 1.43E-14 | 6.75E-12 |
| MP:0001463_abnormal_spatial_learning | 115/162 | 6.87E-14 | 2.96E-11 |
| MP:0011091_prenatal_lethality,_complete_penetrance | 174/272 | 1.81E-13 | 7.22E-11 |
| MP:0011098_embryonic_lethality_during_organogenesis,_complete_penetrance | 319/559 | 2.9E-13 | 1.07E-10 |
| MP:0011088_neonatal_lethality,_incomplete_penetrance | 163/255 | 1.15E-12 | 3.97E-10 |
| MP:0001698_decreased_embryo_size | 273/472 | 1.96E-12 | 6.36E-10 |
| MP:0000267_abnormal_heart_development | 109/157 | 3.11E-12 | 9.48E-10 |
| MP:0011109_lethality_throughout_fetal_growth_and_development,_incomplete_penetrance | 116/170 | 3.92E-12 | 1.13E-09 |
| MP:0002152_abnormal_brain_morphology | 104/152 | 3.9E-11 | 1.06E-08 |
| MP:0011110_preweaning_lethality,_incomplete_penetrance | 222/381 | 8.85E-11 | 2.29E-08 |
| MP:0001473_reduced_long_term_potentiation | 77/106 | 1.52E-10 | 3.75E-08 |
| MP:0001923_reduced_female_fertility | 142/227 | 3.06E-10 | 7.2E-08 |
| MP:0004811_abnormal_neuron_physiology | 67/90 | 4.06E-10 | 8.85E-08 |
| MP:0000788_abnormal_cerebral_cortex_morphology | 98/145 | 4.1E-10 | 8.85E-08 |
| MP:0001406_abnormal_gait | 180/302 | 4.48E-10 | 9.28E-08 |
| MP:0001469_abnormal_contextual_conditioning_behavior | 45/54 | 4.96E-10 | 9.82E-08 |
| MP:0000849_abnormal_cerebellum_morphology | 68/92 | 5.12E-10 | 9.82E-08 |
| MP:0000266_abnormal_heart_morphology | 139/224 | 1.03E-09 | 1.91E-07 |
| MP:0000852_small_cerebellum | 52/66 | 1.15E-09 | 2.05E-07 |
| MP:0001953_respiratory_failure | 98/147 | 1.29E-09 | 2.23E-07 |
| MP:0011089_perinatal_lethality,_complete_penetrance | 135/217 | 1.41E-09 | 2.35E-07 |
| MP:0002206_abnormal_CNS_synaptic_transmission | 66/90 | 1.8E-09 | 2.59E-07 |
| MP:0000438_abnormal_cranium_morphology | 90/133 | 1.91E-09 | 2.99E-07 |
| MP:0001302_eyelids_open_at_birth | 43/52 | 1.97E-09 | 3E-07 |
| MP:0000807_abnormal_hippocampus_morphology | 63/86 | 4.04E-09 | 5.97E-07 |
| MP:0006009_abnormal_neuronal_migration | 57/76 | 5.14E-09 | 7.28E-07 |
| MP:0001954_respiratory_distress | 111/174 | 5.2E-09 | 7.28E-07 |
| MP:0001899_absent_long_term_depression | 25/26 | 5.88E-09 | 7.95E-07 |
| MP:0002906_increased_susceptibility_to_pharmacologically_induced_seizures | 65/90 | 5.99E-09 | 7.95E-07 |
| MP:0003633_abnormal_nervous_system_physiology | 74/106 | 6.33E-09 | 8.12E-07 |

## MGI Mammalian Phenotype Level 4 2019: 8.405 genes

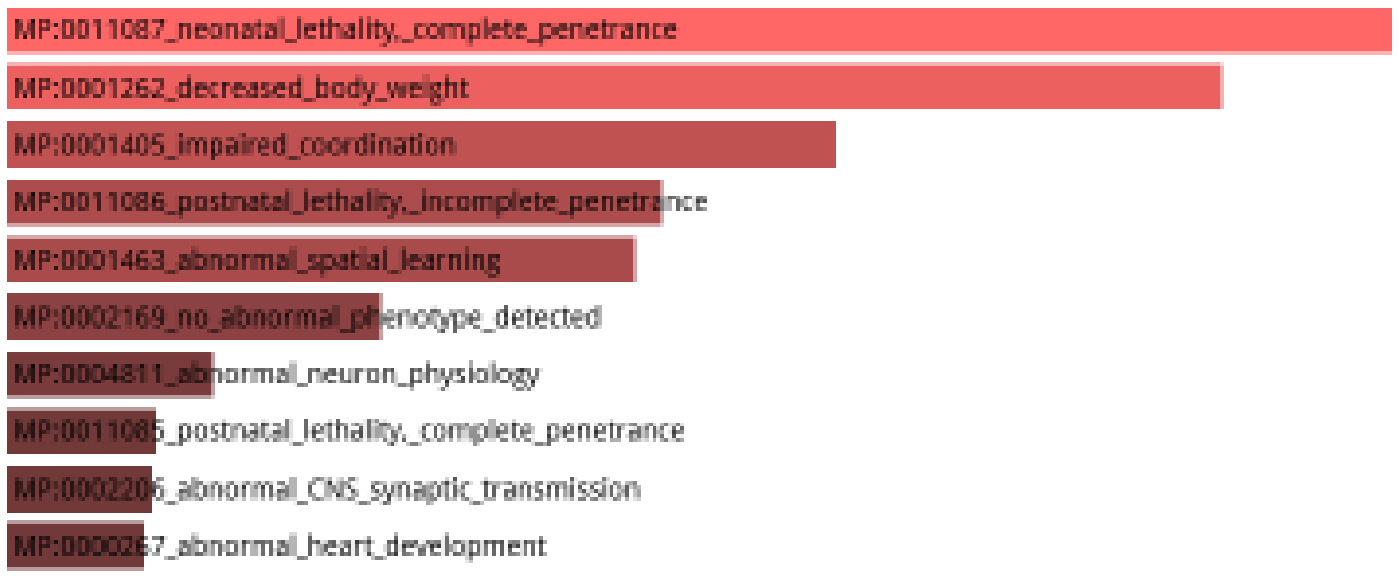

MGI Mammalian Phenotype Level 4 2019: 8.405 genes

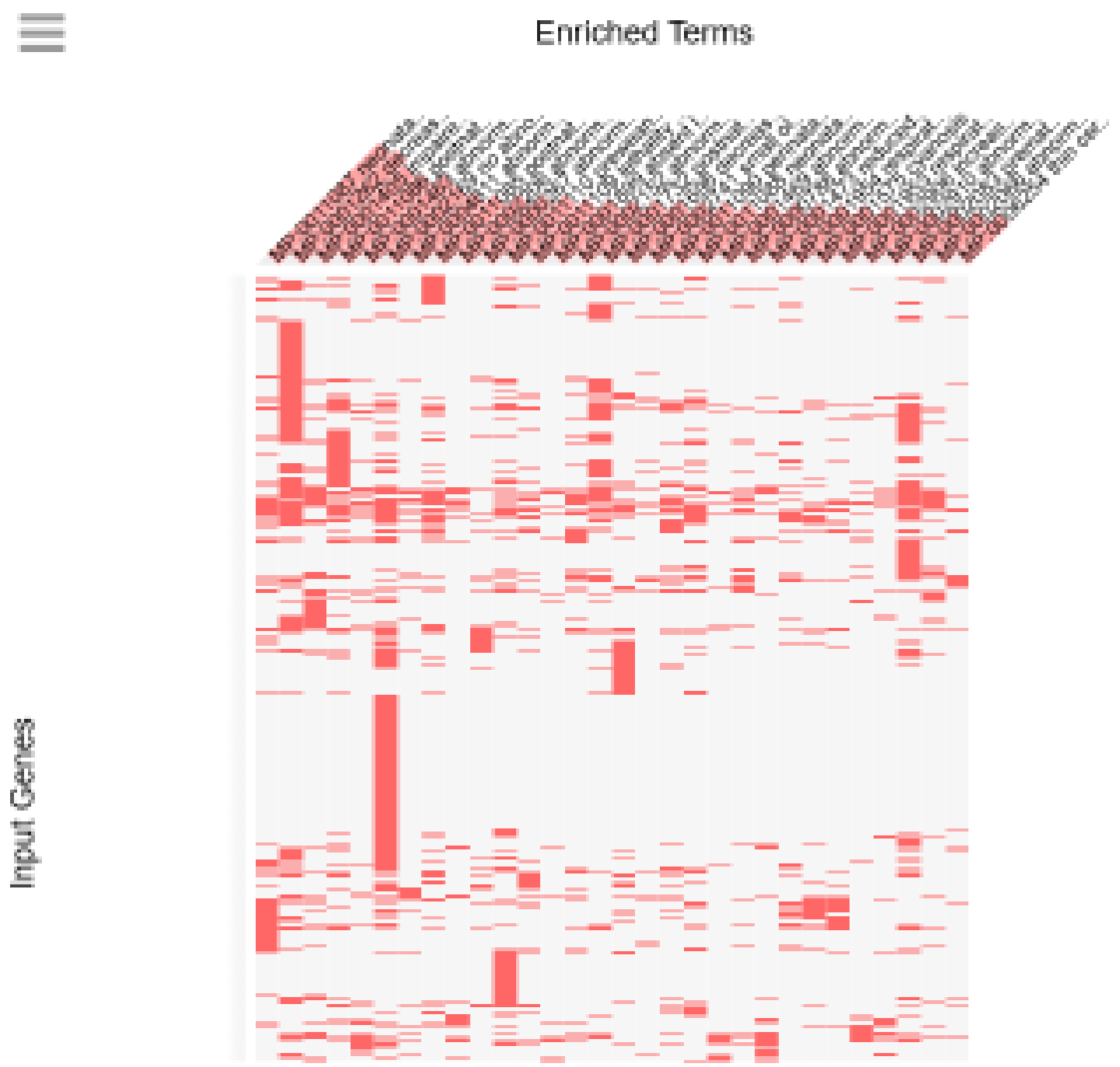

MGI Mammalian Phenotype Level 4 2019 (8,405 genes): top 40 of 407 significant records

| Term | Overlap | P-value | Adjusted P-value |
|---|---|---|---|
| MP:0011087_neonatal_lethality,_complete_penetrance | 315/517 | 1.55E-18 | 8.15E-15 |
| MP:0001262_decreased_body_weight | 773/1471 | 2.04E-17 | 5.36E-14 |
| MP:0001405_impaired_coordination | 247/405 | 6.90E-15 | 1.21E-11 |
| MP:0011086_postnatal_lethality,_incomplete_penetrance | 362/643 | 9.65E-14 | 1.27E-10 |
| MP:0001463_abnormal_spatial_learning | 120/172 | 1.46E-13 | 1.54E-10 |
| MP:0002169_no_abnormal_phenotype_detected | 958/1944 | 6.87E-12 | 6.02E-09 |
| MP:0004811_abnormal_neuron_physiology | 78/107 | 8.69E-11 | 6.53E-08 |
| MP:0002206_abnormal_CNS_synaptic_transmission | 75/103 | 2.18E-10 | 1.27E-07 |
| MP:0000267_abnormal_heart_development | 111/168 | 2.43E-10 | 1.28E-07 |
| MP:0011085_postnatal_lethality,_complete_penetrance | 246/432 | 2.04E-10 | 1.34E-07 |
| MP:0011110_preweaning_lethality,_incomplete_penetrance | 360/669 | 2.90E-10 | 1.39E-07 |
| MP:0011109_lethality_throughout_fetal_growth_and_development,_incomplete_penetrance | 122/191 | 8.21E-10 | 3.60E-07 |
| MP:0001899_absent_long_term_depression | 27/28 | 1.11E-09 | 4.50E-07 |
| MP:0001732_postnatal_growth_retardation | 360/677 | 1.85E-09 | 6.48E-07 |
| MP:0001698_decreased_embryo_size | 293/537 | 2.09E-09 | 6.88E-07 |
| MP:0002152_abnormal_brain_morphology | 119/187 | 1.84E-09 | 6.92E-07 |
| MP:0011090_perinatal_lethality,_incomplete_penetrance | 154/256 | 3.23E-09 | 9.44E-07 |
| MP:0002741_small_olfactory_bulb | 35/40 | 3.07E-09 | 9.51E-07 |
| MP:0001469_abnormal_contextual_conditioning_behavior | 48/61 | 5.45E-09 | 1.43E-06 |
| MP:0001473_reduced_long_term_potentiation | 84/124 | 6.05E-09 | 1.45E-06 |
| MP:0000788_abnormal_cerebral_cortex_morphology | 104/161 | 5.79E-09 | 1.45E-06 |
| MP:0011088_neonatal_lethality,_incomplete_penetrance | 172/293 | 5.33E-09 | 1.48E-06 |
| MP:0001954_respiratory_distress | 121/194 | 7.90E-09 | 1.81E-06 |
| MP:0001575_cyanosis | 129/210 | 1.00E-08 | 2.19E-06 |
| MP:0001953_respiratory_failure | 103/161 | 1.47E-08 | 3.09E-06 |
| MP:0002906_increased_susceptibility_to_pharmacologically_induced_seizures | 70/101 | 2.60E-08 | 5.25E-06 |
| MP:0002910_abnormal_excitatory_postsynaptic_currents | 59/83 | 7.68E-08 | 1.50E-05 |
| MP:0002083_premature_death | 499/997 | 9.80E-08 | 1.84E-05 |
| MP:0002066_abnormal_motor_capabilities/coordination/movement | 102/164 | 1.41E-07 | 2.55E-05 |
| MP:0006254_thin_cerebral_cortex | 53/74 | 2.36E-07 | 4.14E-05 |
| MP:0011098_embryonic_lethality_during_organogenesis,_complete_penetrance | 339/656 | 2.60E-07 | 4.28E-05 |
| MP:0000807_abnormal_hippocampus_morphology | 65/92 | 2.56E-07 | 4.34E-05 |
| MP:0006009_abnormal_neuronal_migration | 59/85 | 2.94E-07 | 4.69E-05 |
| MP:0011108_embryonic_lethality_during_organogenesis,_incomplete_penetrance | 143/247 | 3.20E-07 | 4.95E-05 |
| MP:0000031_abnormal_cochlea_morphology | 44/59 | 3.84E-07 | 5.77E-05 |
| MP:0000852_small_cerebellum | 58/84 | 4.91E-07 | 7.17E-05 |
| MP:0002063_abnormal_learning/memory/conditioning | 42/56 | 5.43E-07 | 7.52E-05 |
| MP:0003633_abnormal_nervous_system_physiology | 79/123 | 5.36E-07 | 7.62E-05 |
| MP:0010025_decreased_total_body_fat_amount | 263/498 | 5.94E-07 | 8.01E-05 |
| MP:0009937_abnormal_neuron_differentiation | 75/116 | 7.01E-07 | 9.22E-05 |

**Figure 10.** Interrogation of MGI Mammalian Phenotype databases identifies genes associated with human-specific SNCs and implicated in premature death and embryonic, perinatal, neonatal, and postnatal lethality phenotypes.

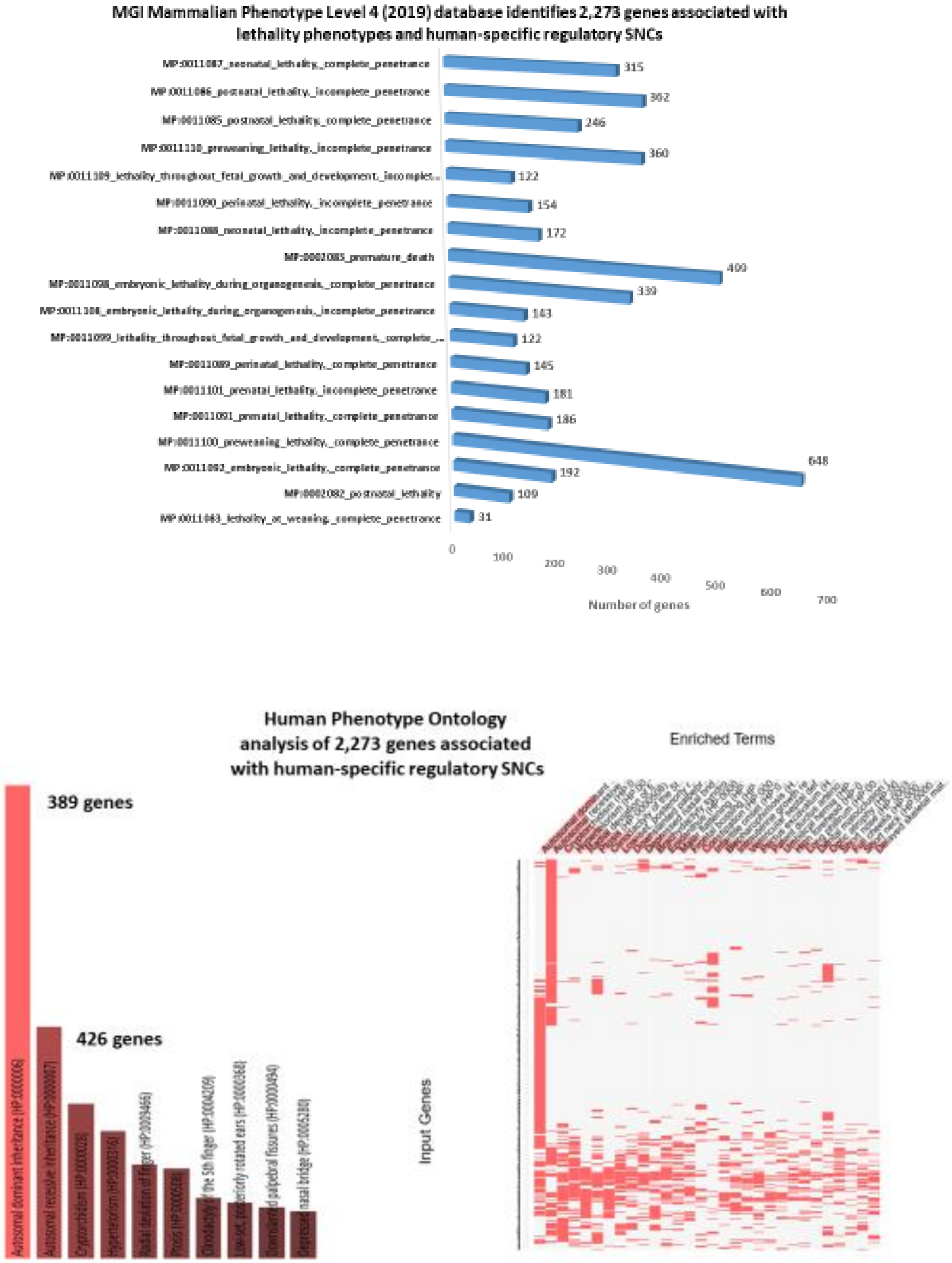

**Figure 11.** Identification of 2,273 genes (offspring survival genes) associated with human-specific SNCs and implicated in premature death and embryonic, perinatal, neonatal, and postnatal lethality phenotypes. Bottom figure shows identification of offspring survival genes that were implicated in the autosomal dominant (389 genes) and recessive (426 genes) inheritance in Modern Humans.